%% file: ie-gtb.tex
\documentclass{article}

\usepackage[margin=1in]{geometry}
\usepackage{setspace}
\usepackage{amsmath,amsfonts,amssymb,amscd,xspace}
\usepackage{subcaption}
\usepackage[export]{adjustbox}
\usepackage{graphicx}
\usepackage[english]{babel}
\usepackage{xcolor}
\usepackage[utf8]{inputenc} 
\usepackage{rotfloat}
\usepackage[authoryear]{natbib}
\usepackage{booktabs}

\usepackage{bibunits}
\defaultbibliographystyle{chicago}
\defaultbibliography{ie-gtb}

\usepackage{tikz}
\usetikzlibrary{positioning, arrows,shapes,arrows.meta, calc}

\usepackage[pdfpagemode={UseOutlines},bookmarks=true,bookmarksopen=true,
bookmarksopenlevel=0,bookmarksnumbered=true,hypertexnames=false,
colorlinks,linkcolor={blue},citecolor={blue},urlcolor={red},
pdfstartview={FitV},unicode,breaklinks=true]{hyperref}

\floatstyle{ruled}
\newfloat{algorithm}{tbp}{loa}
\providecommand{\algorithmname}{Algorithm}
\floatname{algorithm}{\protect\algorithmname}

\newcommand{\loss}{\ensuremath{l}}
\newcommand{\response}{\ensuremath{y}}
\newcommand{\responseindep}{\ensuremath{y^0}}

\newcommand{\features}{\ensuremath{\mathbf{x}}}
\newcommand{\featuresindep}{\ensuremath{\mathbf{x}^0}}

\newcommand{\pred}{\ensuremath{\hat{y}}}
\newcommand{\data}{\ensuremath{\mathcal{D}}}
\newcommand{\E}{\ensuremath{E}}
\newcommand{\Var}{\ensuremath{Var}}
\newcommand{\Cov}{\ensuremath{Cov}}
\newcommand{\tr}{\mathrm{tr}}

\newcommand{\setleafnodes}{\ensuremath{\mathcal{L}}}
\newcommand\norm[1]{\left\lVert#1\right\rVert}

\usepackage{mathtools}

\DeclarePairedDelimiter{\nint}\lfloor\rceil

\begin{document}
	
	\title{An information criterion for automatic gradient tree boosting\\
	}
	
	\author{Berent Å. S. Lunde \footnote{
			Department of Mathematics and Physics, University of Stavanger, 4036 Stavanger, Norway.
			Tel.: +47-47258605.
			\href{mailto:berent.a.lunde@uis.no}{berent.a.lunde@uis.no}  }
		\and
		Tore S. Kleppe \footnote{
			Department of Mathematics and Physics, University of Stavanger, 4036 Stavanger, Norway
		}
		\and
		Hans J. Skaug \footnote{ 
			Department of Mathematics, University of Bergen, 5020 Bergen, Norway
		}
	}
	
	\maketitle
	
	\begin{abstract}
		
		An information theoretic approach to learning the complexity of classification and regression trees and the number of trees in gradient tree boosting is proposed.
		The optimism (test loss minus training loss) of the greedy leaf splitting procedure is shown to be the maximum of a Cox-Ingersoll-Ross process, from which a generalization-error based information criterion is formed. The proposed procedure allows fast local model selection without cross validation based hyper parameter tuning, and hence efficient and automatic comparison among the large number of models performed during each boosting iteration.
		Relative to \texttt{xgboost}, speedups on numerical experiments ranges from around 10 to about 1400, at similar predictive-power measured in terms of test-loss.
	\end{abstract}
	\begin{bibunit}
	\input{Introduction.tex}
	\input{Background.tex}	
	\input{Information_theoretic_approach.tex}
	\input{Simulation_experiments.tex}	
	\input{Benchmark_datasets.tex}	
	\input{Discussion.tex}

	\putbib
	\end{bibunit}
	\begin{bibunit}
	\input{Appendix.tex}

	\putbib
	\end{bibunit}
	
	
\end{document}

%% file: Introduction.tex
\section{Introduction}\label{sec:introduction}

This article is motivated by the problem of selecting the functional form of trees and ensemble size in gradient tree boosting \citep{friedman2001greedy, mason2000boosting}.
Gradient tree boosting (GTB) has become extremely popular in recent years, both in academia and industry:
At present, an increase in the size of datasets, both in the number of observations and the richness of the data, or number of features, is seen.
This, coupled with an exponential increase in computational power and a growing revelation and acceptance for data-driven decisions in the industry makes for an increasing interest in statistical learning \citep{friedman2001elements}.
For these new datasets, standard statistical methods such as generalized linear models \citep{mccullagh1989generalized} that have a fixed learning rate due to their constrained functional form with bounded complexity, struggle in terms of predictive power, as they stop learning at certain information thresholds.
The interest is therefore geared towards more flexible approaches such as ensembles of learners.

GTB has recently risen to prominence for structured or tabular data, and previous to this, the related random forest algorithm \citep{ho1995random, breiman2001random} was the ``off-the-shelf'' machine learning algorithm of choice for many practitioners.
They both perform automatic variable selection, there is a natural measure of feature importance, they are easy to combine, and simple decision trees are often easy to interpret.
In fact, gradient tree boosting has dominated in machine-learning competitions for structured data since around 2014 when the \texttt{xgboost} implementation \citep{chen2018xgbpackage, chen2016xgboost} was made popular.
Recent years have seen the introduction of rivalling implementations such as \texttt{LightGBM} \citep{ke2017lightgbm} and \texttt{CatBoost} \citep{dorogush2018catboost}. 

A difficulty with GTB is that it is prone to overfitting: The functional form changes for every split in a tree, and for every tree that is added.
Hence, it is necessary to constrain the ensemble size and the complexity
of each individual tree.
Standard practice is either the use of a validation set, cross-validation \citep{stone1974cross}, or regularization to target a bias-variance trade-off \citep{friedman2001elements}.
\cite{friedman2001greedy} suggested a constant penalisation of each split, while later implementations have also introduced L2 and L1 regularisation.
All the above mentioned GTB implementations have many hyper-parameters, which must be tuned in a computationally expensive manner, typically involving cross-validation. We will collectively view these measures to avoid overfitting as solutions to a model selection problem.	

In this article we take an information theoretic approach to GTB model selection, as an alternative to cross-validation.
Building on the seminal work of \citet{akaike1974new} and \citet{takeuchi1976distribution} we approximate the difference between
test and training error for each split in the tree growing process.
This difference, known as the ``optimism'' \citep{friedman2001elements}, is used to formulate
new stopping criteria in the GTB algorithm, both for tree growing
and for the boosting algorithm itself.
The resulting algorithm selects its model complexity in a single run, and does not require manual tuning.
We show that it is considerable faster than existing GTB implementations, and we argue that it lowers the bar for applications by non-expert users.

The following section introduces gradient tree boosting. 
We then discuss model selection and develop an information theoretic approach to gradient boosted trees, and comment on evaluation using asymptotic theory together with modifications of the GTB algorithm.
Section \ref{sec:simulation experiments} is concerned with validation through simulation experiments of the theoretical results in section \ref{sec:information theoretic approach}.
Section \ref{sec:comparisons on benchmark datasets} sees applications to real-data and comparisons with competing methodologies.
Proofs of the theoretical results in section \ref{sec:information theoretic approach} may be found in the Appendix.

%% file: Background.tex
\section{Gradient tree boosting}\label{subsec:gradient tree boosting}

Let $\features\in\mathbb R^m$  be a feature vector and $\response\in \mathbb R$ a corresponding response variable. 
The objective of supervised learning in general is to determine the 
function $f(\features)$ that minimises the expected loss,
\begin{align}\label{eq:supervised-learning}
\hat{f} = \arg\min_{f}\E_{\features,\response} \left[ \loss(\response, f(\features)) \right],
\end{align}
given a loss function $\loss(\cdot,\cdot)$. In practice, the expectation
over the joint distribution of $\features$  and $\response$ must be 
replaced by an empirical average over a finite dataset, $\data_n = \left\{ (\features_i,\, \response_i)  \right\},\; \left|\data_n \right| = n,\, \features_i \in \mathbb{R}^m,\; \response\in\mathbb{R}$.
The loss, $l$, is a function that measures the difference between a prediction $\pred_i=f(\features_i)$ and its target $y_i$. We will assume that $l$ is both differentiable and convex in its second argument.

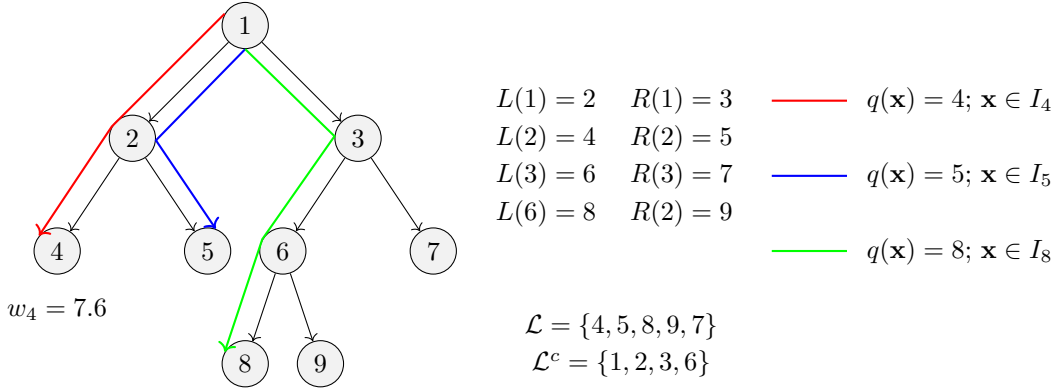
\begin{figure}\centering
	\begin{tikzpicture}
	\tikzstyle{custbox} = [circle, align=left, minimum width=0.5cm, minimum height=0.5cm,text centered, draw=black, fill=gray!10]
	
	\node (r) at (0,0) [custbox]{1};
	
	\node (rl) at (-1.5,-1.5) [custbox]{2}; 
	\node (rll) at (-2.5,-3) [custbox]{4}; \node[below=0.2cm of rll] {	$w_4 = 7.6$	}; 
	\node (rlr) at (-0.5,-3) [custbox]{5}; \node[below=0.2cm of rlr] {}; 
	
	\node (rr) at (1.5,-1.5) [custbox]{3};
	\node (rrl) at (0.5,-3) [custbox]{6};
	\node (rrll) at (0,-4.5) [custbox]{8}; \node[below=0.2cm of rrll] {}; 
	\node (rrlr) at (1,-4.5) [custbox]{9}; \node[below=0.2cm of rrlr] {}; 
	\node (rrr) at (2.5,-3) [custbox]{7}; \node[below=0.2cm of rrr] {}; 
	
	\draw[->,black] (r) -- (rl);
	\draw[->,black] (r) -- (rr);
	\draw[->,black] (rl) -- (rll);
	\draw[->,black] (rl) -- (rlr);
	\draw[->,black] (rr) -- (rrl);
	\draw[->,black] (rr) -- (rrr);
	\draw[->,black] (rrl) -- (rrll);
	\draw[->,black] (rrl) -- (rrlr);

	\draw[->, thick,red] (r.150) -- (rl.150) -- (rll.140);
	\draw[->, thick,blue]  (r.-90) -- (rl.-5) -- (rlr.70);
	\draw[->, thick,green] (r.-90) -- (rr.175) -- (rrl.150) -- (rrll.150);
	
	\node at (5.8, -1) {$R(1)=3$};
	\node at (4, -1) {$L(1)=2$};
	\node at (5.8, -1.5) {$R(2)=5$};
	\node at (4, -1.5) {$L(2)=4$};
	\node at (5.8, -2) {$R(3)=7$};
	\node at (4, -2) {$L(3)=6$};
	\node at (5.8, -2.5) {$R(2)=9$};
	\node at (4, -2.5) {$L(6)=8$};
	
	\node at (5,-4){$\setleafnodes = \{ 4,5,8,9,7 \}$};
	\node at (5,-4.5){$\setleafnodes^{c} = \{ 1,2,3,6 \}$};
			
	\draw[ thick,red] (7,-1) -- (8,-1); \node at (9.5,-1) {$q(\features)=4$; $\features\in I_4$};
	\draw[ thick,blue] (7,-2) -- (8,-2); \node at (9.5,-2) {$q(\features)=5$; $\features\in I_5$};
	\draw[ thick,green] (7,-3) -- (8,-3); \node at (9.5,-3) {$q(\features)=8$; $\features\in I_8$};
	
	\end{tikzpicture}
	\caption{Example of a CART with $T=5$ leaf nodes ($\setleafnodes$) and $4$ internal nodes ($\setleafnodes^c$). $\mathbf{w}=(w_4,w_5,w_7,w_8,w_9)$ is the vector of possible predictions. The operator $q(\features)$ maps different instance sets ($I_t,\,t\in\setleafnodes$) to leaf nodes. The mappings $L(t)$ and $R(t)$ yield the left and right descendants of each internal node 
	$t\in	\setleafnodes^{c}$.}
	\label{fig: binary tree}
\end{figure}

In GTB, $f$ is taken to be an ensemble model, with ensemble members $f_k(\features)$ being classification and regression trees
(CARTs; see Figure \ref{fig: binary tree} for notation).
A prediction from $f$ has the following form:
\begin{align}\label{eq:model-prediction}
\pred_i = f^{(K)}(\features_i) = 
\sum_{k=1}^{K} f_k(\features_i),\quad 
\text{where } f_k(\features_i) = w_{q_k(\features_i),k}.
\end{align}
Here,
$q_k:\mathbb{R}^m\to \setleafnodes_k$  (where $\setleafnodes_k$ is the set of leaf nodes)
is the feature mapping of the $k$'th tree, which assigns every feature vector to a unique leaf node (see Figure \ref{fig: binary tree}).
The predictions associated with each leaf node are contained in a vector
$\mathbf{w}_k = \left\{w_{t,k},\,t\in\setleafnodes_k\right\}\in\mathbb{R}^{T_k}$,
where $T_k$ is the number of leaf nodes in the $k$-th tree (i.e. the cardinality of $\setleafnodes_k$).
Moreover, any  internal node $t$ (i.e. $t\in\setleafnodes_k^c$) has exactly two descendants whose
labels are denoted by $L(t)$ (left descendant) and $R(t)$ (right descendant).
Figure \ref{fig: binary tree} illustrates these concepts graphically for three different input feature-vectors. 

Suppose an ensemble model with $k-1$ trees, $f^{(k-1)}$, has already been selected. In order to sequentially improve the ensemble prediction by adding another member $f_k$, the theoretical objective $\E_{\features,\response} \left[ \loss(\response, f^{(k)}(\features)) \right]$ reduces to
\begin{align}\label{eq: sequential target}
\E_{\features,\response}\left[ \loss\left(\response, f^{(k-1)}(\features)+f_k(\features)\right) \right],
\end{align}
which should be minimized with respect to the $q_k$ and $\mathbf{w}_k$ associated with $f_k$. 
To gain analytical tractability we perform a second order Taylor expansion around~$\hat y=f^{(k-1)}(\features)$:
\begin{equation}
\hat l(y,\hat y + f_k(\features)) =  l(y,\hat y) + g(y,\hat y)f_k(\features)  + \frac{1}{2} h(y,\hat y) f_k^2(\features),
\label{taylor2}
\end{equation}
where $g(y,\hat y) = \frac{\partial }{\partial \hat y}l(y,\hat y)$ and $h(y,\hat y) = \frac{\partial^2 }{\partial (\hat y)^2}l(y,\hat y)$.

As the joint distribution of $(\features,\response)$ is generally unknown, 
the expectation in (\ref{eq: sequential target}) 
is approximated by the training data empirical counterpart:
	\begin{align}\label{eq loss taylor approx}
\frac{1}{n}\sum_{i=1}^{n}
l\left(\response_i, \pred_i^{(k-1)} + f_k(\features_i) \right)
&\approx 
\frac{1}{n}
\sum_{i=1}^{n} 
\left[
\loss\left(y_i,\pred_i^{(k-1)}\right)+g_{ik} f_k(\features_i)
+\frac{1}{2}h_{ik}f_k(\features_i)^2
\right]
\notag\\
&= \frac{1}{n}
\sum_{i=1}^{n}
\loss\left(y_i,\pred_i^{(k-1)}\right)
+ \frac{1}{n}
\sum_{t\in\setleafnodes_k} \left[\sum_{i\in I_{tk}} g_{ik} w_{tk} + \frac{1}{2}h_{ik} w_{tk}^2
\right]\\
&=: \ell_k(q_k,\mathbf w_k).
\end{align}
where 
\begin{equation}\label{eq:loss derivatives g h}
g_{ik}=g(y_i,f^{(k-1)}(\features_i))
\qquad\text{and}\qquad
h_{ik}=h(y_i,f^{(k-1)}(\features_i)).
\end{equation}
and $I_{tk}$ is the instance set of leaf $t$: $I_{tk} = \{i: q_k(\features_i)= t \}$, (see Figure \ref{fig: binary tree}).
Hence, $\ell_k$ is the \textit{training} loss approximation of the theoretical objective (\ref{eq: sequential target}), to be optimized in the $k$-th boosting iteration.
This second order approximation-based boosting strategy was originally proposed by \citet{friedman2000additive} and first implemented for CARTs in \texttt{xgboost} \citet{chen2016xgboost}. Further, notice that for the quadratic loss $l(y,\hat y)=(y-\hat y)^2$, the Taylor expansion is exact.
	
For a \textit{given} feature mapping $q_k$ (and hence instance sets $I_{tk},\;t \in\setleafnodes_k$), the weight estimates $\hat{\mathbf w}_k$ minimizing $\mathbf w_k \mapsto \ell_k(q_k,\mathbf w_k)$ are given by 
\begin{align}\label{eq:leaf-prediction}
\hat{w}_{tk}
=-\frac{G_{tk}}{H_{tk}},\; G_{tk}=\sum_{i \in I_{tk}} g_{ik},\; H_{tk}=\sum_{i \in I_{tk}} h_{ik}.
\end{align}
Further, the improvement in training loss resulting from using weights (\ref{eq:leaf-prediction}) is given by
\begin{align}\label{eq:biased-loss-gtb}
\ell_k(q_k,\hat{ \mathbf w}) - \frac{1}{n} \sum_{i=1}^n l(y_i,\hat{y}_i^{(k-1)}) = 
-\frac{1}{2n}\sum_{t=1}^{T_k}\frac{G_{tk}^2}{H_{tk}}
.
\end{align}
The explicit expressions for leaf weights \eqref{eq:leaf-prediction} and loss reduction \eqref{eq:biased-loss-gtb} allow comparison of a large number of different candidate feature maps $q_k$.
Still, to consider every possible tree structure leads to combinatorial explosion, and it is therefore customary to do recursive binary splitting in a greedy fashion \citep[p.~307][]{friedman2001elements, chen2016xgboost}:
\begin{enumerate}
	\item Begin with a constant prediction for all features, i.e. $\hat{w}=-\frac{\sum_{i=1}^n g_{ik}}{\sum_{i=1}^n h_{ik}}$, in a root node. 
	
	\item Choose a leaf node $t$. For each feature $j$, compute the training loss reduction 
	\begin{align}\label{eq:loss-reduction-gtb}
	\mathcal R_t(j,s_j)
	=
	\frac{1}{2n}
	\left[
	\frac{\left(\sum_{i\in I_{L}(j,s:j)}g_{ik}\right)^2}{\sum_{i\in I_{L}(j,s_j)}h_{ik}}
	+ \frac{\left(\sum_{i\in I_{R}(j,s_j)}g_{ik}\right)^2}{\sum_{i\in I_{R}(j,s_j)}h_{ik}}	
	-\frac{\left(\sum_{i\in I_{tk}}g_{ik}\right)^2}{\sum_{i\in I_{tk}}h_{ik}}
	\right],
	\end{align}
	for different split-points $s_j$, and where $I_{L}(j,s_j) = \{ i\in I_{tk}: x_{ij}\leq s_j \}$ and $I_{R}(j, s_j) = \{ i\in I_{tk} : x_{ij} > s_j \}$.
		The values of $j$ and $s_j$ maximizing $\mathcal R_t(j,s_j)$ are chosen as the next split, creating two new leaves from the old leaf $t$. 
	
	\item Continue step 2 iteratively, until some threshold on tree-complexity is reached.
\end{enumerate}
Notice that $\mathcal R_t(j,s_j)$ is the difference in training loss reduction (\ref{eq:biased-loss-gtb}) between 1) a tree where $t$ is a leaf node and 2) otherwise the same tree, but where $t$ is the ancestor to two leaf nodes $L(t)$, $R(t)$ split on the $j$th feature. In particular, $\mathcal R_t(j,s_j)$ depends only on the data that are passed to node $t$.

The measures of tree-complexity in step 3 vary, and multiple criteria can be used at the same time, such as a maximum depth, maximum terminal nodes, minimum number of instances in node, or a regularized objective. Also, several alternative strategies for choosing candidate $t$ and proposal $s_j$s in step 2 exist, \citep[see e.g][]{chen2016xgboost, ke2017lightgbm}. 
A typical strategy is to build a very large tree, and then \textit{prune} it back to a subtree using cost complexity pruning \citep[p.~308]{friedman2001elements}. 

Algorithm \ref{alg:gradient-tree-boosting} illustrates the full second order gradient tree boosting process with CART trees and several split-stopping criteria. 
Note an until now unmentioned hyperparameter, the "learning rate" $\delta\in(0,1]$. The learning rate (or shrinkage \citep{friedman2002stochastic}) shrinks the effect of each new tree with a constant factor in step $2.iv)$, and thereby opens up space for feature trees to learn. This significantly improves the predictive power of the ensemble, but comes at the cost of more boosting iterations until convergence.
Note how the special case of $\delta=1$ and $K=1$ gives a decision tree, and $\delta\to 0$ and $K\to\infty$ potentially gives a continuous model.
	
\begin{algorithm*}[h!]
	\begin{tabbing}
		\hspace{2em} \= \hspace{2em} \= \hspace{2em} \= \\
		{\bfseries Input}: \\
		\> - A training set $\data_n=\{(x_i, y_{i})\}_{i=1}^n$\\
		\> - A differentiable loss function $l(\cdot,\cdot)$\\
		\> - A learning rate $\delta\in(0,1]$\\
		\> \colorbox{blue!20}{- Number of boosting iterations $K$}\\
		\> \colorbox{blue!20}{- One or more tree-complexity regularization criteria}\\
		
		1. Initialize model with a constant value:\\
		\>	$f^{(0)}(\features) \equiv \underset{\eta}{\arg\min} \sum_{i=1}^n \loss(y_i, \eta)$\\
		
		2. \colorbox{blue!20}{{\bfseries for} $k = 1$ to $K$:} \colorbox{orange!30}{{\bfseries while} the inequality \eqref{eq:stopping criterion boosting} evaluates to \textbf{\texttt{false}} } \\
		
		\>	$i)$ Compute derivatives \eqref{eq:loss derivatives g h}\\
		
		\> $ii)$ Determine the structure $q_k$ by iteratively selecting the binary split that maximizes \eqref{eq:loss-reduction-gtb} until\\
		\>\> \colorbox{blue!20}{a regularization criterion is reached.} \colorbox{orange!30}{ the inequality \eqref{eq:stopping criterion splitting} evaluates to \textbf{\texttt{true}}  for all leaf nodes $t$ }\\ 
		
		\> $iii)$ Determine leaf weights \eqref{eq:leaf-prediction}, given $q_k$\\
		
		\> $iv)$ Scale the tree with the learning rate\\
		\>\> $f_k(\features)= \delta w_{q_k(\features)}$ \\
		
		\>	$v)$ Update the model:\\
		\>\>	$f^{(k)}(\features) = f^{(k-1)}(\features) + f_k(\features)$\\
		{\bfseries end \colorbox{blue!20}{for} \colorbox{orange!30}{while}}\\
		
		3. Output the model: {\bfseries Return} $f^{(K)}$
		
	\end{tabbing}
	
	\colorbox{blue!20}{Blue} background colour signifies steps unique to the original algorithm, while \colorbox{orange!30}{orange} signifies steps unique to the modified algorithm proposed here.
	\vspace{0.5cm}
	
	\caption{\label{alg:gradient-tree-boosting} Original \citep{friedman2001elements, chen2016xgboost} and modified second order gradient tree boosting.}
\end{algorithm*}

%% file: Information_theoretic_approach.tex
\section{Information theoretic approach to gradient boosted trees}\label{sec:information theoretic approach}
	
\subsection{Model selection problem}\label{subsec:model selection problem}

In the GTB Algorithm \ref{alg:gradient-tree-boosting}, there are two places where decisions are made with respect to the functional form of $f^{(k)}$: 
\begin{itemize}
\item in step $2.ii)$, decisions must be made whether to perform the proposed leaf splits, i.e.~sequential decisions with respect to the feature map $q_k$.
\item in step $2.v)$ a decision must be made whether to add $f_k$ to $f^{(k-1)}$, or
otherwise to terminate the algorithm, i.e.~selecting the number of boosting iterations $K$.
\end{itemize}

The overarching aim of this paper is to develop automatic and computationally fast methodology for performing such decisions while minimizing the generalization error. Suppose the model $f(\features;\theta)$ depends on some parameters $\theta$, and a procedure for fitting $\theta$ to the training data, say $\hat{\theta}=\hat{\theta}(\data_n)$ is given. Further, let $(\featuresindep,\responseindep)$ be a test-data realization with the same distribution as each $(\features_i,\response_i)\in \data_n$, unseen in the training phase and hence independent from $\hat{\theta}$. We will use
\begin{align}\label{eq:expected test loss}
Err 
= \E_{\hat{\mathbf{\theta}}}\E_{\featuresindep,\responseindep}\left[\loss \left( \responseindep, f(\featuresindep;\hat{\mathbf{\theta}})  \right)\right].
\end{align}
as our measure of generalization error, as it is well suited for analytical purposes.

In GTB described above, it is not the generalization error that is used when comparing possible splits in step 2 in the greedy binary splitting procedure.
Equations (\ref{eq:biased-loss-gtb},\ref{eq:loss-reduction-gtb}) are estimators (modulo errors introduced by the Taylor expansions) of \textit{reduction in training loss}, where the training loss is given by:
\begin{align}\label{eq:training loss}
\overline{err} = 
\frac{1}{n}\sum_{i=1}^n l(\response_i, f(\features_i;\hat{\theta})).
\end{align}
As is well known, $\overline{err}$ as an estimator for $Err$ is biased downwards in expectation, favouring complex models which leads to overfitting. 

The bias of~\eqref{eq:training loss} relative to~\eqref{eq:expected test loss} is 
commonly referred to as the \textit{optimism} of the estimation procedure \citep{friedman2001elements}. 
The reminder of this section is devoted to deriving estimators of such optimism in the GTB context, 
and subsequently using these to obtain optimism-corrected estimators of $\overline{err}$. 

\subsection{Correcting the training loss for optimism}
Define the conditional on feature $j$ reduction in training loss
\begin{equation}\label{eq:max_loss_reduction}
\mathcal R_t(j) = \max_{s_j} \mathcal R_t(j,s_j),\; j=1,\dots,m,
\end{equation}
and unconditional reduction in training loss
\begin{equation}
\mathcal R_t = \max_{j\in (1,\dots,m)} \mathcal R_t(j),
\end{equation}
where the reduction in training loss $\mathcal R_t(j,s_j)$ for given ancestor node $t$, feature $j$ and split point $s_j$ is given in (\ref{eq:loss-reduction-gtb}).
A key part of  our approach is to derive estimators of the generalization-loss based counterparts of $\mathcal R_t(j)$ and $\mathcal R_t$ to~\eqref{eq:training loss}, which we denote by 
$\mathcal R^0_t(j)$ and $\;\mathcal R^0_t$, respectively.
In the current section we focus on $\mathcal R^0_t(j)$, 
while $\mathcal R^0_t$ will be considered in Section \ref{sec:many_features}.

The proposed estimator of $\mathcal R^0_t(j)$, and hence that of $\;\mathcal R^0_t$, does not rely on cross validation or bootstrapping, but rather on analytical results adapted from traditional information theory.
The approach enables learning of the feature maps $q_k$, and also suggests a natural stopping criterion for boosting iterations. 
The algorithm is terminated when splitting the root node is not beneficial. This is automatic and with minimal worries of overfitting.

As should be clear from Algorithm \ref{alg:gradient-tree-boosting}, only (local) splitting decisions on a single leaf node are performed in each step. Moreover, the splitting decisions on two distinct leaf nodes do not influence each other as different subsets of the data are passed to the respective leafs. In the presentation that follows, we therefore focus on estimating  $\mathcal R^0_t(j)$ for a split/no-split decision of a single leaf node. 
To avoid overly complicated notation, we consider the \textit{root node only}, i.e.~$t=1$, and subsequently suppress the ancestor index $t$. This simplification introduces no loss of generality, as the split/no-split decisions at any  leaf node are exactly the same, except that they only operate on the subsets of the original data passed to that leaf node.

With the understanding that $j$ is fixed in this section, we suppress the index $j$ from our notation, except when strictly needed for future reference.
The no-split decision involves a \textit{root} tree, consisting of a single node with prediction $\hat{w}_1=-\sum_{i=1}^n g_i/\sum_{i=1}^n h_i$.
The do-split decision involves a \textit{stump} tree, with two leaf nodes and parameters
$\hat{\theta}=\{\hat{s}, \hat{w}_l, \hat{w}_r \}$. Here, $s$ is the split point (for the $j$th feature) and $\hat{w}_l$ and $\hat{w}_r$ are the leaf
weights of the left and right leaf nodes, respectively, given by~\eqref{eq:leaf-prediction}.

The subsequent theory is derived using the 2nd order Taylor approximation $\hat l$, given by~\eqref {taylor2}, instead of the original loss $l$. 

In what follows, we seek an adjustment of the training loss reduction $\mathcal R(j)$ defined in (\ref{eq:max_loss_reduction}) to represent $\mathcal R^0(j)$ in expectation over the training data.
The optimism $C$ for the constant (root) model is defined as
\newcommand{\yvektor}{\ensuremath{\mathbf{y}}}
\begin{align}\label{eq:optimism_root}
C_\text{root} = 
\E_{\hat{w}}\E_{\responseindep}
\left[\hat \loss \left( \responseindep, \hat{w}  \right)\right ]
- E_\yvektor\left[ \hat l(y_1,\hat{w})  \right],
\end{align}
where $\yvektor=(y_1,\ldots,y_n)$ and $\hat{w}=\hat{w}(\yvektor)=-\sum_{i=1}^n g_i/\sum_{i=1}^n h_i$.	
The use of $y_1$ in the last term above is justified by the fact that the $(y_i,\features_i)$
are identically distributed for $i=1,\ldots,n$.
Note that $C_\text{root}$ does not depend $j$ as the root model does not utilize any feature information.
For the tree-stump (stump) we get
\begin{align}\label{eq:optimism_stump}
C_\text{stump}(j) = 
\E_{\hat{\theta}}\E_{\featuresindep,\responseindep}\left[ \hat l(y^0, f(\featuresindep;\hat{\theta})) \right]
- E_{\features_1,\ldots,\features_n,\yvektor}\left[ \hat l(y_1, f(\features_1;\hat{\theta})) \right].
\end{align}
where  $\hat{\theta}=\hat{\theta}(\features_1,\ldots,\features_n,\yvektor)=\{\hat{s}, \hat{w}_l, \hat{w}_r \}$.
When interpreting~\eqref{eq:optimism_stump} it should be kept in mind that we are currently
only using the $j$th component of the feature vector $\features$.

Equations (\ref{eq:optimism_root}) and (\ref{eq:optimism_stump}) may be combined to get an equivalent representation of $\mathcal R^0(j)$ expressed in terms of expected reduction in training loss (i.e. \eqref{eq:max_loss_reduction} in expectation), namely
\begin{align}\label{eq:reduction generalization training adjusted}
E_{y^0,x^0}[\mathcal R^0(j)] = E_{\features,\response}[\mathcal R(j)] +  C_\text{root} - C_\text{stump}(j).
\end{align}
Under the assumption that the $j$-th feature is independent of the response $y$, each term on the right hand side of (\ref{eq:reduction generalization training adjusted}) may be estimated efficiently and consequently allows us to correct the training loss reduction. In practice, $E_{\features,\response}[\mathcal R(j)]$ is estimated using the observed training loss reduction $\mathcal R(j)$. Hence, similarly to conventional hypothesis testing, 
if the estimated version of (\ref{eq:reduction generalization training adjusted}) is negative we retain root model, whereas if the estimated (\ref{eq:reduction generalization training adjusted}) as a consequence of a large training loss difference is positive, we opt for the stump model. The next few sections are devoted to derive approximations for the optimisms $C_\text{root}$ and $C_\text{stump}(j)$, and constitute the main methodological contribution of the paper.

\subsection{Optimism for loss differentiable in parameters}		   

In the standard case where some loss function $l$ is differentiable in its parameters,
say $\eta$, and adhere to the regularity conditions in \ref{app:regularity conditions}
the optimism may be estimated by \citep[Eqn. 7.32]{burnham2003model}:
\begin{align}\label{eq:gradient boosting bias approximate}
\tilde{C}
&= 
\tr
\left(
\E_{\features,\response} \left[ \nabla_{\mathbf{\eta}_0}^2 \loss (\response,f(\features;\mathbf{\eta}_0)) \right]
\Cov\left[\hat{\mathbf{\eta}}\right]
\right),
\end{align}
where $\eta_0 = \lim_{n\to\infty}\hat{\eta}$.
If $\Cov\left[\hat{\mathbf{\eta}}\right]$ is estimated using the Sandwich Estimator \citep{huber1967behavior, white1982maximum} one obtains the network information criterion (NIC) \citep{murata1994network}.
The training loss of the stump model is discontinuous in $s_j$s for finite $n$, and hence \eqref{eq:gradient boosting bias approximate} is not applicable in the $s_j$-direction. 

Again taking the local perspective, we omit dependence on being in node $t$. 
We start off with considering an optimism approximation for $C_{root}$, say $\tilde{C}_{root}$,
which subsequently will be used in \eqref{eq:gradient boosting bias approximate}.
Moreover $\tilde C_{root}$ constitute a building block for our approximation to $C_{stump}$.
The root-model does not involve any split-points, and hence when
\eqref{eq:gradient boosting bias approximate} is applied we obtain:
\begin{align}\label{eq:optimism constant model}
\tilde C_\text{root} = \E_{\features,\response}\left[ \frac{1}{n}\sum_{i=1}^{n}h_i \right]\Var\left(\hat w_1\right).
\end{align}

Turning to the stump-model, suppose momentarily that the split-point $s_j$ is given
a-priori. Then we may compute the optimism approximation \eqref{eq:optimism constant model} for the tree-stump model 
(conditioned on $s_j$), say $\tilde C_\text{stump}(j|s_j)$, which is given by
\begin{align}\label{eq:conditional tree stump optimism}
\tilde C_\text{stump}(j|s_j) = 
\tilde C_{\text{root},L} P(x_j\leq s_j) + \tilde C_{\text{root},R} P(x_j>s_j).
\end{align}
Here $\tilde C_{\text{root},L}$ and $\tilde C_{\text{root},R}$ are as in 
\eqref{eq:optimism constant model} but computed from sub-datasets corresponding to the child leaf nodes $L$ (i.e. $i : x_{i,j} \leq s_j$) and $R$ (i.e. $i : x_{i,j} > s_j$)) respectively.
$\tilde C_\text{stump}(j|s_j)$, however, cannot be substituted in \eqref{eq:reduction generalization training adjusted} directly, since the optimism induced by optimizing over $s_j$ is not accounted for in
(\ref{eq:conditional tree stump optimism}). Consequently $\tilde C_\text{stump}(j|s_j)$ will be downward biased relative to $C_\text{stump}(j)$. The next section attempts to account for this bias by providing an approximate correction factor.

As a side-note, notice that \eqref{eq:gradient boosting bias approximate} may
be applied more generally to a full tree, if the structure $q$ is given a-priori.
In this case, the optimism of the full tree may be approximated by
\begin{align}\label{eq:optimism full tree without profiling}
\sum_{t=1}^T
\tilde{C}_{root,t} P(q(x)=t)
\end{align}
Again, $\tilde{C}_{root,t}$ is computed as in \eqref{eq:optimism constant model},
based on the data that is passed in leaf-node $t$, i.e. $i\in I_t$.
Of course, \eqref{eq:optimism full tree without profiling} is also biased downward relative
to the optimism of the tree when $q$ is learned from training data.

\subsection{Optimism from greedy-splitting over one feature}\label{subsec:optimism from greedy-splitting one feature}

\begin{figure}
\centering
\includegraphics[width=1\textwidth,height=5cm]{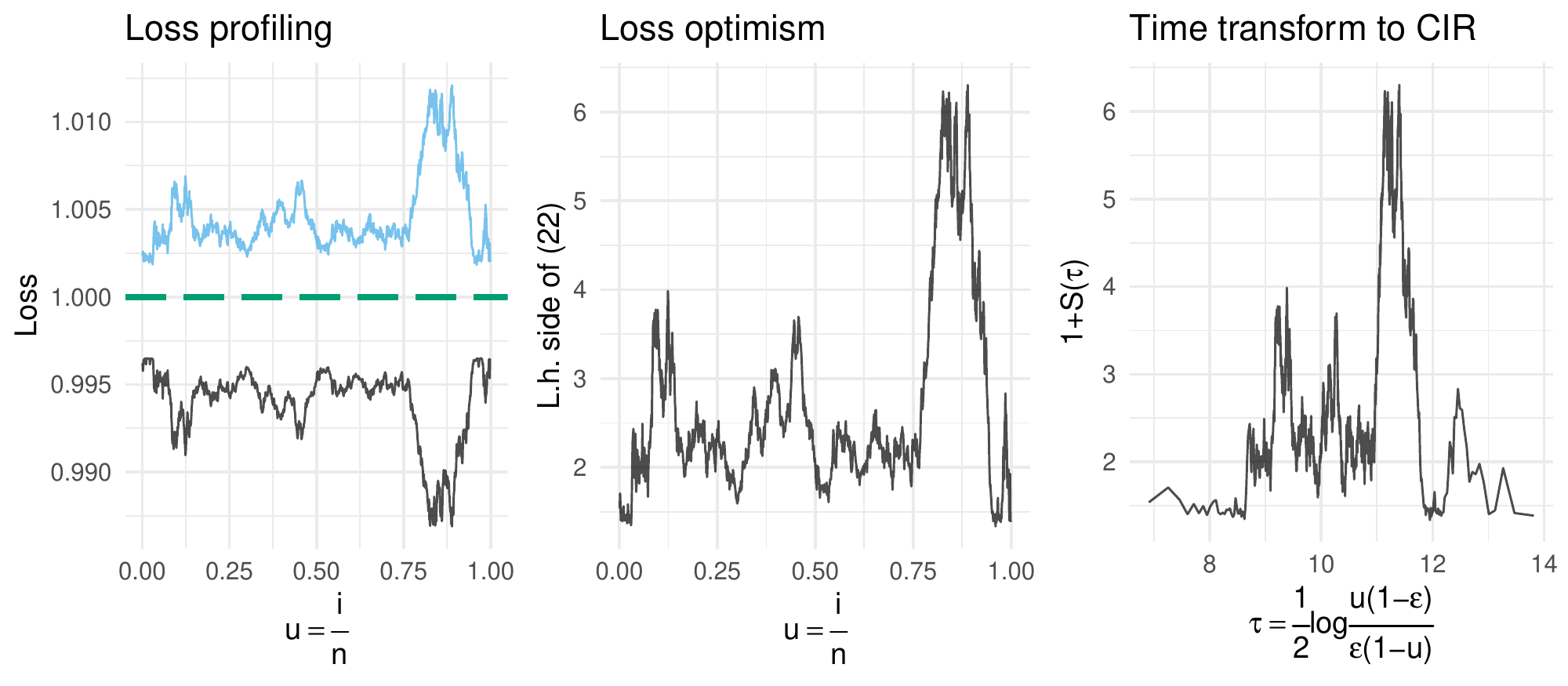} 
\rule{25em}{0.5pt}
\caption[]{Left: Training loss $\hat{l}(y,f(x;\hat{\theta}))$ (black) and generalization loss $E_{y^0,x^0}\left[\hat{l}(y^0,f(x^0;\hat{\theta}))\right]$ (blue) as a function of $u=\frac{i}{n}$, defined from the sorted order of $x_j$. Green long-dashed line is the expected loss-value at $\theta_0=\lim_{n\to\infty}\hat{\theta}_n$ if $n\to\infty$ for the training data, constant as there is no information in $x_j$ for this instance. Right plot: The transformation of distance between generalization loss and training loss into a CIR process, with $\epsilon = 10^{-7}$ which is used throughout. In this case with no information in feature $x_j$, choosing the value of $s$ giving the smallest value of training loss in the left plot induces an optimism at the value of $1$ plus the expected maximum of the CIR-process, illustrated in the right plot. }
\label{fig: split sequence to cir}
\end{figure}

In order to resolve $C_\text{stump}(j)$ appearing in \eqref{eq:reduction generalization training adjusted}, this section provides an approximation $\tilde{C}_\text{stump}(j)$ which in general is biased upward relative to $C_\text{stump}(j)$.  Consequently, the approximation of $\mathcal R^0(j)$ resulting from substituting $C_\text{stump}(j)$ with $\tilde{C}_\text{stump}(j)$ is biased downward, in practice favouring the constant model. However, it is illustrated in the simulation experiments in Section \ref{sec:simulation experiments} that this bias is rather small.
In order to construct $\tilde{C}_\text{stump}(j)$, we first assume that $y$ is independent of $x_j=x_{\cdot,j}$. This assumption appears to be necessary to get an asymptotic approximation to the joint distribution of the difference in test and training loss (which in expectation over training data is the conditional optimism) for different values of split points. This distribution obtains as the limiting distribution of an empirical process.
The argument leading to this limiting distribution has similarities to the one originally presented in \citet{miller1982maximally} regarding maximally selected chi-square statistics, and generalized with refinements in \citet{gombay1990asymptotic}.

Suppose the training data $(y_i,x_{i,j})$ has been sorted in ascending order over $i$ according to the $j$-th feature. If $x_j$ contains repeated values, the ordering (in $i$) of observations with identical $x_{i,j}$ is arbitrary. 
Further, define $u_i:= i/n $, and let 
$f(\cdot; \hat{w}_l(u_i),\hat{w}_r(u_i))$, $i=1,\dots,n-1$,
be the tree stump with left node containing $x_{1:i,j}$, right node containing 
$x_{(i+1):n,j}$ (and hence split point $s=x_{i,j}$).
Notice that $\lim_{n\to\infty} u_i = p(x_{.,j}\leq x_{i,j})$.
Under the independence assumption, the difference between generalization loss and training loss as a function of $i$ converges in distribution as
\begin{multline}\label{eq:varying optimism as cir}
n\left[E_{y^0,x^0}\left[\hat{\loss}(y^0, f(\mathbf{x}^0; \hat{w}_l(u_i),\hat{w}_r(u_i)))\right]
- \hat{\loss}(y, f(\mathbf{x}; \hat{w}_l(u_i),\hat{w}_r(u_i)))\right]  \overset{D}{\underset{n\to\infty}{\longrightarrow}} 
n \hat C_\text{root}\left(1+S(\tau(u))\right),
\end{multline}
where $n \hat C_\text{root} = O(1)$.
Here $S(\tau)$ is defined through the stochastic differential equation
\begin{align}\label{eq: cir process sde, param (2,1,2)}
dS(\tau) = 2(1-S(\tau))d \tau + 2\sqrt{2S(\tau)} dW(\tau).
\end{align}
Moreover, $W(\tau)$ is a Wiener process with time $\tau$ following $\tau=\frac{1}{2}\log\frac{u(1-\epsilon)}{\epsilon(1-u)}$, $u= \min\left\{ 1-\epsilon, \max \left\{ \epsilon, \frac{i}{n}   \right\} \right\} $ and $0<\epsilon<< 1$. 
The diffusion specified by \eqref{eq: cir process sde, param (2,1,2)} is recognized as a Cox-Ingersoll-Ross process \citep{cox1985theory}, with unconditional mean $E[S_\tau] = 1$. Appendix \ref{app:regularity conditions} gives the details underlying this result.

Figure \ref{fig: split sequence to cir} is included to illustrate this result for a known distribution on $(y,\mathbf x)$, $m=1$, $y \perp \mathbf x$, and a simulated training data set of size $n=1000$. The left hand side panel displays both the training loss $\hat{\loss}(y, f(\mathbf{x};  \hat{w}_l(u_i),\hat{w}_r(u_i)))$ (black) and the test loss    $E_{y^0,x^0}\left[\hat{\loss}(y^0, f(\mathbf{x}^0;  \hat{w}_l(u_i),\hat{w}_r(u_i)))\right]$ (blue, resolved approximately using 100000 Monte Carlo simulation from the true data generating process) as functions of $u=i/n$. Also included in the left panel is the asymptotic limit (green, dashed) which coincides for both types of loss, and is constant as the feature is uninformative w.r.t. to the response. The paths of the training- and expected test-loss are almost mirror images
about the asymptotic line, and asymptotically they are indeed exactly that. This becomes clear upon inspection of \eqref{eq:varying optimism as cir}: The only source of randomness in the expected test-loss, is the estimator based upon the (random) training data -- the source of randomness for the training loss. 
The middle panel shows the difference in losses (left hand side of (\ref{eq:varying optimism as cir}) scaled with conditional optimism), also as a function of $u=i/n$. Finally, the right hand side panel depicts the same curve as the middle panel, but with transformed horizontal axis conforming with the "time" $\tau$ of (\ref{eq: cir process sde, param (2,1,2)}).

\begin{figure}
\centering
\includegraphics[width=0.6\textwidth,height=5cm]{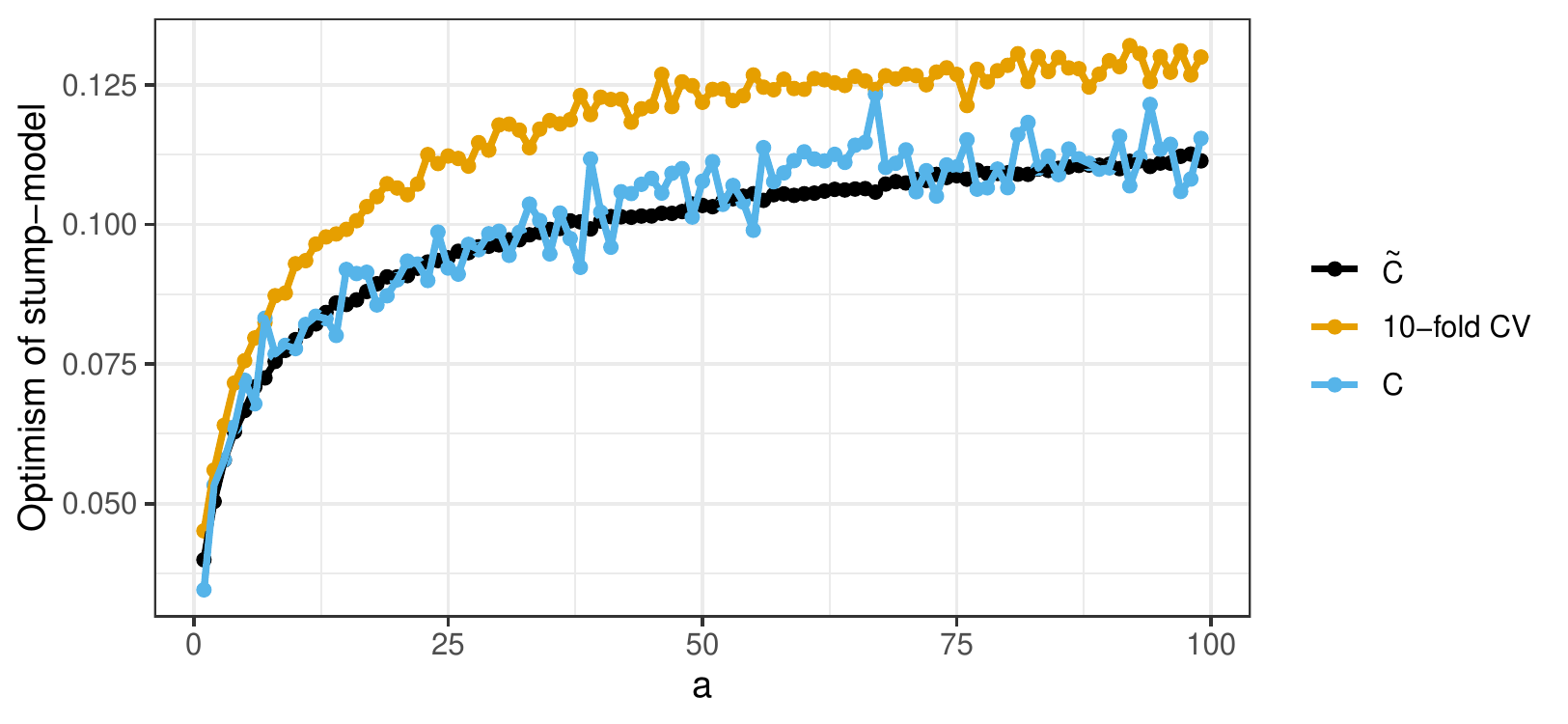}
\caption[]{
	Illustration of Equation \eqref{eq:optimism splitting bound} by repeatedly 
	simulating $n=100$ observations in training data, for each value of $a$
	possible split-points, under the assumption $y\perp x,~y\sim N(0,1)$, 
	for which $\tilde{C}_{stump}$ is asymptotically exact.
	The feature $x$ is constructed to have the same expected number of observations in each 
	group, and each group is also guaranteed to have at least one observation.
	The simulation experiment is repeated 1000 times, and the average values are reported.
	The black line shows the value of $\tilde{C}_{stump}$, which aligns with the intuition
	that more number of split-points should correspond to increased optimism.
	The blue line is the Monte-Carlo estimated generalization loss $C_{stump}$,
	using the average of 1000 test-loss datasets. 
	It verifies the above intuition and values of $\tilde{C}_{stump}$ as it 
	fluctuates mildly about the optimism approximation.
	Also included is the 10-fold CV optimism (orange), that retains the shape of $C$ and 
	$\tilde{C}_{stump}$, but is upward biased, resulting from using 9/10'ths of the 
	data in the estimator fitting procedure.
}\label{fig:loss reduction versus increase in number of splits}
\end{figure}

Now suppose $x_j$ takes $a+1$ distinct values, then there are $a$ different split-points $s_k$, $k=1,\dots,a$, which are compared in terms of training loss during the greedy profiling procedure. These correspond to the $i$s such that $i/n = u_k = P(x_j\leq s_k)$ and $\tau_k = \tau(u_k)$ for the right hand side. Equation \ref{eq:varying optimism as cir} provides the joint distribution of the differences in test and training loss in terms of the joint distribution of 
$\{ \hat C_\text{root}(1+S(\tau_k)) \}_{k=1}^a$.
Consequently, the expected maximum of
$\{ \hat C_\text{root}(1+S_{\tau_k}) \}_{k=1}^a$
is upward biased relative to $C_\text{stump}(j)$. In the proceeding, we will use this expected maximum, i.e. 
\begin{align}\label{eq:optimism splitting bound}
\tilde{C}_\text{stump}(j)
= \hat C_\text{root}\left(1+E\left[\max_{1\leq k\leq a}S(\tau_k)\right]\right)
\end{align}
as the (somewhat conservative in favor of the root model) approximation of $C_\text{stump}(j)$.
As shown in in Appendix \ref{app:max cir as a bound on optimism}, under assumption $y\perp x$, $\tilde{C}_{\text{stump}}(j)$ converges to $C_{\text{stump}}(j)$ as $n\to\infty$.
The corresponding finite-sample behaviour, for different numbers of split-points $a$, is illustrated in Figure \ref{fig:loss reduction versus increase in number of splits}. In the Figure, the exact value of $C_{\text{stump}}(j)$, estimated using Monte-Carlo simulations of the true data-generating process, slightly fluctuates (due to being a simulation estimate) about the value of $\tilde{C}_{\text{stump}}(j)$. The optimism implied by 10-fold CV has the 
exact same shape, but is upward biased relative to $C_{\text{stump}}(j)$ and $\tilde{C}_{\text{stump}}(j)$
as it only employes 9/10'ths of the data in its fitting procedure.

The scaling factor $E\left[ \max_{1\leq k \leq a} S(\tau_k)  \right]$ depends on the nature of the $j$-th feature. In particular for a feature taking only two values, e.g. one-hot encoding, we have $\tilde{C}_\text{stump}(j) = \tilde{C}_\text{root}(1+E(S_{\tau_1}))=2\tilde{C}_\text{root}$ as no optimization over the split point is performed, which agrees with AIC-type criteria when the number of parameters are doubled. At the other extreme, for a feature with absolutely continuous marginal distribution, the scaling factor converges to the expected maximum of a CIR process over the "time"-interval obtained by applying $\tau(u)$ to each $u\in\left(\max \left\{\frac{1}{n},\epsilon\right\}, \min\left\{\frac{n-1}{n}, 1-\epsilon \right\} \right)$. Setting $\epsilon = 10^{-7}$ gives $E\left[ \max_{0\leq \tau \leq \frac{1}{2}\log\frac{(1-\epsilon)^2}{\epsilon^2}} S(\tau)  \right] \approx 7.5$ as $n\rightarrow \infty$. In general, the expectation is bounded as long as $\epsilon > 0$, as the CIR process is positively recurrent.
\citet{linetsky2004computing} give an exact analytical expression for its distribution in terms of special functions, but is not applied here as evaluation is computationally costly, generally not straight forward, and would only apply to continuous features when $a=n-1$.

\subsection{Optimism over several features}\label{sec:many_features}
In general, the greedy binary splitting procedure profiles both over features $j$ and within feature split-point $s_j$. Let $B_j = \tilde{C}_\text{root} \max_{1\leq k \leq a_j}(1+ S_j(\tau_k)) $ where $\{\tau_k\}$ correspond to the potential split points on the $j$-th feature with $a_j$ possible split-points, so that $E_{S_\tau}(B_j)=\tilde{C}_\text{stump}(j)$. Following a similar logic as leading to \eqref{eq:optimism splitting bound}, an upward biased approximation of the unconditional (over feature $j$) optimism $C_\text{stump}$ obtains as
\begin{align}\label{eq:theoretical_uncond_bound}
\E\left[ \max_{j\in \{1,\dots,m\}} B_j  \right]
\end{align}
However, for typical values of $m>>1$, characterization of the dependence structure among the $B_j$s appears difficult. Hence, in order to get a practical approximation to (\ref{eq:theoretical_uncond_bound}), we calculate as if the $B_j$s are independent, to get the approximation 
\begin{align}\label{eq:final approximate stump optimism}
\tilde C_\text{stump}=
\int_{0}^{\infty}
1 - \prod_{j=1}^{m}
P\left(B_j\leq z \right) dz,
\end{align}
where the integral is over a single dimension and hence efficiently calculated numerically. In Section \ref{sec:sim_multi_features}, the errors incurred by using the independence simplification on data sets with correlated features are studied.

\subsection{Applications to gradient tree boosting}
\label{subsec:applications to gradient tree boosting}

Returning attention to the application of the above theory in the GTB context, the ancestor node subscript $t$ is re-introduced. All quantities, e.g. $\mathcal R_t$ and $\tilde C_{\text{stump},t}$ are calculated as if node $t$ was the root node in the above theory, and in particular based only on the data passed to node $t$.
The previous sections gives us the needed approximation to adjust the training loss reduction $\mathcal R_t$ according to the unconditional (over $j$) counterpart to (\ref{eq:reduction generalization training adjusted}), namely
\begin{align}\label{eq:reduction in training loss adjusted}
\tilde{\mathcal R}^0_t = 
\mathcal R_t + \tilde C_{\text{root},t} - \tilde C_{\text{stump},t}.
\end{align}
The approximation of generalization loss reduction $\tilde{\mathcal R}^0_t$ has at least two important applications to the tree boosting algorithm. Firstly, it provides a natural criterion on whether to split a node or not, with the stopping criterion for splitting a leaf node $t$ becomes
\begin{align}\label{eq:stopping criterion splitting}
\tilde{\mathcal R}_t^0 < 0.
\end{align}
If no leaf node $t$ in the tree $f_k$ has positive $\tilde{\mathcal R}_t^0$, the tree building process in boosting iteration $k$ is stopped. 
Note that due to the usage of an upward biased optimism approximation for the stump model, this criterion will slightly favour less complex models. In principle, (\ref{eq:stopping criterion splitting}) can be augmented to read $\tilde{\mathcal R}_t^0 < \rho$ where $\rho$ is a tuning parameter controlling individual tree complexity in a coherent manner. However, this option is not pursued further as the default $\rho=0$ produces good results in practice.

Further, the proposed approximate optimism may also be applied within a stopping-rule for the gradient boosting iteration -- often referred to as "early stopping".
When a tree-stump, scaled by the learning rate $\delta$, no longer gives a positive reduction in approximate generalization loss relative to the previous boosting iterate, we terminate the algorithm.
Care must be taken as the learning rate $\delta$ scales the training loss and the optimism differently.
Recalculating the training loss \eqref{eq:biased-loss-gtb}, with $\delta f_k$ as the predictive function, we obtain that the training loss associated with $f_k$ should be scaled with a factor $\delta\left(2 - \delta\right)$.
The optimism, on the other hand scales linearly, as is seen from expressing optimism as a covariance, $C=\frac{2}{n}\sum_{i=1}^{n}\Cov(\response_i, \pred_i)$, \citep[p.~229]{friedman2001elements} and recalling that $\pred_i$ is linear in $\delta$.
The boosting stopping criterion hence becomes (with ancestor index $t=1$):
\begin{align}\label{eq:stopping criterion boosting}
\tilde{\mathcal R}_\delta^0 = 
\delta(2-\delta) \mathcal R_1 + \delta 
\left(   C_{\text{root},1} - \tilde C_{\text{stump},1}
\right) > 0.
\end{align}
When \eqref{eq:stopping criterion boosting} evaluates to true, there is no more information left in data for another member added to the ensemble $f^{(k-1)}$ to learn, in the generalization error sense, using the boosting iteration of Algorithm \ref{alg:gradient-tree-boosting}.

Algorithm \ref{alg:gradient-tree-boosting} with orange markers (and not blue markers) gives the proposed modified algorithm. 
The early stopping criterion saves one hyperparameter. 
The adaptive tree complexity on the other hand alleviate the need for the multiple hyperparameters typically used to fine-tune the tree complexities.
E.g. the popular \texttt{xgboost} implementation has 4 such hyperparameters: a constant minimum reduction in loss, a maximum depth, a minimum child weight and a maximum number of leaves.
These computational-reductions stemming from not having to tune the original algorithm are explored and measured in more detail in Section \ref{subsec:real-data-results}. 

\subsection{Implementation}
\label{subsec: evaluation and implementation}
Recall that the basic building block of the above theory is the root optimism approximation (\ref{eq:optimism constant model}). However, this approximation also depends on moments (Expected loss Hessian and parameter variance) which must be estimated empirically in the numerical implementation.
As previously mentioned, \eqref{eq:optimism constant model} is a special case of theoretical optimism of \citet{murata1994network}. Further, \citet{murata1994network} estimated the parameter variance using the conventional Sandwich Estimator \citep[see e.g.][Section 5.3]{vanDerVaart}, as the estimated leaf weights (\ref{eq:leaf-prediction}) are M-estimators. This approach is also
taken here, and results in the root optimism estimator:
\begin{align}\label{eq:GBTIC}
\hat C_{\text{root},t}\approx\frac{\sum_{i\in I_t} \left(g_i+h_i\hat{w}_t\right)^2}{n_t\sum_{i\in I_t}h_i}
\end{align}
where $n_t=|I_t|$ is the number of observations passed to leaf $t$. 

The same estimator is also used for evaluating conditional stump optimisms $\hat C_{\text{stump},t}(j|\hat s_j)$ in (\ref{eq:conditional tree stump optimism}), but of course then based on the on the subsets of data falling into the left and right child nodes of $t$. The probabilities in (\ref{eq:conditional tree stump optimism}) are simply estimated as the corresponding relative frequencies in the training data.

When \eqref{eq:optimism constant model} is evaluated using \eqref{eq:GBTIC}, 
adding evaluation of $\hat{\mathcal R}^0$ to the greedy-binary-splitting procedure does not change the computational complexity of the overall algorithm, as the only cost is to keep track of sum of squares and cross multiplication among the $g$ and $h$ vectors. 

The expected maximums over CIR processes (\ref{eq:final approximate stump optimism}) are resolved based on a combination of Monte Carlo simulations and approximating the $B_j$-s by a parametric distribution. 
First of all, \eqref{eq:theoretical_uncond_bound} is approximated by assuming independence, obtaining \eqref{eq:final approximate stump optimism}. We then \textit{only} need knowledge of the CDF of the maximum of the CIR process observed on time-points associated with the split-points of feature $j$. \citet{linetsky2004computing} gives expressions for the maximum of the CIR on an interval, however, the expressions are not easily calculated and comes to a non-negligible computational cost, and would also penalize non-continuous features too much. We therefore consider an alternative approach: For the case with only one possible split, the Gamma distribution with shape 0.5 and scale 2 is used, which is exact. For the cases with more than one split a Monte Carlo simulation procedure is used to simulate the expected maximum of the CIR over the split-points on feature $j$. In principle we could simulate indefinitely to obtain exact estimates of the CDF. However, this quickly becomes infeasible when the number of features grows large, and \eqref{eq:final approximate stump optimism} will be concerned with the tail-behaviour of the CIR maximums. We therefore do an asymptotic approximation, by fitting the CIR to the Gumbel distribution, which it is in the maximum domain of attraction of, as it has a Gamma stationary distribution. The approximation is asymptotic in the number of observations-points, and will be expected to perform increasingly well in the number of split points. 

%% file: Simulation_experiments.tex
\section{Simulation experiments}\label{sec:simulation experiments}

The theory developed in the previous section involves multiple approximations. This section studies the performance of the proposed training  loss reduction estimator when the data generating process is known a-priori. All computations involving the proposed methodology were done using the associated R-package \texttt{aGTBoost} which can be downloaded from \url{https://github.com/Blunde1/aGTBoost}, and scripts that re-create the below results can be found at the same place. \texttt{aGTBoost} is written mainly in C++, and computing times are therefore directly comparable to those of e.g. \texttt{xgboost}.

\subsection{Simulations in the single feature case}
\label{subsec:validity of lemmas}

\begin{table}
{\small 
\centering 
\begin{tabular}{lllllllllllll} 
\hline 
DGP & \multicolumn{2}{c}{$y\sim N(0,1)$} & \multicolumn{2}{c}{$y\sim N(0,5^2)$} & \multicolumn{2}{c}{$y\sim N(\nint{x},1)$} & \multicolumn{2}{c}{$y\sim N(\nint{x},5^2)$} & \multicolumn{2}{c}{$y\sim N(x,1)$} & \multicolumn{2}{c}{$y\sim N(x,5^2)$} \\ 
\cmidrule(lr){2-3}  \cmidrule(lr){4-5}  \cmidrule(lr){6-7}  \cmidrule(lr){8-9}  \cmidrule(lr){10-11}  \cmidrule(lr){12-13}  
& $E$ & $P$ & $E$ & $P$ & $E$ & $P$ & $E$ & $P$ & $E$ & $P$ &$E$ & $P$ \\ 
\hline 
\multicolumn{13}{c}{$a+1=2$} \\ \hline 
$\mathcal{R}$ & 0.969 & 1 & 24.1 & 1 & 25.6 & 1 & 48.4 & 1 & 26 & 1 & 47.8 & 1 \\ 
$\mathcal{R}^0$ & -0.966 & 0.016 & -24.1 & 0.024 & 24 & 1 & .212 & 0.684 & 23.9 & 1 & .47 & 0.691 \\ 
$\tilde{\mathcal{R}}^0$ & -1.03 & 0.154 & -25.5 & 0.157 & 22.7 & 0.998 & -2.82 & 0.332 & 22.9 & 1 & -2.65 & 0.35 \\ 
10-fold CV & -1.16 & 0.165 & -29.9 & 0.162 & 24 & 0.998 & -4.93 & 0.342 & 24.3 & 1 & -3.68 & 0.365 \\ 
100-fold CV & -1.06 & 0.159 & -26.3 & 0.157 & 24.1 & 0.999 & -2.17 & 0.335 & 24.4 & 1 & -2.03 & 0.352 \\ 
\hline 
\multicolumn{13}{c}{$a+1=10$} \\ \hline 
$\mathcal{R}$ & 2.99 & 1 & 72.4 & 1 & 26.6 & 1 & 95 & 1 & 12.3 & 1 & 84.9 & 1 \\ 
$\mathcal{R}^0$ & -2.99 & 0 & -72.5 & 0 & 22.5 & 0.996 & -52.5 & 0.185 & 4.86 & 0.947 & -61.9 & 0.046 \\ 
$\tilde{\mathcal{R}}^0$ & -2.82 & 0.086 & -75 & 0.084 & 17.8 & 0.992 & -53.3 & 0.164 & 4.4 & 0.763 & -62.3 & 0.136 \\ 
10-fold CV & -3.46 & 0.202 & -90.3 & 0.199 & 22.7 & 0.982 & -65.7 & 0.266 & 4.54 & 0.682 & -73.7 & 0.243 \\ 
100-fold CV & -3.18 & 0.316 & -81.2 & 0.286 & 23.1 & 0.98 & -60.3 & 0.37 & 4.77 & 0.727 & -60.6 & 0.364 \\ 
\hline 
\multicolumn{13}{c}{$a+1=100$} \\ \hline 
$\mathcal{R}$ & 4.58 & 1 & 115 & 1 & 28 & 1 & 136 & 1 & 12.9 & 1 & 124 & 1 \\ 
$\mathcal{R}^0$ & -4.58 & 0 & -115 & 0 & 20.7 & 0.995 & -96.9 & 0.052 & 2.46 & 0.799 & -106 & 0 \\ 
$\tilde{\mathcal{R}}^0$ & -4.73 & 0.048 & -116 & 0.057 & 14 & 0.957 & -103 & 0.092 & .582 & 0.489 & -112 & 0.061 \\ 
10-fold CV & -5.41 & 0.158 & -141 & 0.157 & 20.3 & 0.975 & -119 & 0.21 & 2.14 & 0.586 & -142 & 0.183 \\ 
100-fold CV & -5.31 & 0.226 & -130 & 0.233 & 20.7 & 0.965 & -111 & 0.301 & 2.53 & 0.672 & -127 & 0.258 \\ 
\hline 
\end{tabular} 
} 
\caption{Single feature root versus stump loss reduction simulation study with $n=100$ observations. The contending methods are cross validation (CV) and the proposed test loss reduction estimator $\tilde{\mathcal R}^0$. In addition, the test loss $\mathcal R^0$ and training loss $\mathcal R$ were included for reference. Columns $E$ give the expected loss reduction, multiplied by a factor $100$ for readability, for the different estimators, and columns $P$ give the probability of a positive loss reduction (i.e. probability of choosing the stump model). In all cases, the feature was simulated on $(0,1)$ and with $a+1$ distinct values. The results are based on 1000 simulated data sets in each case. The test loss $\mathcal R^0$, was estimated using 1000 simulated test responses for each simulated data set.  } 
\label{tab:amazingtab}
\end{table} 

\begin{table}
{\small 
\centering 
\begin{tabular}{lllllllllllll} 
\hline 
DGP & \multicolumn{2}{c}{$y\sim N(0,1)$} & \multicolumn{2}{c}{$y\sim N(0,5^2)$} & \multicolumn{2}{c}{$y\sim N(\nint{x},1)$} & \multicolumn{2}{c}{$y\sim N(\nint{x},5^2)$} & \multicolumn{2}{c}{$y\sim N(x,1)$} & \multicolumn{2}{c}{$y\sim N(x,5^2)$} \\ 
\cmidrule(lr){2-3}  \cmidrule(lr){4-5}  \cmidrule(lr){6-7}  \cmidrule(lr){8-9}  \cmidrule(lr){10-11}  \cmidrule(lr){12-13}  
& $E$ & $P$ & $E$ & $P$ & $E$ & $P$ & $E$ & $P$ & $E$ & $P$ &$E$ & $P$ \\ 
\hline 
\multicolumn{13}{c}{$a+1=2$} \\ \hline 
$\mathcal{R}$ & 0.904 & 1 & 23.3 & 1 & 251 & 1 & 280 & 1 & 253 & 1 & 277 & 1 \\ 
$\mathcal{R}^0$ & -0.903 & 0.023 & -23.3 & 0.022 & 249 & 1 & 226 & 1 & 249 & 1 & 222 & 0.998 \\ 
$\tilde{\mathcal{R}}^0$ & -1.09 & 0.138 & -26.7 & 0.15 & 248 & 1 & 229 & 0.974 & 250 & 1 & 226 & 0.956 \\ 
10-fold CV & -1.22 & 0.157 & -29.7 & 0.166 & 250 & 1 & 228 & 0.962 & 252 & 1 & 225 & 0.947 \\ 
100-fold CV & -1.1 & 0.134 & -27.3 & 0.146 & 250 & 1 & 231 & 0.971 & 252 & 1 & 227 & 0.957 \\ 
\hline 
\multicolumn{13}{c}{$a+1=10$} \\ \hline 
$\mathcal{R}$ & 2.89 & 1 & 75.3 & 1 & 251 & 1 & 301 & 1 & 84 & 1 & 174 & 1 \\ 
$\mathcal{R}^0$ & -2.9 & 0 & -75.3 & 0 & 249 & 1 & 162 & 0.935 & 71.9 & 1 & 14.3 & 0.709 \\ 
$\tilde{\mathcal{R}}^0$ & -2.94 & 0.087 & -70.6 & 0.104 & 242 & 1 & 152 & 0.828 & 76.1 & 1 & 25.8 & 0.52 \\ 
10-fold CV & -3.4 & 0.204 & -81.4 & 0.226 & 250 & 1 & 166 & 0.784 & 71.7 & 1 & 5.04 & 0.471 \\ 
100-fold CV & -2.97 & 0.379 & -75.1 & 0.383 & 250 & 1 & 169 & 0.797 & 71.4 & 0.992 & 15.4 & 0.608 \\ 
\hline 
\multicolumn{13}{c}{$a+1=100$} \\ \hline 
$\mathcal{R}$ & 4.58 & 1 & 114 & 1 & 252 & 1 & 328 & 1 & 74.6 & 1 & 207 & 1 \\ 
$\mathcal{R}^0$ & -4.57 & 0 & -114 & 0 & 248 & 1 & 137 & 0.872 & 58.8 & 1 & -39.9 & 0.382 \\ 
$\tilde{\mathcal{R}}^0$ & -4.73 & 0.051 & -119 & 0.041 & 238 & 1 & 90.9 & 0.694 & 62.1 & 1 & -28.7 & 0.322 \\ 
10-fold CV & -5.52 & 0.176 & -139 & 0.176 & 248 & 1 & 107 & 0.675 & 57.7 & 0.999 & -43.5 & 0.362 \\ 
100-fold CV & -5.08 & 0.325 & -129 & 0.33 & 249 & 1 & 120 & 0.726 & 58.9 & 0.982 & -28.4 & 0.542 \\ 
\hline 
\multicolumn{13}{c}{$a+1=1000$} \\ \hline 
$\mathcal{R}$ & 5.7 & 1 & 143 & 1 & 254 & 1 & 365 & 1 & 75.6 & 1 & 220 & 1 \\ 
$\mathcal{R}^0$ & -5.71 & 0 & -143 & 0 & 246 & 1 & 117 & 0.833 & 57.4 & 1 & -64.6 & 0.284 \\ 
$\tilde{\mathcal{R}}^0$ & -5.78 & 0.03 & -144 & 0.033 & 236 & 1 & 71.9 & 0.626 & 60.3 & 1 & -71.3 & 0.208 \\ 
10-fold CV & -6.57 & 0.16 & -162 & 0.149 & 246 & 1 & 109 & 0.668 & 57.2 & 1 & -82.9 & 0.316 \\ 
100-fold CV & -6.1 & 0.288 & -154 & 0.298 & 247 & 1 & 131 & 0.732 & 57.7 & 0.985 & -78.3 & 0.455 \\ 
\hline 
\end{tabular} 
} 
\caption{Single feature root versus stump loss reduction simulation study with $n=1000$ observations. The contending methods are cross validation (CV) and the proposed test loss reduction estimator $\tilde{\mathcal R}^0$. In addition, the test loss $\mathcal R^0$ and training loss $\mathcal R$ were included for reference. Columns $E$ give the expected loss reduction, multiplied by a factor $1000$ for readability, for the different estimators, and columns $P$ give the probability of a positive loss reduction (i.e. probability of choosing the stump model). In all cases, the feature was simulated on $(0,1)$ and with $a+1$ distinct values. The results are based on 1000 simulated data sets in each case. The test loss $\mathcal R^0$, was estimated using 1000 simulated test responses for each simulated data set. } 
\label{tab:amazingtab2} 
\end{table}  

\begin{figure}
\centering
\input{loss_reduction_sim_simple.tex}
\caption[]{Histograms of single feature root versus stump loss reduction, with $n=\{30,100,1000\}$, $a=1$ and the DGP being $y\sim N(0,1)$. The results are based on 1000 simulation replica in each case. The two values taken by the feature were simulated uniformly on $(0,1)$.
}\label{fig:indep_response_hists}
\end{figure}
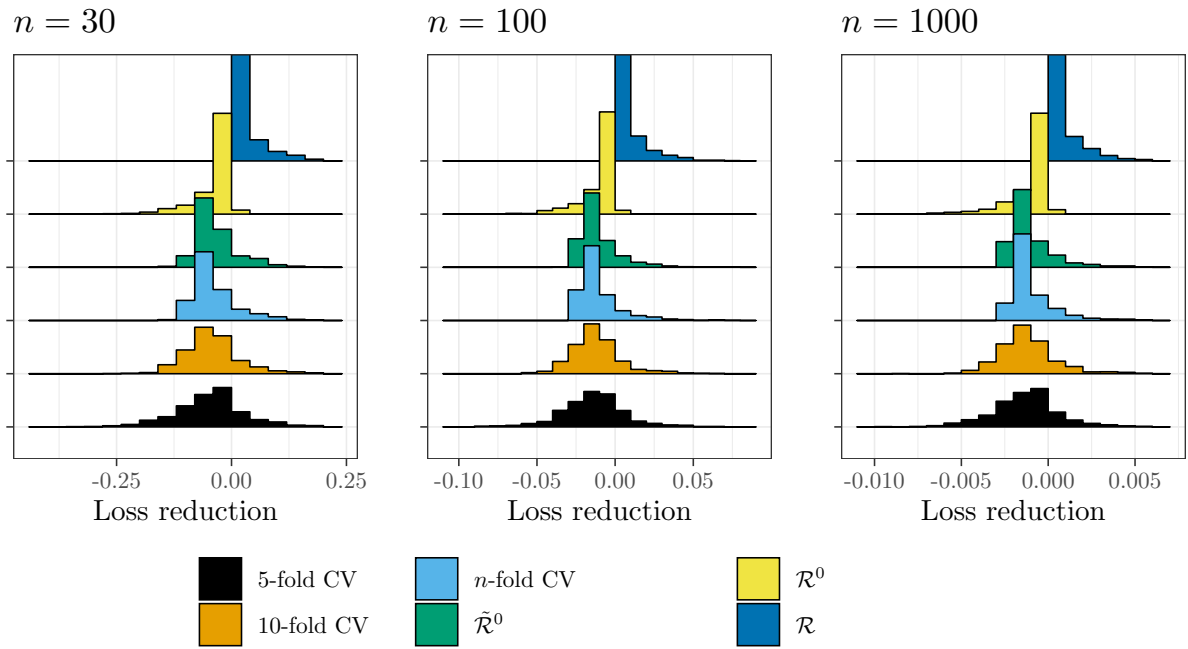

In the first batch of simulation experiments, the single feature estimator of the test loss reduction in the root versus stump situation, developed in Section \ref{subsec:optimism from greedy-splitting one feature} is considered. The results are summarized in Tables \ref{tab:amazingtab} and \ref{tab:amazingtab2} for $n=100$ and $n=1000$ respectively. In the experiments, the test loss reduction estimator $\tilde{ \mathcal{R}}^0$ is compared to two fidelities of cross validation, and in addition test loss and training loss are provided as references. For both sample sizes, a range of numbers of potential split points $a$ are considered, including binary feature ($a=1$) and continuous feature ($a+1=n$). In the tables, "$E$" corresponds to the mean the loss reductions, and $P$ is the probability of rejecting the root model.

Six data generating process (DGP) cases were considered. For the former two DGPs ($y\sim N(0,\sigma^2)$ with $\sigma=1,5$) the feature is un-informative with respect to $y$. The rejection rate of the (true) root model for non-binary features is around 5-10 \% for $n=100$ observations and around 5 \% for $n=1000$. It is seen from Tables \ref{tab:amazingtab}, \ref{tab:amazingtab2} the proposed methodology does on par (the $a=1$ case) or better (the $a>1$ cases) than cross validation.
For binary features ($a=1$), the expectation of $\tilde{ \mathcal{R}}^0$ is very close to that of $\mathcal R^0$, but the root model rejection rate is higher. To better understand this phenomenon, the $a=1$, $y \sim N(0,1)$ case is further explored in Figure \ref{fig:indep_response_hists}. As expected in the $a=1$ case, the training losses and test losses are close to being mirror images around the asymptotic loss reduction, which in this independent response case of course is 0. This effect is a consequence of the training- and test loss empirical processes (see left panel of Figure \ref{fig: split sequence to cir}) themselves are close to being symmetric around zero loss reduction. Specifically, in the $a=1$ case, these losses obtains as evaluations of the empirical processes at single point on the horizontal axis, which gives rise to the symmetry.  It is also seen that the shapes of the right hand side tails of $\mathcal R$ and $\tilde{\mathcal R}^0$ are very similar, but with $\tilde{\mathcal R}^0$ shifted to have expectation close to that of $\mathcal R^0$ (see Tables \ref{tab:amazingtab} and \ref{tab:amazingtab2}). In this case, the heavy right hand side tail of $\tilde{\mathcal R}^0$ leads to non-negligible rate of rejection of the (appropriate) root model, even if the mean is essentially that of $\mathcal R^0$.

In the next two DGPs ($y\sim N(\nint{x},\sigma^2)$ with $\sigma=1,5$), the stump model with split point $s=0.5$ is the true model. In the high signal-to-noise ratio case $\sigma = 1$, both the proposed estimator $\tilde{ \mathcal{R}}^0$ and cross validation select the true model almost perfectly in both sample sizes. Interestingly, in the $\sigma=5$ case, the test loss reduction $\mathcal R^0$ selects the root model rather often, and the proposed test loss reduction estimator and cross validation largely follow this behavior. 

Finally, the last two DGPs ($y\sim N(x,\sigma^2)$ with $\sigma=1,5$), the feature is also informative with respect to the response, but the dependence is linear rather discontinuous. It is seen also in this case that the proposed test loss reduction estimator and cross validation estimators behaves similarly to the test loss with respect to rejection probabilities.

Recall that stump optimism estimator (\ref{eq:optimism splitting bound}) was derived based on an independence assumption between the feature and response. The simulation studies do not provide evidence that optimism estimators calculated from informative features somehow overwhelms the reduction in training loss. Further, notice in the $a=1$ case where no optimization over split-points is performed, it is still not expected that $\tilde{ \mathcal{R}}^0$ is exactly equal to $\mathcal{R}^0$. This is as there is approximation error in (\ref{eq:gradient boosting bias approximate}), and that involved moments are estimated from data in $\tilde{ \mathcal{R}}^0$. 

The initial conclusion to be drawn from from these simulations is that the proposed estimator has small sample performance at least on par with cross validation in the root vs stump situation with one feature and mean squared error losses, but at a much lower computational cost.

\subsection{Simulations of the multiple feature case}\label{sec:sim_multi_features}

\begin{figure}
\centering
\input{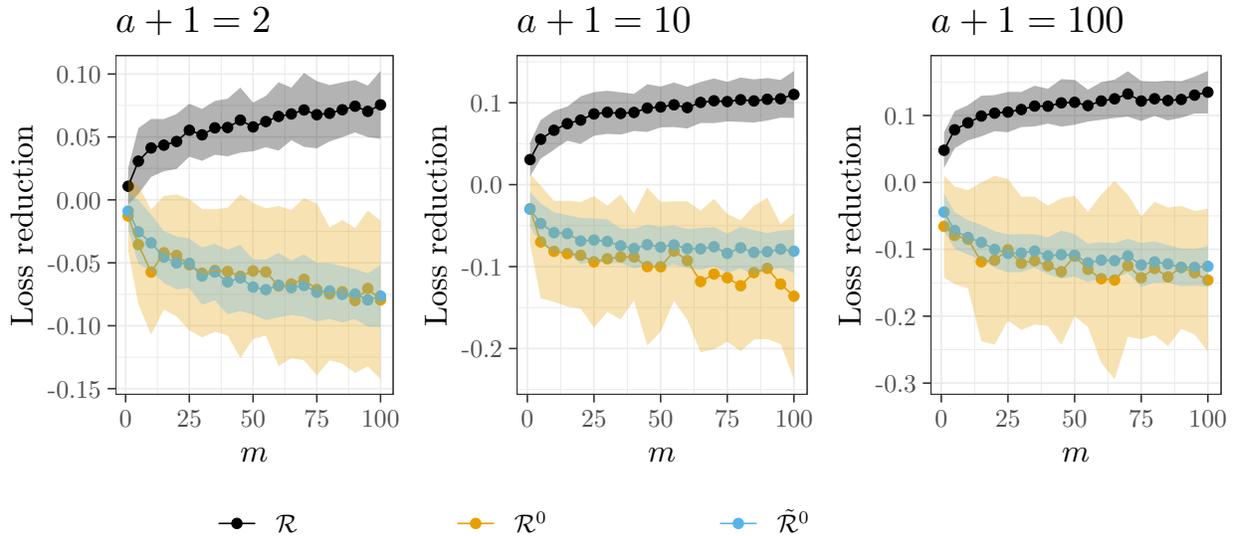}
\caption[]{Root versus stump loss reduction as a function of the number of features $m$, when all features are uninformative. The DGP is $y_i \sim N(0,1)$, and the $m$ features are simulated conditional on the given number $a$ of potential split points. Dots give the average loss reductions over 100 simulated data sets of size $n=100$ for each considered $m$, and shaded areas are the averages $\pm$ one standard
deviation. The test loss $\mathcal{R}^0$ was obtained from a single simulated test set for each simulation replica.}
\label{fig:loss reduction versus increase in features}
\end{figure}
This subsection explores the performance of the proposed methodology in the presense of more than one feature. Recall from Section \ref{sec:many_features} that $\tilde{\mathcal{R}}^0$ in this case is derived under the assumption that the (multiple) features are mutually independent and also independent of the response. Figure \ref{fig:loss reduction versus increase in features} depicts average $\tilde{\mathcal{R}}^0$, along with simulated test loss reduction $\mathcal R^0$ for different numbers of uninformative and independent features and standard normal responses. Also included in the Figure is the corresponding training loss reduction, $\mathcal R$. 

It is seen that the $\tilde{\mathcal{R}}^0$ and $\mathcal R^0$ have very similar behavior. However, small deviations still exist, stemming both from the deliberate (downward) bias introduced in equations \ref{eq:optimism splitting bound}, \ref{eq:theoretical_uncond_bound} for $a>1$, and also the simulation based algorithm used to estimate the expected maxima of (\ref{eq:theoretical_uncond_bound}). Still, it does not seem that these approximations introduces an undue amount of bias towards the root model in this particular setting.

\begin{table} \centering
\begin{tabular}{lccccccccc}
\hline \\
Method 	& \multicolumn{3}{c}{Case 1 ($m=1$)} & \multicolumn{3}{c}{Case 2 ($m=10000$)} & \multicolumn{3}{c}{Case 3 ($m=10000$)}\\
\cmidrule(lr){2-4} \cmidrule(lr){5-7} \cmidrule(lr){8-10}
& Loss & $K$ & CPU-Time & Loss & $K$ & CPU-Time & Loss & $K$ &  CPU-Time \\
\hline \\[-1.8ex] 
linear model & 0.977 &  & 0.0293 & 1.01 &  & 16 & 1.07 &  & 43 \\
\texttt{aGTBoost} & 1.01 & 365 & 0.162 & 1.05 & 294 & 723 & 1.04 & 348 & 821 \\
\texttt{xgboost}: cv & 1.11 & 275 & 4.28 & 1.07 & 357 & 3447 & 1.08 & 370 & 3908 \\
\texttt{xgboost}: val & 1.16 & 311 & 0.507 & 1.16 & 371 & 258 & 1.09 & 249 & 171 \\
\hline \\[-1.8ex] 
\end{tabular}
\caption{Test losses, (where relevant) number of trees $K$ and associated computing times for the linear model (\ref{eq:linear_test_model}) with the different cases corresponding to different design matrices described in Section \ref{sec:sim_multi_features}. Single core CPU-times are measured in seconds. Test losses are evaluated on a test data set of size $n=1000$. As a reference, a constant model corresponds to a test loss of $\approx 2.34$. 
}
\label{tab:simulated data results} 
\end{table}

\begin{figure}
\centering
\includegraphics[width=1\textwidth,height=7cm]{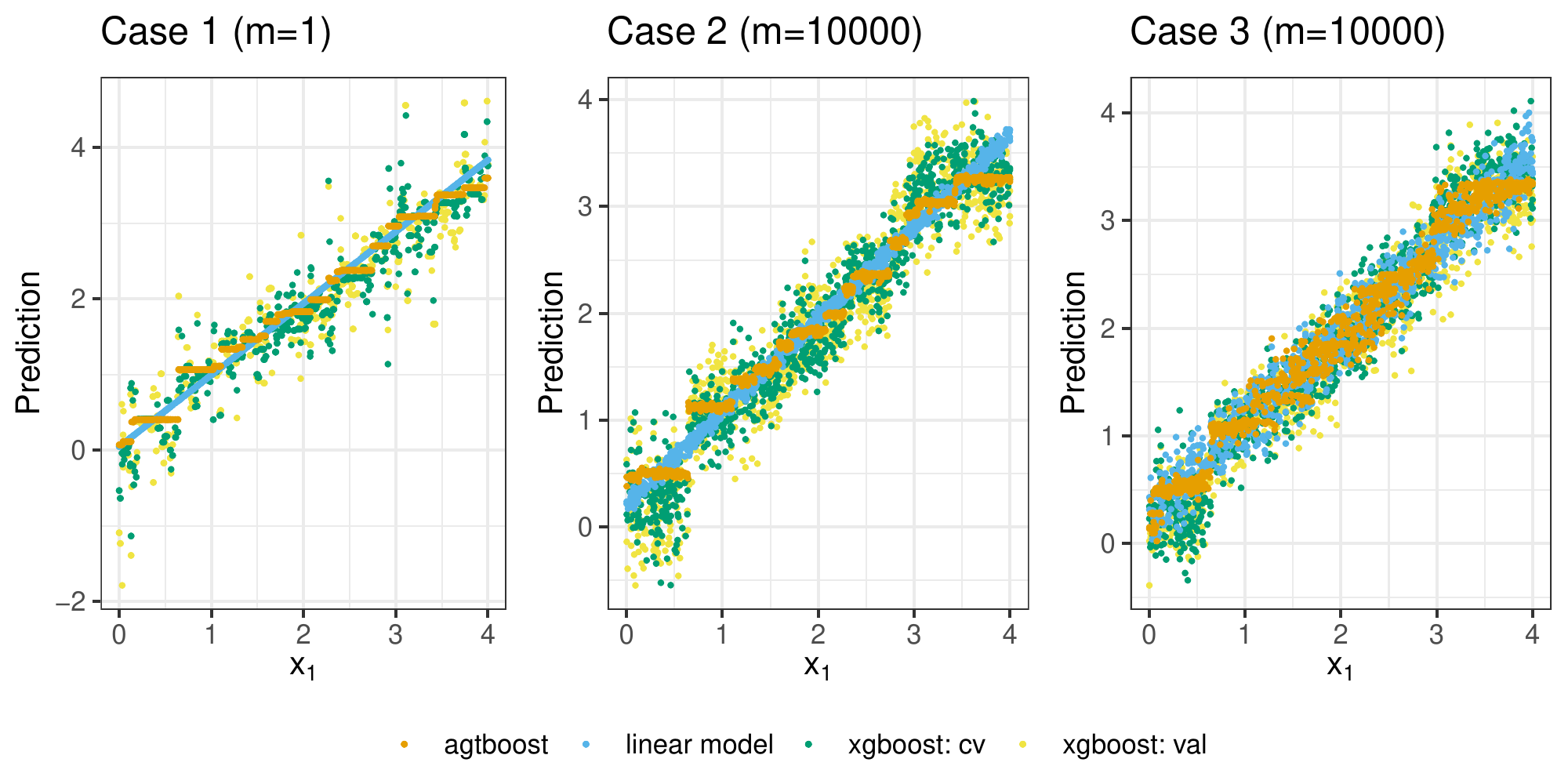} 
\caption[]{
Predictions for the linear model (\ref{eq:linear_test_model}) with the different cases corresponding to different design matrices described in Section \ref{sec:sim_multi_features}. The predictions are evaluated on the test data set features and plotted as function of $x_1$. 
}
\label{fig:boosting lm xgb fit}
\end{figure}
\begin{figure}
\centering
\includegraphics[width=1\textwidth,height=6cm]{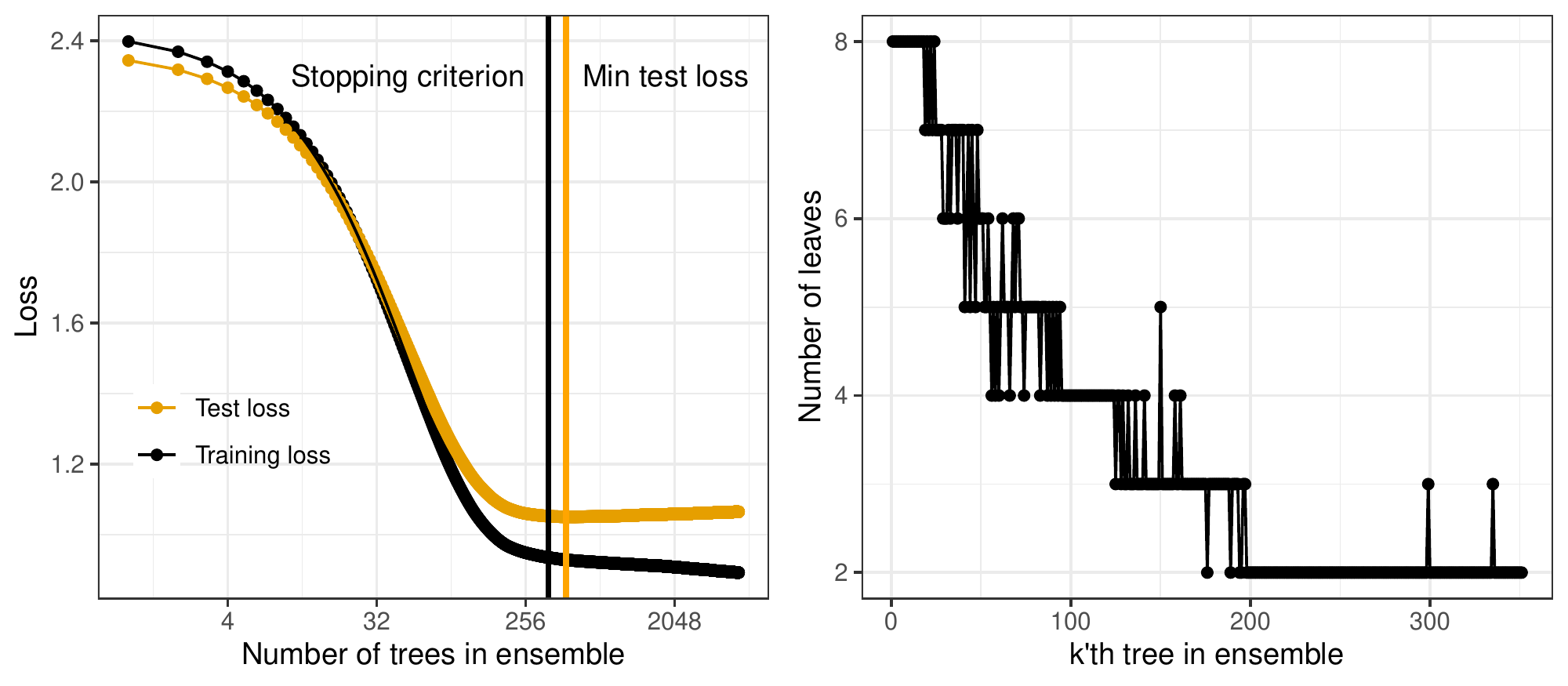} 
\caption[]{
(left) Training loss (black) and test loss (orange) plotted versus the number of boosting iterations or the number of trees in the ensemble (note the $\log_{10}$ axis). Included is a vertical line (orange) representing the iteration number that the stopping criterion \eqref{eq:stopping criterion boosting} terminates the procedure and another (black) representing the minimum test-loss. (right) The number of leaves for each tree in the ensemble until termination by the stopping criterion. The data is simulated with a linear relationship between the response and the first feature, $y_i\sim N(x_{i,1},1)$ and additional 9 noisy Gaussian features. The informative feature $x_{\cdot,i}$ is sampled uniformly on $[0,4]$. Both training and test loss consists of $n=1000$ examples, the ensemble had a learning rate of 0.01.
}
\label{fig:boosting convergence}
\end{figure}

As more realistic, but still simulated situation, we considered $n=1000$ training data observations with data generating process being 
\begin{equation}
y_i \sim N(x_{i,1},1),\;i=1,\dots,n,\; x_{i,1}\sim \text{ iid } U(0,4).\label{eq:linear_test_model}
\end{equation}
This situation tests the recursive usage of the proposed methodology in a full application of gradient tree boosting, including tree building- and boosting iteration termination criteria. Three cases, with appropriate linear model benchmarks, where considered:
\begin{description}
\item[Case 1:] $m=1$, where $x_{\cdot,1}$ is the only feature. The benchmark linear model was an un-regularized linear regression model.
\item[Case 2:] $m=10000$ with $x_{i,k},\;i=1,\dots,n,\;k=2,\dots,m$ being iid $U(0,4)$ noise independent of $x_{\cdot,1}$. The benchmark linear model was the Lasso regression with regularization determined by 10-fold cross validation, implemented in the \texttt{glmnet} R-package.
\item[Case 3:] $m=10000$ with dependent features $x_{i,k} = \frac{m-k}{m} x_{i,k-1} + N(0,(k/m)^2),\;i=1,\dots, n,\;k=2,\dots,m$. The benchmark linear model was the Ridge regression with regularization determined by 10-fold cross validation, implemented in the \texttt{glmnet} R-package.
\end{description}
As additional benchmarks, gradient boosted tree ensembles were obtained using \texttt{xgboost}. Default settings were used, and number of boosting iterations were learned using cross validation (\texttt{xgboost}:CV) and a $30\%$ validation set (\texttt{xgboost}:VAL). 

The linear model (\ref{eq:linear_test_model}) constitutes a substantial model selection challenge for tree-based predictors, as a rather complex tree ensembles are required to faithfully represent the linear functional form. Table \ref{tab:simulated data results} provides test losses for the proposed methodology and the benchmarks obtained from test data sets with 1000 observations. 

From the Table, it is seen that \texttt{aGTBoost} provides better test losses than the \texttt{xgboost}-based benchmarks, and also better test loss than for Ridge regression in Case 3. Further, in all cases, the test loss obtained by \texttt{aGTBoost} is quite close to the benchmark linear models, indicating a close to optimal behavior given that the linear functional form cannot be represented exactly by finite tree ensembles. Further, \texttt{aGTBoost} produces marginally better test losses than \texttt{xgboost}:CV, whereas \texttt{xgboost}:Val is not competitive. The computing time associated with \texttt{aGTBoost} is about an order of magnitude smaller than that of \texttt{xgboost}:CV.

Figure \ref{fig:boosting lm xgb fit} gives a graphical illustration of the predictions made by the contending methods. It is seen that \texttt{aGTBoost} produces substantially more parsimonious fits than both \texttt{xgboost} methods. In particular in Case 2, the \texttt{aGTBoost} boosting iterations stop criterion is meet before the algorithm starts utilizing the noise features $x_{\cdot,k},\;k=2,\dots,m$. This is in contrast to the Lasso regression, which as can be seen from the noisy predictions in the plot, assigns non-zero predictive power to some of the noise features. In Case 3, some of the dependent noise features $x_{\cdot,k},\;k=2,\dots,m$ are used by \texttt{aGTBoost}, but the fit is still substantially less variable than for the contenting tree boosting methods.

The left panel of Figure \ref{fig:boosting convergence} depicts the test- and training losses of \texttt{aGTBoost} as function of the boosting iterations in Case 1. Also indicated with an orange vertical line is the boosting iteration where stop criterion (\ref{eq:stopping criterion boosting}) becomes negative. More precisely, the \texttt{aGTBoost} results reported in Table \ref{tab:simulated data results} and Figure \ref{fig:boosting lm xgb fit} are based on the boosting iterate immediately before the vertical line (but more iterations were carried out for the purpose of Figure \ref{fig:boosting convergence}). It is seen that the stop criterion becomes active very close to the global minimum of the training loss (also indicated by black vertical line in the Figure). 

From the right panel of Figure \ref{fig:boosting convergence}, it is seen that \texttt{aGTBoost} builds deep trees (relative to stumps) at early iterations. As information is learned by the ensemble, subsequent trees become smaller until they are stumps, and the algorithm terminates shortly thereafter.

To summarize; the application of the proposed methodology in actual gradient tree boosting results in highly competitive tree ensemble fits in the example model cases 1-3. This appears to be a consequence of both the adaptive selection of the number of leaf nodes in each individual tree, and also that such adaptive features enable the (automatic) selection of quite few (and hence computationally cheap) boosting iterations.

%% file: loss_reduction_sim_simple.tex
\begin{tikzpicture}[x=1pt,y=1pt]
\definecolor{fillColor}{RGB}{255,255,255}
\path[use as bounding box,fill=fillColor,fill opacity=0.00] (0,0) rectangle (469.76,252.94);
\begin{scope}
\path[clip] (  0.00, 45.69) rectangle (156.58,252.94);
\definecolor{drawColor}{RGB}{255,255,255}
\definecolor{fillColor}{RGB}{255,255,255}

\path[draw=drawColor,line width= 0.6pt,line join=round,line cap=round,fill=fillColor] (  0.00, 45.69) rectangle (156.59,252.95);
\end{scope}
\begin{scope}
\path[clip] ( 21.01, 76.94) rectangle (151.09,229.98);
\definecolor{fillColor}{RGB}{255,255,255}

\path[fill=fillColor] ( 21.01, 76.94) rectangle (151.08,229.98);
\definecolor{drawColor}{gray}{0.92}

\path[draw=drawColor,line width= 0.3pt,line join=round] ( 38.22, 76.94) --
	( 38.22,229.98);

\path[draw=drawColor,line width= 0.3pt,line join=round] ( 81.70, 76.94) --
	( 81.70,229.98);

\path[draw=drawColor,line width= 0.3pt,line join=round] (125.17, 76.94) --
	(125.17,229.98);

\path[draw=drawColor,line width= 0.6pt,line join=round] ( 21.01, 89.02) --
	(151.09, 89.02);

\path[draw=drawColor,line width= 0.6pt,line join=round] ( 21.01,109.16) --
	(151.09,109.16);

\path[draw=drawColor,line width= 0.6pt,line join=round] ( 21.01,129.30) --
	(151.09,129.30);

\path[draw=drawColor,line width= 0.6pt,line join=round] ( 21.01,149.44) --
	(151.09,149.44);

\path[draw=drawColor,line width= 0.6pt,line join=round] ( 21.01,169.57) --
	(151.09,169.57);

\path[draw=drawColor,line width= 0.6pt,line join=round] ( 21.01,189.71) --
	(151.09,189.71);

\path[draw=drawColor,line width= 0.6pt,line join=round] ( 59.96, 76.94) --
	( 59.96,229.98);

\path[draw=drawColor,line width= 0.6pt,line join=round] (103.44, 76.94) --
	(103.44,229.98);

\path[draw=drawColor,line width= 0.6pt,line join=round] (146.91, 76.94) --
	(146.91,229.98);
\definecolor{drawColor}{RGB}{0,0,0}
\definecolor{fillColor}{RGB}{0,114,178}

\path[draw=drawColor,line width= 0.6pt,line join=round,line cap=round,fill=fillColor] ( 26.92,189.71) --
	( 33.87,189.71) --
	( 33.87,189.71) --
	( 40.83,189.71) --
	( 40.83,189.71) --
	( 47.79,189.71) --
	( 47.79,189.71) --
	( 54.74,189.71) --
	( 54.74,189.71) --
	( 61.70,189.71) --
	( 61.70,189.71) --
	( 68.65,189.71) --
	( 68.65,189.71) --
	( 75.61,189.71) --
	( 75.61,189.71) --
	( 82.57,189.71) --
	( 82.57,189.71) --
	( 89.52,189.71) --
	( 89.52,189.71) --
	( 96.48,189.71) --
	( 96.48,189.71) --
	(103.44,189.71) --
	(103.44,229.98) --
	(110.39,229.98) --
	(110.39,197.71) --
	(117.35,197.71) --
	(117.35,193.30) --
	(124.30,193.30) --
	(124.30,191.86) --
	(131.26,191.86) --
	(131.26,190.37) --
	(138.22,190.37) --
	(138.22,189.71) --
	(145.17,189.71) --
	(145.17,189.71) --
	(138.22,189.71) --
	(138.22,189.71) --
	(131.26,189.71) --
	(131.26,189.71) --
	(124.30,189.71) --
	(124.30,189.71) --
	(117.35,189.71) --
	(117.35,189.71) --
	(110.39,189.71) --
	(110.39,189.71) --
	(103.44,189.71) --
	(103.44,189.71) --
	( 96.48,189.71) --
	( 96.48,189.71) --
	( 89.52,189.71) --
	( 89.52,189.71) --
	( 82.57,189.71) --
	( 82.57,189.71) --
	( 75.61,189.71) --
	( 75.61,189.71) --
	( 68.65,189.71) --
	( 68.65,189.71) --
	( 61.70,189.71) --
	( 61.70,189.71) --
	( 54.74,189.71) --
	( 54.74,189.71) --
	( 47.79,189.71) --
	( 47.79,189.71) --
	( 40.83,189.71) --
	( 40.83,189.71) --
	( 33.87,189.71) --
	( 33.87,189.71) --
	( 26.92,189.71) --
	cycle;
\definecolor{fillColor}{RGB}{240,228,66}

\path[draw=drawColor,line width= 0.6pt,line join=round,line cap=round,fill=fillColor] ( 26.92,169.57) --
	( 33.87,169.57) --
	( 33.87,169.57) --
	( 40.83,169.57) --
	( 40.83,169.57) --
	( 47.79,169.57) --
	( 47.79,169.63) --
	( 54.74,169.63) --
	( 54.74,169.74) --
	( 61.70,169.74) --
	( 61.70,169.85) --
	( 68.65,169.85) --
	( 68.65,170.34) --
	( 75.61,170.34) --
	( 75.61,171.65) --
	( 82.57,171.65) --
	( 82.57,172.96) --
	( 89.52,172.96) --
	( 89.52,177.83) --
	( 96.48,177.83) --
	( 96.48,207.73) --
	(103.44,207.73) --
	(103.44,171.10) --
	(110.39,171.10) --
	(110.39,169.57) --
	(117.35,169.57) --
	(117.35,169.57) --
	(124.30,169.57) --
	(124.30,169.57) --
	(131.26,169.57) --
	(131.26,169.57) --
	(138.22,169.57) --
	(138.22,169.57) --
	(145.17,169.57) --
	(145.17,169.57) --
	(138.22,169.57) --
	(138.22,169.57) --
	(131.26,169.57) --
	(131.26,169.57) --
	(124.30,169.57) --
	(124.30,169.57) --
	(117.35,169.57) --
	(117.35,169.57) --
	(110.39,169.57) --
	(110.39,169.57) --
	(103.44,169.57) --
	(103.44,169.57) --
	( 96.48,169.57) --
	( 96.48,169.57) --
	( 89.52,169.57) --
	( 89.52,169.57) --
	( 82.57,169.57) --
	( 82.57,169.57) --
	( 75.61,169.57) --
	( 75.61,169.57) --
	( 68.65,169.57) --
	( 68.65,169.57) --
	( 61.70,169.57) --
	( 61.70,169.57) --
	( 54.74,169.57) --
	( 54.74,169.57) --
	( 47.79,169.57) --
	( 47.79,169.57) --
	( 40.83,169.57) --
	( 40.83,169.57) --
	( 33.87,169.57) --
	( 33.87,169.57) --
	( 26.92,169.57) --
	cycle;
\definecolor{fillColor}{RGB}{0,158,115}

\path[draw=drawColor,line width= 0.6pt,line join=round,line cap=round,fill=fillColor] ( 26.92,149.44) --
	( 33.87,149.44) --
	( 33.87,149.44) --
	( 40.83,149.44) --
	( 40.83,149.44) --
	( 47.79,149.44) --
	( 47.79,149.44) --
	( 54.74,149.44) --
	( 54.74,149.44) --
	( 61.70,149.44) --
	( 61.70,149.44) --
	( 68.65,149.44) --
	( 68.65,149.44) --
	( 75.61,149.44) --
	( 75.61,149.82) --
	( 82.57,149.82) --
	( 82.57,153.81) --
	( 89.52,153.81) --
	( 89.52,175.70) --
	( 96.48,175.70) --
	( 96.48,163.83) --
	(103.44,163.83) --
	(103.44,153.76) --
	(110.39,153.76) --
	(110.39,152.77) --
	(117.35,152.77) --
	(117.35,150.42) --
	(124.30,150.42) --
	(124.30,149.87) --
	(131.26,149.87) --
	(131.26,149.60) --
	(138.22,149.60) --
	(138.22,149.44) --
	(145.17,149.44) --
	(145.17,149.44) --
	(138.22,149.44) --
	(138.22,149.44) --
	(131.26,149.44) --
	(131.26,149.44) --
	(124.30,149.44) --
	(124.30,149.44) --
	(117.35,149.44) --
	(117.35,149.44) --
	(110.39,149.44) --
	(110.39,149.44) --
	(103.44,149.44) --
	(103.44,149.44) --
	( 96.48,149.44) --
	( 96.48,149.44) --
	( 89.52,149.44) --
	( 89.52,149.44) --
	( 82.57,149.44) --
	( 82.57,149.44) --
	( 75.61,149.44) --
	( 75.61,149.44) --
	( 68.65,149.44) --
	( 68.65,149.44) --
	( 61.70,149.44) --
	( 61.70,149.44) --
	( 54.74,149.44) --
	( 54.74,149.44) --
	( 47.79,149.44) --
	( 47.79,149.44) --
	( 40.83,149.44) --
	( 40.83,149.44) --
	( 33.87,149.44) --
	( 33.87,149.44) --
	( 26.92,149.44) --
	cycle;
\definecolor{fillColor}{RGB}{86,180,233}

\path[draw=drawColor,line width= 0.6pt,line join=round,line cap=round,fill=fillColor] ( 26.92,129.30) --
	( 33.87,129.30) --
	( 33.87,129.30) --
	( 40.83,129.30) --
	( 40.83,129.30) --
	( 47.79,129.30) --
	( 47.79,129.30) --
	( 54.74,129.30) --
	( 54.74,129.30) --
	( 61.70,129.30) --
	( 61.70,129.30) --
	( 68.65,129.30) --
	( 68.65,129.30) --
	( 75.61,129.30) --
	( 75.61,129.63) --
	( 82.57,129.63) --
	( 82.57,136.91) --
	( 89.52,136.91) --
	( 89.52,155.32) --
	( 96.48,155.32) --
	( 96.48,141.02) --
	(103.44,141.02) --
	(103.44,133.57) --
	(110.39,133.57) --
	(110.39,131.87) --
	(117.35,131.87) --
	(117.35,130.72) --
	(124.30,130.72) --
	(124.30,129.74) --
	(131.26,129.74) --
	(131.26,129.57) --
	(138.22,129.57) --
	(138.22,129.30) --
	(145.17,129.30) --
	(145.17,129.30) --
	(138.22,129.30) --
	(138.22,129.30) --
	(131.26,129.30) --
	(131.26,129.30) --
	(124.30,129.30) --
	(124.30,129.30) --
	(117.35,129.30) --
	(117.35,129.30) --
	(110.39,129.30) --
	(110.39,129.30) --
	(103.44,129.30) --
	(103.44,129.30) --
	( 96.48,129.30) --
	( 96.48,129.30) --
	( 89.52,129.30) --
	( 89.52,129.30) --
	( 82.57,129.30) --
	( 82.57,129.30) --
	( 75.61,129.30) --
	( 75.61,129.30) --
	( 68.65,129.30) --
	( 68.65,129.30) --
	( 61.70,129.30) --
	( 61.70,129.30) --
	( 54.74,129.30) --
	( 54.74,129.30) --
	( 47.79,129.30) --
	( 47.79,129.30) --
	( 40.83,129.30) --
	( 40.83,129.30) --
	( 33.87,129.30) --
	( 33.87,129.30) --
	( 26.92,129.30) --
	cycle;
\definecolor{fillColor}{RGB}{230,159,0}

\path[draw=drawColor,line width= 0.6pt,line join=round,line cap=round,fill=fillColor] ( 26.92,109.16) --
	( 33.87,109.16) --
	( 33.87,109.16) --
	( 40.83,109.16) --
	( 40.83,109.16) --
	( 47.79,109.16) --
	( 47.79,109.16) --
	( 54.74,109.16) --
	( 54.74,109.22) --
	( 61.70,109.22) --
	( 61.70,109.33) --
	( 68.65,109.33) --
	( 68.65,109.60) --
	( 75.61,109.60) --
	( 75.61,112.56) --
	( 82.57,112.56) --
	( 82.57,118.15) --
	( 89.52,118.15) --
	( 89.52,126.75) --
	( 96.48,126.75) --
	( 96.48,123.51) --
	(103.44,123.51) --
	(103.44,114.15) --
	(110.39,114.15) --
	(110.39,111.74) --
	(117.35,111.74) --
	(117.35,110.26) --
	(124.30,110.26) --
	(124.30,109.87) --
	(131.26,109.87) --
	(131.26,109.49) --
	(138.22,109.49) --
	(138.22,109.16) --
	(145.17,109.16) --
	(145.17,109.16) --
	(138.22,109.16) --
	(138.22,109.16) --
	(131.26,109.16) --
	(131.26,109.16) --
	(124.30,109.16) --
	(124.30,109.16) --
	(117.35,109.16) --
	(117.35,109.16) --
	(110.39,109.16) --
	(110.39,109.16) --
	(103.44,109.16) --
	(103.44,109.16) --
	( 96.48,109.16) --
	( 96.48,109.16) --
	( 89.52,109.16) --
	( 89.52,109.16) --
	( 82.57,109.16) --
	( 82.57,109.16) --
	( 75.61,109.16) --
	( 75.61,109.16) --
	( 68.65,109.16) --
	( 68.65,109.16) --
	( 61.70,109.16) --
	( 61.70,109.16) --
	( 54.74,109.16) --
	( 54.74,109.16) --
	( 47.79,109.16) --
	( 47.79,109.16) --
	( 40.83,109.16) --
	( 40.83,109.16) --
	( 33.87,109.16) --
	( 33.87,109.16) --
	( 26.92,109.16) --
	cycle;
\definecolor{fillColor}{RGB}{0,0,0}

\path[draw=drawColor,line width= 0.6pt,line join=round,line cap=round,fill=fillColor] ( 26.92, 89.02) --
	( 33.87, 89.02) --
	( 33.87, 89.02) --
	( 40.83, 89.02) --
	( 40.83, 89.13) --
	( 47.79, 89.13) --
	( 47.79, 89.19) --
	( 54.74, 89.19) --
	( 54.74, 89.46) --
	( 61.70, 89.46) --
	( 61.70, 89.96) --
	( 68.65, 89.96) --
	( 68.65, 91.56) --
	( 75.61, 91.56) --
	( 75.61, 92.88) --
	( 82.57, 92.88) --
	( 82.57, 97.00) --
	( 89.52, 97.00) --
	( 89.52,101.29) --
	( 96.48,101.29) --
	( 96.48,103.93) --
	(103.44,103.93) --
	(103.44, 94.75) --
	(110.39, 94.75) --
	(110.39, 91.78) --
	(117.35, 91.78) --
	(117.35, 90.90) --
	(124.30, 90.90) --
	(124.30, 89.69) --
	(131.26, 89.69) --
	(131.26, 89.52) --
	(138.22, 89.52) --
	(138.22, 89.02) --
	(145.17, 89.02) --
	(145.17, 89.02) --
	(138.22, 89.02) --
	(138.22, 89.02) --
	(131.26, 89.02) --
	(131.26, 89.02) --
	(124.30, 89.02) --
	(124.30, 89.02) --
	(117.35, 89.02) --
	(117.35, 89.02) --
	(110.39, 89.02) --
	(110.39, 89.02) --
	(103.44, 89.02) --
	(103.44, 89.02) --
	( 96.48, 89.02) --
	( 96.48, 89.02) --
	( 89.52, 89.02) --
	( 89.52, 89.02) --
	( 82.57, 89.02) --
	( 82.57, 89.02) --
	( 75.61, 89.02) --
	( 75.61, 89.02) --
	( 68.65, 89.02) --
	( 68.65, 89.02) --
	( 61.70, 89.02) --
	( 61.70, 89.02) --
	( 54.74, 89.02) --
	( 54.74, 89.02) --
	( 47.79, 89.02) --
	( 47.79, 89.02) --
	( 40.83, 89.02) --
	( 40.83, 89.02) --
	( 33.87, 89.02) --
	( 33.87, 89.02) --
	( 26.92, 89.02) --
	cycle;
\definecolor{drawColor}{gray}{0.20}

\path[draw=drawColor,line width= 0.6pt,line join=round,line cap=round] ( 21.01, 76.94) rectangle (151.08,229.98);
\end{scope}
\begin{scope}
\path[clip] (  0.00,  0.00) rectangle (469.76,252.94);
\definecolor{drawColor}{gray}{0.20}

\path[draw=drawColor,line width= 0.6pt,line join=round] ( 18.26, 89.02) --
	( 21.01, 89.02);

\path[draw=drawColor,line width= 0.6pt,line join=round] ( 18.26,109.16) --
	( 21.01,109.16);

\path[draw=drawColor,line width= 0.6pt,line join=round] ( 18.26,129.30) --
	( 21.01,129.30);

\path[draw=drawColor,line width= 0.6pt,line join=round] ( 18.26,149.44) --
	( 21.01,149.44);

\path[draw=drawColor,line width= 0.6pt,line join=round] ( 18.26,169.57) --
	( 21.01,169.57);

\path[draw=drawColor,line width= 0.6pt,line join=round] ( 18.26,189.71) --
	( 21.01,189.71);
\end{scope}
\begin{scope}
\path[clip] (  0.00,  0.00) rectangle (469.76,252.94);
\definecolor{drawColor}{gray}{0.20}

\path[draw=drawColor,line width= 0.6pt,line join=round] ( 59.96, 74.19) --
	( 59.96, 76.94);

\path[draw=drawColor,line width= 0.6pt,line join=round] (103.44, 74.19) --
	(103.44, 76.94);

\path[draw=drawColor,line width= 0.6pt,line join=round] (146.91, 74.19) --
	(146.91, 76.94);
\end{scope}
\begin{scope}
\path[clip] (  0.00,  0.00) rectangle (469.76,252.94);
\definecolor{drawColor}{gray}{0.30}

\node[text=drawColor,anchor=base,inner sep=0pt, outer sep=0pt, scale=  0.88] at ( 59.96, 65.93) {-0.25};

\node[text=drawColor,anchor=base,inner sep=0pt, outer sep=0pt, scale=  0.88] at (103.44, 65.93) {0.00};

\node[text=drawColor,anchor=base,inner sep=0pt, outer sep=0pt, scale=  0.88] at (146.91, 65.93) {0.25};
\end{scope}
\begin{scope}
\path[clip] (  0.00,  0.00) rectangle (469.76,252.94);
\definecolor{drawColor}{RGB}{0,0,0}

\node[text=drawColor,anchor=base,inner sep=0pt, outer sep=0pt, scale=  1.10] at ( 86.05, 53.62) {Loss reduction};
\end{scope}
\begin{scope}
\path[clip] (  0.00,  0.00) rectangle (469.76,252.94);
\definecolor{drawColor}{RGB}{0,0,0}

\node[text=drawColor,anchor=base west,inner sep=0pt, outer sep=0pt, scale=  1.32] at ( 21.01,238.35) {$n=30$};
\end{scope}
\begin{scope}
\path[clip] (156.59, 45.69) rectangle (313.17,252.94);
\definecolor{drawColor}{RGB}{255,255,255}
\definecolor{fillColor}{RGB}{255,255,255}

\path[draw=drawColor,line width= 0.6pt,line join=round,line cap=round,fill=fillColor] (156.59, 45.69) rectangle (313.17,252.95);
\end{scope}
\begin{scope}
\path[clip] (177.59, 76.94) rectangle (307.67,229.98);
\definecolor{fillColor}{RGB}{255,255,255}

\path[fill=fillColor] (177.59, 76.94) rectangle (307.67,229.98);
\definecolor{drawColor}{gray}{0.92}

\path[draw=drawColor,line width= 0.3pt,line join=round] (204.20, 76.94) --
	(204.20,229.98);

\path[draw=drawColor,line width= 0.3pt,line join=round] (233.76, 76.94) --
	(233.76,229.98);

\path[draw=drawColor,line width= 0.3pt,line join=round] (263.32, 76.94) --
	(263.32,229.98);

\path[draw=drawColor,line width= 0.3pt,line join=round] (292.89, 76.94) --
	(292.89,229.98);

\path[draw=drawColor,line width= 0.6pt,line join=round] (177.59, 89.02) --
	(307.67, 89.02);

\path[draw=drawColor,line width= 0.6pt,line join=round] (177.59,109.16) --
	(307.67,109.16);

\path[draw=drawColor,line width= 0.6pt,line join=round] (177.59,129.30) --
	(307.67,129.30);

\path[draw=drawColor,line width= 0.6pt,line join=round] (177.59,149.44) --
	(307.67,149.44);

\path[draw=drawColor,line width= 0.6pt,line join=round] (177.59,169.57) --
	(307.67,169.57);

\path[draw=drawColor,line width= 0.6pt,line join=round] (177.59,189.71) --
	(307.67,189.71);

\path[draw=drawColor,line width= 0.6pt,line join=round] (189.42, 76.94) --
	(189.42,229.98);

\path[draw=drawColor,line width= 0.6pt,line join=round] (218.98, 76.94) --
	(218.98,229.98);

\path[draw=drawColor,line width= 0.6pt,line join=round] (248.54, 76.94) --
	(248.54,229.98);

\path[draw=drawColor,line width= 0.6pt,line join=round] (278.11, 76.94) --
	(278.11,229.98);
\definecolor{drawColor}{RGB}{0,0,0}
\definecolor{fillColor}{RGB}{0,114,178}

\path[draw=drawColor,line width= 0.6pt,line join=round,line cap=round,fill=fillColor] (183.50,189.71) --
	(189.42,189.71) --
	(189.42,189.71) --
	(195.33,189.71) --
	(195.33,189.71) --
	(201.24,189.71) --
	(201.24,189.71) --
	(207.15,189.71) --
	(207.15,189.71) --
	(213.07,189.71) --
	(213.07,189.71) --
	(218.98,189.71) --
	(218.98,189.71) --
	(224.89,189.71) --
	(224.89,189.71) --
	(230.80,189.71) --
	(230.80,189.71) --
	(236.72,189.71) --
	(236.72,189.71) --
	(242.63,189.71) --
	(242.63,189.71) --
	(248.54,189.71) --
	(248.54,229.98) --
	(254.46,229.98) --
	(254.46,199.14) --
	(260.37,199.14) --
	(260.37,193.67) --
	(266.28,193.67) --
	(266.28,191.86) --
	(272.19,191.86) --
	(272.19,190.99) --
	(278.11,190.99) --
	(278.11,190.00) --
	(284.02,190.00) --
	(284.02,190.00) --
	(289.93,190.00) --
	(289.93,189.83) --
	(295.84,189.83) --
	(295.84,189.71) --
	(301.76,189.71) --
	(301.76,189.71) --
	(295.84,189.71) --
	(295.84,189.71) --
	(289.93,189.71) --
	(289.93,189.71) --
	(284.02,189.71) --
	(284.02,189.71) --
	(278.11,189.71) --
	(278.11,189.71) --
	(272.19,189.71) --
	(272.19,189.71) --
	(266.28,189.71) --
	(266.28,189.71) --
	(260.37,189.71) --
	(260.37,189.71) --
	(254.46,189.71) --
	(254.46,189.71) --
	(248.54,189.71) --
	(248.54,189.71) --
	(242.63,189.71) --
	(242.63,189.71) --
	(236.72,189.71) --
	(236.72,189.71) --
	(230.80,189.71) --
	(230.80,189.71) --
	(224.89,189.71) --
	(224.89,189.71) --
	(218.98,189.71) --
	(218.98,189.71) --
	(213.07,189.71) --
	(213.07,189.71) --
	(207.15,189.71) --
	(207.15,189.71) --
	(201.24,189.71) --
	(201.24,189.71) --
	(195.33,189.71) --
	(195.33,189.71) --
	(189.42,189.71) --
	(189.42,189.71) --
	(183.50,189.71) --
	cycle;
\definecolor{fillColor}{RGB}{240,228,66}

\path[draw=drawColor,line width= 0.6pt,line join=round,line cap=round,fill=fillColor] (183.50,169.57) --
	(189.42,169.57) --
	(189.42,169.63) --
	(195.33,169.63) --
	(195.33,169.69) --
	(201.24,169.69) --
	(201.24,169.69) --
	(207.15,169.69) --
	(207.15,169.86) --
	(213.07,169.86) --
	(213.07,169.80) --
	(218.98,169.80) --
	(218.98,170.91) --
	(224.89,170.91) --
	(224.89,171.84) --
	(230.80,171.84) --
	(230.80,173.52) --
	(236.72,173.52) --
	(236.72,178.86) --
	(242.63,178.86) --
	(242.63,208.27) --
	(248.54,208.27) --
	(248.54,171.02) --
	(254.46,171.02) --
	(254.46,169.57) --
	(260.37,169.57) --
	(260.37,169.57) --
	(266.28,169.57) --
	(266.28,169.57) --
	(272.19,169.57) --
	(272.19,169.57) --
	(278.11,169.57) --
	(278.11,169.57) --
	(284.02,169.57) --
	(284.02,169.57) --
	(289.93,169.57) --
	(289.93,169.57) --
	(295.84,169.57) --
	(295.84,169.57) --
	(301.76,169.57) --
	(301.76,169.57) --
	(295.84,169.57) --
	(295.84,169.57) --
	(289.93,169.57) --
	(289.93,169.57) --
	(284.02,169.57) --
	(284.02,169.57) --
	(278.11,169.57) --
	(278.11,169.57) --
	(272.19,169.57) --
	(272.19,169.57) --
	(266.28,169.57) --
	(266.28,169.57) --
	(260.37,169.57) --
	(260.37,169.57) --
	(254.46,169.57) --
	(254.46,169.57) --
	(248.54,169.57) --
	(248.54,169.57) --
	(242.63,169.57) --
	(242.63,169.57) --
	(236.72,169.57) --
	(236.72,169.57) --
	(230.80,169.57) --
	(230.80,169.57) --
	(224.89,169.57) --
	(224.89,169.57) --
	(218.98,169.57) --
	(218.98,169.57) --
	(213.07,169.57) --
	(213.07,169.57) --
	(207.15,169.57) --
	(207.15,169.57) --
	(201.24,169.57) --
	(201.24,169.57) --
	(195.33,169.57) --
	(195.33,169.57) --
	(189.42,169.57) --
	(189.42,169.57) --
	(183.50,169.57) --
	cycle;
\definecolor{fillColor}{RGB}{0,158,115}

\path[draw=drawColor,line width= 0.6pt,line join=round,line cap=round,fill=fillColor] (183.50,149.44) --
	(189.42,149.44) --
	(189.42,149.44) --
	(195.33,149.44) --
	(195.33,149.44) --
	(201.24,149.44) --
	(201.24,149.44) --
	(207.15,149.44) --
	(207.15,149.44) --
	(213.07,149.44) --
	(213.07,149.44) --
	(218.98,149.44) --
	(218.98,149.44) --
	(224.89,149.44) --
	(224.89,149.61) --
	(230.80,149.61) --
	(230.80,160.23) --
	(236.72,160.23) --
	(236.72,177.58) --
	(242.63,177.58) --
	(242.63,159.59) --
	(248.54,159.59) --
	(248.54,153.79) --
	(254.46,153.79) --
	(254.46,151.35) --
	(260.37,151.35) --
	(260.37,150.65) --
	(266.28,150.65) --
	(266.28,149.96) --
	(272.19,149.96) --
	(272.19,149.67) --
	(278.11,149.67) --
	(278.11,149.55) --
	(284.02,149.55) --
	(284.02,149.55) --
	(289.93,149.55) --
	(289.93,149.49) --
	(295.84,149.49) --
	(295.84,149.44) --
	(301.76,149.44) --
	(301.76,149.44) --
	(295.84,149.44) --
	(295.84,149.44) --
	(289.93,149.44) --
	(289.93,149.44) --
	(284.02,149.44) --
	(284.02,149.44) --
	(278.11,149.44) --
	(278.11,149.44) --
	(272.19,149.44) --
	(272.19,149.44) --
	(266.28,149.44) --
	(266.28,149.44) --
	(260.37,149.44) --
	(260.37,149.44) --
	(254.46,149.44) --
	(254.46,149.44) --
	(248.54,149.44) --
	(248.54,149.44) --
	(242.63,149.44) --
	(242.63,149.44) --
	(236.72,149.44) --
	(236.72,149.44) --
	(230.80,149.44) --
	(230.80,149.44) --
	(224.89,149.44) --
	(224.89,149.44) --
	(218.98,149.44) --
	(218.98,149.44) --
	(213.07,149.44) --
	(213.07,149.44) --
	(207.15,149.44) --
	(207.15,149.44) --
	(201.24,149.44) --
	(201.24,149.44) --
	(195.33,149.44) --
	(195.33,149.44) --
	(189.42,149.44) --
	(189.42,149.44) --
	(183.50,149.44) --
	cycle;
\definecolor{fillColor}{RGB}{86,180,233}

\path[draw=drawColor,line width= 0.6pt,line join=round,line cap=round,fill=fillColor] (183.50,129.30) --
	(189.42,129.30) --
	(189.42,129.30) --
	(195.33,129.30) --
	(195.33,129.30) --
	(201.24,129.30) --
	(201.24,129.30) --
	(207.15,129.30) --
	(207.15,129.30) --
	(213.07,129.30) --
	(213.07,129.30) --
	(218.98,129.30) --
	(218.98,129.30) --
	(224.89,129.30) --
	(224.89,129.30) --
	(230.80,129.30) --
	(230.80,140.90) --
	(236.72,140.90) --
	(236.72,157.61) --
	(242.63,157.61) --
	(242.63,139.05) --
	(248.54,139.05) --
	(248.54,133.01) --
	(254.46,133.01) --
	(254.46,131.16) --
	(260.37,131.16) --
	(260.37,130.63) --
	(266.28,130.63) --
	(266.28,129.88) --
	(272.19,129.88) --
	(272.19,129.65) --
	(278.11,129.65) --
	(278.11,129.36) --
	(284.02,129.36) --
	(284.02,129.47) --
	(289.93,129.47) --
	(289.93,129.36) --
	(295.84,129.36) --
	(295.84,129.30) --
	(301.76,129.30) --
	(301.76,129.30) --
	(295.84,129.30) --
	(295.84,129.30) --
	(289.93,129.30) --
	(289.93,129.30) --
	(284.02,129.30) --
	(284.02,129.30) --
	(278.11,129.30) --
	(278.11,129.30) --
	(272.19,129.30) --
	(272.19,129.30) --
	(266.28,129.30) --
	(266.28,129.30) --
	(260.37,129.30) --
	(260.37,129.30) --
	(254.46,129.30) --
	(254.46,129.30) --
	(248.54,129.30) --
	(248.54,129.30) --
	(242.63,129.30) --
	(242.63,129.30) --
	(236.72,129.30) --
	(236.72,129.30) --
	(230.80,129.30) --
	(230.80,129.30) --
	(224.89,129.30) --
	(224.89,129.30) --
	(218.98,129.30) --
	(218.98,129.30) --
	(213.07,129.30) --
	(213.07,129.30) --
	(207.15,129.30) --
	(207.15,129.30) --
	(201.24,129.30) --
	(201.24,129.30) --
	(195.33,129.30) --
	(195.33,129.30) --
	(189.42,129.30) --
	(189.42,129.30) --
	(183.50,129.30) --
	cycle;
\definecolor{fillColor}{RGB}{230,159,0}

\path[draw=drawColor,line width= 0.6pt,line join=round,line cap=round,fill=fillColor] (183.50,109.16) --
	(189.42,109.16) --
	(189.42,109.16) --
	(195.33,109.16) --
	(195.33,109.16) --
	(201.24,109.16) --
	(201.24,109.16) --
	(207.15,109.16) --
	(207.15,109.16) --
	(213.07,109.16) --
	(213.07,109.57) --
	(218.98,109.57) --
	(218.98,110.09) --
	(224.89,110.09) --
	(224.89,113.80) --
	(230.80,113.80) --
	(230.80,120.24) --
	(236.72,120.24) --
	(236.72,128.02) --
	(242.63,128.02) --
	(242.63,121.99) --
	(248.54,121.99) --
	(248.54,113.92) --
	(254.46,113.92) --
	(254.46,110.79) --
	(260.37,110.79) --
	(260.37,110.32) --
	(266.28,110.32) --
	(266.28,110.09) --
	(272.19,110.09) --
	(272.19,109.45) --
	(278.11,109.45) --
	(278.11,109.34) --
	(284.02,109.34) --
	(284.02,109.28) --
	(289.93,109.28) --
	(289.93,109.16) --
	(295.84,109.16) --
	(295.84,109.16) --
	(301.76,109.16) --
	(301.76,109.16) --
	(295.84,109.16) --
	(295.84,109.16) --
	(289.93,109.16) --
	(289.93,109.16) --
	(284.02,109.16) --
	(284.02,109.16) --
	(278.11,109.16) --
	(278.11,109.16) --
	(272.19,109.16) --
	(272.19,109.16) --
	(266.28,109.16) --
	(266.28,109.16) --
	(260.37,109.16) --
	(260.37,109.16) --
	(254.46,109.16) --
	(254.46,109.16) --
	(248.54,109.16) --
	(248.54,109.16) --
	(242.63,109.16) --
	(242.63,109.16) --
	(236.72,109.16) --
	(236.72,109.16) --
	(230.80,109.16) --
	(230.80,109.16) --
	(224.89,109.16) --
	(224.89,109.16) --
	(218.98,109.16) --
	(218.98,109.16) --
	(213.07,109.16) --
	(213.07,109.16) --
	(207.15,109.16) --
	(207.15,109.16) --
	(201.24,109.16) --
	(201.24,109.16) --
	(195.33,109.16) --
	(195.33,109.16) --
	(189.42,109.16) --
	(189.42,109.16) --
	(183.50,109.16) --
	cycle;
\definecolor{fillColor}{RGB}{0,0,0}

\path[draw=drawColor,line width= 0.6pt,line join=round,line cap=round,fill=fillColor] (183.50, 89.02) --
	(189.42, 89.02) --
	(189.42, 89.02) --
	(195.33, 89.02) --
	(195.33, 89.32) --
	(201.24, 89.32) --
	(201.24, 89.43) --
	(207.15, 89.43) --
	(207.15, 89.61) --
	(213.07, 89.61) --
	(213.07, 90.30) --
	(218.98, 90.30) --
	(218.98, 91.06) --
	(224.89, 91.06) --
	(224.89, 95.29) --
	(230.80, 95.29) --
	(230.80, 98.60) --
	(236.72, 98.60) --
	(236.72,102.43) --
	(242.63,102.43) --
	(242.63,101.50) --
	(248.54,101.50) --
	(248.54, 95.23) --
	(254.46, 95.23) --
	(254.46, 91.23) --
	(260.37, 91.23) --
	(260.37, 90.36) --
	(266.28, 90.36) --
	(266.28, 89.78) --
	(272.19, 89.78) --
	(272.19, 89.66) --
	(278.11, 89.66) --
	(278.11, 89.20) --
	(284.02, 89.20) --
	(284.02, 89.20) --
	(289.93, 89.20) --
	(289.93, 89.02) --
	(295.84, 89.02) --
	(295.84, 89.02) --
	(301.76, 89.02) --
	(301.76, 89.02) --
	(295.84, 89.02) --
	(295.84, 89.02) --
	(289.93, 89.02) --
	(289.93, 89.02) --
	(284.02, 89.02) --
	(284.02, 89.02) --
	(278.11, 89.02) --
	(278.11, 89.02) --
	(272.19, 89.02) --
	(272.19, 89.02) --
	(266.28, 89.02) --
	(266.28, 89.02) --
	(260.37, 89.02) --
	(260.37, 89.02) --
	(254.46, 89.02) --
	(254.46, 89.02) --
	(248.54, 89.02) --
	(248.54, 89.02) --
	(242.63, 89.02) --
	(242.63, 89.02) --
	(236.72, 89.02) --
	(236.72, 89.02) --
	(230.80, 89.02) --
	(230.80, 89.02) --
	(224.89, 89.02) --
	(224.89, 89.02) --
	(218.98, 89.02) --
	(218.98, 89.02) --
	(213.07, 89.02) --
	(213.07, 89.02) --
	(207.15, 89.02) --
	(207.15, 89.02) --
	(201.24, 89.02) --
	(201.24, 89.02) --
	(195.33, 89.02) --
	(195.33, 89.02) --
	(189.42, 89.02) --
	(189.42, 89.02) --
	(183.50, 89.02) --
	cycle;
\definecolor{drawColor}{gray}{0.20}

\path[draw=drawColor,line width= 0.6pt,line join=round,line cap=round] (177.59, 76.94) rectangle (307.67,229.98);
\end{scope}
\begin{scope}
\path[clip] (  0.00,  0.00) rectangle (469.76,252.94);
\definecolor{drawColor}{gray}{0.20}

\path[draw=drawColor,line width= 0.6pt,line join=round] (174.84, 89.02) --
	(177.59, 89.02);

\path[draw=drawColor,line width= 0.6pt,line join=round] (174.84,109.16) --
	(177.59,109.16);

\path[draw=drawColor,line width= 0.6pt,line join=round] (174.84,129.30) --
	(177.59,129.30);

\path[draw=drawColor,line width= 0.6pt,line join=round] (174.84,149.44) --
	(177.59,149.44);

\path[draw=drawColor,line width= 0.6pt,line join=round] (174.84,169.57) --
	(177.59,169.57);

\path[draw=drawColor,line width= 0.6pt,line join=round] (174.84,189.71) --
	(177.59,189.71);
\end{scope}
\begin{scope}
\path[clip] (  0.00,  0.00) rectangle (469.76,252.94);
\definecolor{drawColor}{gray}{0.20}

\path[draw=drawColor,line width= 0.6pt,line join=round] (189.42, 74.19) --
	(189.42, 76.94);

\path[draw=drawColor,line width= 0.6pt,line join=round] (218.98, 74.19) --
	(218.98, 76.94);

\path[draw=drawColor,line width= 0.6pt,line join=round] (248.54, 74.19) --
	(248.54, 76.94);

\path[draw=drawColor,line width= 0.6pt,line join=round] (278.11, 74.19) --
	(278.11, 76.94);
\end{scope}
\begin{scope}
\path[clip] (  0.00,  0.00) rectangle (469.76,252.94);
\definecolor{drawColor}{gray}{0.30}

\node[text=drawColor,anchor=base,inner sep=0pt, outer sep=0pt, scale=  0.88] at (189.42, 65.93) {-0.10};

\node[text=drawColor,anchor=base,inner sep=0pt, outer sep=0pt, scale=  0.88] at (218.98, 65.93) {-0.05};

\node[text=drawColor,anchor=base,inner sep=0pt, outer sep=0pt, scale=  0.88] at (248.54, 65.93) {0.00};

\node[text=drawColor,anchor=base,inner sep=0pt, outer sep=0pt, scale=  0.88] at (278.11, 65.93) {0.05};
\end{scope}
\begin{scope}
\path[clip] (  0.00,  0.00) rectangle (469.76,252.94);
\definecolor{drawColor}{RGB}{0,0,0}

\node[text=drawColor,anchor=base,inner sep=0pt, outer sep=0pt, scale=  1.10] at (242.63, 53.62) {Loss reduction};
\end{scope}
\begin{scope}
\path[clip] (  0.00,  0.00) rectangle (469.76,252.94);
\definecolor{drawColor}{RGB}{0,0,0}

\node[text=drawColor,anchor=base west,inner sep=0pt, outer sep=0pt, scale=  1.32] at (177.59,238.35) {$n=100$};
\end{scope}
\begin{scope}
\path[clip] (313.17, 45.69) rectangle (469.76,252.94);
\definecolor{drawColor}{RGB}{255,255,255}
\definecolor{fillColor}{RGB}{255,255,255}

\path[draw=drawColor,line width= 0.6pt,line join=round,line cap=round,fill=fillColor] (313.17, 45.69) rectangle (469.76,252.95);
\end{scope}
\begin{scope}
\path[clip] (334.18, 76.94) rectangle (464.26,229.98);
\definecolor{fillColor}{RGB}{255,255,255}

\path[fill=fillColor] (334.18, 76.94) rectangle (464.26,229.98);
\definecolor{drawColor}{gray}{0.92}

\path[draw=drawColor,line width= 0.3pt,line join=round] (363.08, 76.94) --
	(363.08,229.98);

\path[draw=drawColor,line width= 0.3pt,line join=round] (395.93, 76.94) --
	(395.93,229.98);

\path[draw=drawColor,line width= 0.3pt,line join=round] (428.78, 76.94) --
	(428.78,229.98);

\path[draw=drawColor,line width= 0.3pt,line join=round] (461.63, 76.94) --
	(461.63,229.98);

\path[draw=drawColor,line width= 0.6pt,line join=round] (334.18, 89.02) --
	(464.26, 89.02);

\path[draw=drawColor,line width= 0.6pt,line join=round] (334.18,109.16) --
	(464.26,109.16);

\path[draw=drawColor,line width= 0.6pt,line join=round] (334.18,129.30) --
	(464.26,129.30);

\path[draw=drawColor,line width= 0.6pt,line join=round] (334.18,149.44) --
	(464.26,149.44);

\path[draw=drawColor,line width= 0.6pt,line join=round] (334.18,169.57) --
	(464.26,169.57);

\path[draw=drawColor,line width= 0.6pt,line join=round] (334.18,189.71) --
	(464.26,189.71);

\path[draw=drawColor,line width= 0.6pt,line join=round] (346.66, 76.94) --
	(346.66,229.98);

\path[draw=drawColor,line width= 0.6pt,line join=round] (379.51, 76.94) --
	(379.51,229.98);

\path[draw=drawColor,line width= 0.6pt,line join=round] (412.35, 76.94) --
	(412.35,229.98);

\path[draw=drawColor,line width= 0.6pt,line join=round] (445.20, 76.94) --
	(445.20,229.98);
\definecolor{drawColor}{RGB}{0,0,0}
\definecolor{fillColor}{RGB}{0,114,178}

\path[draw=drawColor,line width= 0.6pt,line join=round,line cap=round,fill=fillColor] (340.09,189.71) --
	(346.66,189.71) --
	(346.66,189.71) --
	(353.23,189.71) --
	(353.23,189.71) --
	(359.80,189.71) --
	(359.80,189.71) --
	(366.37,189.71) --
	(366.37,189.71) --
	(372.94,189.71) --
	(372.94,189.71) --
	(379.51,189.71) --
	(379.51,189.71) --
	(386.08,189.71) --
	(386.08,189.71) --
	(392.65,189.71) --
	(392.65,189.71) --
	(399.21,189.71) --
	(399.22,189.71) --
	(405.78,189.71) --
	(405.78,189.71) --
	(412.35,189.71) --
	(412.35,229.98) --
	(418.92,229.98) --
	(418.92,198.96) --
	(425.49,198.96) --
	(425.49,194.39) --
	(432.06,194.39) --
	(432.06,191.76) --
	(438.63,191.76) --
	(438.63,190.59) --
	(445.20,190.59) --
	(445.20,190.24) --
	(451.77,190.24) --
	(451.77,189.71) --
	(458.34,189.71) --
	(458.34,189.71) --
	(451.77,189.71) --
	(451.77,189.71) --
	(445.20,189.71) --
	(445.20,189.71) --
	(438.63,189.71) --
	(438.63,189.71) --
	(432.06,189.71) --
	(432.06,189.71) --
	(425.49,189.71) --
	(425.49,189.71) --
	(418.92,189.71) --
	(418.92,189.71) --
	(412.35,189.71) --
	(412.35,189.71) --
	(405.78,189.71) --
	(405.78,189.71) --
	(399.22,189.71) --
	(399.21,189.71) --
	(392.65,189.71) --
	(392.65,189.71) --
	(386.08,189.71) --
	(386.08,189.71) --
	(379.51,189.71) --
	(379.51,189.71) --
	(372.94,189.71) --
	(372.94,189.71) --
	(366.37,189.71) --
	(366.37,189.71) --
	(359.80,189.71) --
	(359.80,189.71) --
	(353.23,189.71) --
	(353.23,189.71) --
	(346.66,189.71) --
	(346.66,189.71) --
	(340.09,189.71) --
	cycle;
\definecolor{fillColor}{RGB}{240,228,66}

\path[draw=drawColor,line width= 0.6pt,line join=round,line cap=round,fill=fillColor] (340.09,169.57) --
	(346.66,169.57) --
	(346.66,169.63) --
	(353.23,169.63) --
	(353.23,169.63) --
	(359.80,169.63) --
	(359.80,169.69) --
	(366.37,169.69) --
	(366.37,169.98) --
	(372.94,169.98) --
	(372.94,170.27) --
	(379.51,170.27) --
	(379.51,170.61) --
	(386.08,170.61) --
	(386.08,171.42) --
	(392.65,171.42) --
	(392.65,174.19) --
	(399.21,174.19) --
	(399.22,178.52) --
	(405.78,178.52) --
	(405.78,207.78) --
	(412.35,207.78) --
	(412.35,171.25) --
	(418.92,171.25) --
	(418.92,169.57) --
	(425.49,169.57) --
	(425.49,169.57) --
	(432.06,169.57) --
	(432.06,169.57) --
	(438.63,169.57) --
	(438.63,169.57) --
	(445.20,169.57) --
	(445.20,169.57) --
	(451.77,169.57) --
	(451.77,169.57) --
	(458.34,169.57) --
	(458.34,169.57) --
	(451.77,169.57) --
	(451.77,169.57) --
	(445.20,169.57) --
	(445.20,169.57) --
	(438.63,169.57) --
	(438.63,169.57) --
	(432.06,169.57) --
	(432.06,169.57) --
	(425.49,169.57) --
	(425.49,169.57) --
	(418.92,169.57) --
	(418.92,169.57) --
	(412.35,169.57) --
	(412.35,169.57) --
	(405.78,169.57) --
	(405.78,169.57) --
	(399.22,169.57) --
	(399.21,169.57) --
	(392.65,169.57) --
	(392.65,169.57) --
	(386.08,169.57) --
	(386.08,169.57) --
	(379.51,169.57) --
	(379.51,169.57) --
	(372.94,169.57) --
	(372.94,169.57) --
	(366.37,169.57) --
	(366.37,169.57) --
	(359.80,169.57) --
	(359.80,169.57) --
	(353.23,169.57) --
	(353.23,169.57) --
	(346.66,169.57) --
	(346.66,169.57) --
	(340.09,169.57) --
	cycle;
\definecolor{fillColor}{RGB}{0,158,115}

\path[draw=drawColor,line width= 0.6pt,line join=round,line cap=round,fill=fillColor] (340.09,149.44) --
	(346.66,149.44) --
	(346.66,149.44) --
	(353.23,149.44) --
	(353.23,149.44) --
	(359.80,149.44) --
	(359.80,149.44) --
	(366.37,149.44) --
	(366.37,149.44) --
	(372.94,149.44) --
	(372.94,149.44) --
	(379.51,149.44) --
	(379.51,149.44) --
	(386.08,149.44) --
	(386.08,149.44) --
	(392.65,149.44) --
	(392.65,159.15) --
	(399.21,159.15) --
	(399.22,178.87) --
	(405.78,178.87) --
	(405.78,159.38) --
	(412.35,159.38) --
	(412.35,154.06) --
	(418.92,154.06) --
	(418.92,151.11) --
	(425.49,151.11) --
	(425.49,150.53) --
	(432.06,150.53) --
	(432.06,149.96) --
	(438.63,149.96) --
	(438.63,149.96) --
	(445.20,149.96) --
	(445.20,149.55) --
	(451.77,149.55) --
	(451.77,149.44) --
	(458.34,149.44) --
	(458.34,149.44) --
	(451.77,149.44) --
	(451.77,149.44) --
	(445.20,149.44) --
	(445.20,149.44) --
	(438.63,149.44) --
	(438.63,149.44) --
	(432.06,149.44) --
	(432.06,149.44) --
	(425.49,149.44) --
	(425.49,149.44) --
	(418.92,149.44) --
	(418.92,149.44) --
	(412.35,149.44) --
	(412.35,149.44) --
	(405.78,149.44) --
	(405.78,149.44) --
	(399.22,149.44) --
	(399.21,149.44) --
	(392.65,149.44) --
	(392.65,149.44) --
	(386.08,149.44) --
	(386.08,149.44) --
	(379.51,149.44) --
	(379.51,149.44) --
	(372.94,149.44) --
	(372.94,149.44) --
	(366.37,149.44) --
	(366.37,149.44) --
	(359.80,149.44) --
	(359.80,149.44) --
	(353.23,149.44) --
	(353.23,149.44) --
	(346.66,149.44) --
	(346.66,149.44) --
	(340.09,149.44) --
	cycle;
\definecolor{fillColor}{RGB}{86,180,233}

\path[draw=drawColor,line width= 0.6pt,line join=round,line cap=round,fill=fillColor] (340.09,129.30) --
	(346.66,129.30) --
	(346.66,129.30) --
	(353.23,129.30) --
	(353.23,129.30) --
	(359.80,129.30) --
	(359.80,129.30) --
	(366.37,129.30) --
	(366.37,129.30) --
	(372.94,129.30) --
	(372.94,129.30) --
	(379.51,129.30) --
	(379.51,129.30) --
	(386.08,129.30) --
	(386.08,129.30) --
	(392.65,129.30) --
	(392.65,135.89) --
	(399.21,135.89) --
	(399.22,162.09) --
	(405.78,162.09) --
	(405.78,138.90) --
	(412.35,138.90) --
	(412.35,133.87) --
	(418.92,133.87) --
	(418.92,131.27) --
	(425.49,131.27) --
	(425.49,130.17) --
	(432.06,130.17) --
	(432.06,129.88) --
	(438.63,129.88) --
	(438.63,129.82) --
	(445.20,129.82) --
	(445.20,129.47) --
	(451.77,129.47) --
	(451.77,129.30) --
	(458.34,129.30) --
	(458.34,129.30) --
	(451.77,129.30) --
	(451.77,129.30) --
	(445.20,129.30) --
	(445.20,129.30) --
	(438.63,129.30) --
	(438.63,129.30) --
	(432.06,129.30) --
	(432.06,129.30) --
	(425.49,129.30) --
	(425.49,129.30) --
	(418.92,129.30) --
	(418.92,129.30) --
	(412.35,129.30) --
	(412.35,129.30) --
	(405.78,129.30) --
	(405.78,129.30) --
	(399.22,129.30) --
	(399.21,129.30) --
	(392.65,129.30) --
	(392.65,129.30) --
	(386.08,129.30) --
	(386.08,129.30) --
	(379.51,129.30) --
	(379.51,129.30) --
	(372.94,129.30) --
	(372.94,129.30) --
	(366.37,129.30) --
	(366.37,129.30) --
	(359.80,129.30) --
	(359.80,129.30) --
	(353.23,129.30) --
	(353.23,129.30) --
	(346.66,129.30) --
	(346.66,129.30) --
	(340.09,129.30) --
	cycle;
\definecolor{fillColor}{RGB}{230,159,0}

\path[draw=drawColor,line width= 0.6pt,line join=round,line cap=round,fill=fillColor] (340.09,109.16) --
	(346.66,109.16) --
	(346.66,109.16) --
	(353.23,109.16) --
	(353.23,109.22) --
	(359.80,109.22) --
	(359.80,109.16) --
	(366.37,109.16) --
	(366.37,109.16) --
	(372.94,109.16) --
	(372.94,109.39) --
	(379.51,109.39) --
	(379.51,110.43) --
	(386.08,110.43) --
	(386.08,113.61) --
	(392.65,113.61) --
	(392.65,120.44) --
	(399.21,120.44) --
	(399.22,127.55) --
	(405.78,127.55) --
	(405.78,121.54) --
	(412.35,121.54) --
	(412.35,113.90) --
	(418.92,113.90) --
	(418.92,111.88) --
	(425.49,111.88) --
	(425.49,109.86) --
	(432.06,109.86) --
	(432.06,109.91) --
	(438.63,109.91) --
	(438.63,109.68) --
	(445.20,109.68) --
	(445.20,109.34) --
	(451.77,109.34) --
	(451.77,109.16) --
	(458.34,109.16) --
	(458.34,109.16) --
	(451.77,109.16) --
	(451.77,109.16) --
	(445.20,109.16) --
	(445.20,109.16) --
	(438.63,109.16) --
	(438.63,109.16) --
	(432.06,109.16) --
	(432.06,109.16) --
	(425.49,109.16) --
	(425.49,109.16) --
	(418.92,109.16) --
	(418.92,109.16) --
	(412.35,109.16) --
	(412.35,109.16) --
	(405.78,109.16) --
	(405.78,109.16) --
	(399.22,109.16) --
	(399.21,109.16) --
	(392.65,109.16) --
	(392.65,109.16) --
	(386.08,109.16) --
	(386.08,109.16) --
	(379.51,109.16) --
	(379.51,109.16) --
	(372.94,109.16) --
	(372.94,109.16) --
	(366.37,109.16) --
	(366.37,109.16) --
	(359.80,109.16) --
	(359.80,109.16) --
	(353.23,109.16) --
	(353.23,109.16) --
	(346.66,109.16) --
	(346.66,109.16) --
	(340.09,109.16) --
	cycle;
\definecolor{fillColor}{RGB}{0,0,0}

\path[draw=drawColor,line width= 0.6pt,line join=round,line cap=round,fill=fillColor] (340.09, 89.02) --
	(346.66, 89.02) --
	(346.66, 89.14) --
	(353.23, 89.14) --
	(353.23, 89.08) --
	(359.80, 89.08) --
	(359.80, 89.20) --
	(366.37, 89.20) --
	(366.37, 89.49) --
	(372.94, 89.49) --
	(372.94, 90.47) --
	(379.51, 90.47) --
	(379.51, 91.74) --
	(386.08, 91.74) --
	(386.08, 93.48) --
	(392.65, 93.48) --
	(392.65, 98.74) --
	(399.21, 98.74) --
	(399.22,102.04) --
	(405.78,102.04) --
	(405.78,103.54) --
	(412.35,103.54) --
	(412.35, 94.63) --
	(418.92, 94.63) --
	(418.92, 91.63) --
	(425.49, 91.63) --
	(425.49, 90.36) --
	(432.06, 90.36) --
	(432.06, 89.83) --
	(438.63, 89.83) --
	(438.63, 89.43) --
	(445.20, 89.43) --
	(445.20, 89.26) --
	(451.77, 89.26) --
	(451.77, 89.02) --
	(458.34, 89.02) --
	(458.34, 89.02) --
	(451.77, 89.02) --
	(451.77, 89.02) --
	(445.20, 89.02) --
	(445.20, 89.02) --
	(438.63, 89.02) --
	(438.63, 89.02) --
	(432.06, 89.02) --
	(432.06, 89.02) --
	(425.49, 89.02) --
	(425.49, 89.02) --
	(418.92, 89.02) --
	(418.92, 89.02) --
	(412.35, 89.02) --
	(412.35, 89.02) --
	(405.78, 89.02) --
	(405.78, 89.02) --
	(399.22, 89.02) --
	(399.21, 89.02) --
	(392.65, 89.02) --
	(392.65, 89.02) --
	(386.08, 89.02) --
	(386.08, 89.02) --
	(379.51, 89.02) --
	(379.51, 89.02) --
	(372.94, 89.02) --
	(372.94, 89.02) --
	(366.37, 89.02) --
	(366.37, 89.02) --
	(359.80, 89.02) --
	(359.80, 89.02) --
	(353.23, 89.02) --
	(353.23, 89.02) --
	(346.66, 89.02) --
	(346.66, 89.02) --
	(340.09, 89.02) --
	cycle;
\definecolor{drawColor}{gray}{0.20}

\path[draw=drawColor,line width= 0.6pt,line join=round,line cap=round] (334.18, 76.94) rectangle (464.26,229.98);
\end{scope}
\begin{scope}
\path[clip] (  0.00,  0.00) rectangle (469.76,252.94);
\definecolor{drawColor}{gray}{0.20}

\path[draw=drawColor,line width= 0.6pt,line join=round] (331.43, 89.02) --
	(334.18, 89.02);

\path[draw=drawColor,line width= 0.6pt,line join=round] (331.43,109.16) --
	(334.18,109.16);

\path[draw=drawColor,line width= 0.6pt,line join=round] (331.43,129.30) --
	(334.18,129.30);

\path[draw=drawColor,line width= 0.6pt,line join=round] (331.43,149.44) --
	(334.18,149.44);

\path[draw=drawColor,line width= 0.6pt,line join=round] (331.43,169.57) --
	(334.18,169.57);

\path[draw=drawColor,line width= 0.6pt,line join=round] (331.43,189.71) --
	(334.18,189.71);
\end{scope}
\begin{scope}
\path[clip] (  0.00,  0.00) rectangle (469.76,252.94);
\definecolor{drawColor}{gray}{0.20}

\path[draw=drawColor,line width= 0.6pt,line join=round] (346.66, 74.19) --
	(346.66, 76.94);

\path[draw=drawColor,line width= 0.6pt,line join=round] (379.51, 74.19) --
	(379.51, 76.94);

\path[draw=drawColor,line width= 0.6pt,line join=round] (412.35, 74.19) --
	(412.35, 76.94);

\path[draw=drawColor,line width= 0.6pt,line join=round] (445.20, 74.19) --
	(445.20, 76.94);
\end{scope}
\begin{scope}
\path[clip] (  0.00,  0.00) rectangle (469.76,252.94);
\definecolor{drawColor}{gray}{0.30}

\node[text=drawColor,anchor=base,inner sep=0pt, outer sep=0pt, scale=  0.88] at (346.66, 65.93) {-0.010};

\node[text=drawColor,anchor=base,inner sep=0pt, outer sep=0pt, scale=  0.88] at (379.51, 65.93) {-0.005};

\node[text=drawColor,anchor=base,inner sep=0pt, outer sep=0pt, scale=  0.88] at (412.35, 65.93) {0.000};

\node[text=drawColor,anchor=base,inner sep=0pt, outer sep=0pt, scale=  0.88] at (445.20, 65.93) {0.005};
\end{scope}
\begin{scope}
\path[clip] (  0.00,  0.00) rectangle (469.76,252.94);
\definecolor{drawColor}{RGB}{0,0,0}

\node[text=drawColor,anchor=base,inner sep=0pt, outer sep=0pt, scale=  1.10] at (399.22, 53.62) {Loss reduction};
\end{scope}
\begin{scope}
\path[clip] (  0.00,  0.00) rectangle (469.76,252.94);
\definecolor{drawColor}{RGB}{0,0,0}

\node[text=drawColor,anchor=base west,inner sep=0pt, outer sep=0pt, scale=  1.32] at (334.18,238.35) {$n=1000$};
\end{scope}
\begin{scope}
\path[clip] (  0.00,  0.00) rectangle (469.76,252.94);
\definecolor{fillColor}{RGB}{255,255,255}

\path[fill=fillColor] ( 79.58,  0.00) rectangle (390.17, 45.69);
\end{scope}
\begin{scope}
\path[clip] (  0.00,  0.00) rectangle (469.76,252.94);
\definecolor{fillColor}{RGB}{255,255,255}

\path[fill=fillColor] ( 90.58, 22.84) rectangle (107.93, 40.19);
\end{scope}
\begin{scope}
\path[clip] (  0.00,  0.00) rectangle (469.76,252.94);
\definecolor{drawColor}{RGB}{0,0,0}
\definecolor{fillColor}{RGB}{0,0,0}

\path[draw=drawColor,line width= 0.6pt,line cap=round,fill=fillColor] ( 91.30, 23.56) rectangle (107.22, 39.48);
\end{scope}
\begin{scope}
\path[clip] (  0.00,  0.00) rectangle (469.76,252.94);
\definecolor{fillColor}{RGB}{255,255,255}

\path[fill=fillColor] ( 90.58,  5.50) rectangle (107.93, 22.84);
\end{scope}
\begin{scope}
\path[clip] (  0.00,  0.00) rectangle (469.76,252.94);
\definecolor{drawColor}{RGB}{0,0,0}
\definecolor{fillColor}{RGB}{230,159,0}

\path[draw=drawColor,line width= 0.6pt,line cap=round,fill=fillColor] ( 91.30,  6.21) rectangle (107.22, 22.13);
\end{scope}
\begin{scope}
\path[clip] (  0.00,  0.00) rectangle (469.76,252.94);
\definecolor{fillColor}{RGB}{255,255,255}

\path[fill=fillColor] (172.38, 22.84) rectangle (189.72, 40.19);
\end{scope}
\begin{scope}
\path[clip] (  0.00,  0.00) rectangle (469.76,252.94);
\definecolor{drawColor}{RGB}{0,0,0}
\definecolor{fillColor}{RGB}{86,180,233}

\path[draw=drawColor,line width= 0.6pt,line cap=round,fill=fillColor] (173.09, 23.56) rectangle (189.01, 39.48);
\end{scope}
\begin{scope}
\path[clip] (  0.00,  0.00) rectangle (469.76,252.94);
\definecolor{fillColor}{RGB}{255,255,255}

\path[fill=fillColor] (172.38,  5.50) rectangle (189.72, 22.84);
\end{scope}
\begin{scope}
\path[clip] (  0.00,  0.00) rectangle (469.76,252.94);
\definecolor{drawColor}{RGB}{0,0,0}
\definecolor{fillColor}{RGB}{0,158,115}

\path[draw=drawColor,line width= 0.6pt,line cap=round,fill=fillColor] (173.09,  6.21) rectangle (189.01, 22.13);
\end{scope}
\begin{scope}
\path[clip] (  0.00,  0.00) rectangle (469.76,252.94);
\definecolor{fillColor}{RGB}{255,255,255}

\path[fill=fillColor] (294.07, 22.84) rectangle (311.42, 40.19);
\end{scope}
\begin{scope}
\path[clip] (  0.00,  0.00) rectangle (469.76,252.94);
\definecolor{drawColor}{RGB}{0,0,0}
\definecolor{fillColor}{RGB}{240,228,66}

\path[draw=drawColor,line width= 0.6pt,line cap=round,fill=fillColor] (294.78, 23.56) rectangle (310.71, 39.48);
\end{scope}
\begin{scope}
\path[clip] (  0.00,  0.00) rectangle (469.76,252.94);
\definecolor{fillColor}{RGB}{255,255,255}

\path[fill=fillColor] (294.07,  5.50) rectangle (311.42, 22.84);
\end{scope}
\begin{scope}
\path[clip] (  0.00,  0.00) rectangle (469.76,252.94);
\definecolor{drawColor}{RGB}{0,0,0}
\definecolor{fillColor}{RGB}{0,114,178}

\path[draw=drawColor,line width= 0.6pt,line cap=round,fill=fillColor] (294.78,  6.21) rectangle (310.71, 22.13);
\end{scope}
\begin{scope}
\path[clip] (  0.00,  0.00) rectangle (469.76,252.94);
\definecolor{drawColor}{RGB}{0,0,0}

\node[text=drawColor,anchor=base west,inner sep=0pt, outer sep=0pt, scale=  0.88] at (113.43, 28.49) {$5$-fold CV};
\end{scope}
\begin{scope}
\path[clip] (  0.00,  0.00) rectangle (469.76,252.94);
\definecolor{drawColor}{RGB}{0,0,0}

\node[text=drawColor,anchor=base west,inner sep=0pt, outer sep=0pt, scale=  0.88] at (113.43, 11.14) {$10$-fold CV};
\end{scope}
\begin{scope}
\path[clip] (  0.00,  0.00) rectangle (469.76,252.94);
\definecolor{drawColor}{RGB}{0,0,0}

\node[text=drawColor,anchor=base west,inner sep=0pt, outer sep=0pt, scale=  0.88] at (195.22, 28.49) {$n$-fold CV};
\end{scope}
\begin{scope}
\path[clip] (  0.00,  0.00) rectangle (469.76,252.94);
\definecolor{drawColor}{RGB}{0,0,0}

\node[text=drawColor,anchor=base west,inner sep=0pt, outer sep=0pt, scale=  0.88] at (195.22, 11.14) {$\tilde{\mathcal{R}}^0$};
\end{scope}
\begin{scope}
\path[clip] (  0.00,  0.00) rectangle (469.76,252.94);
\definecolor{drawColor}{RGB}{0,0,0}

\node[text=drawColor,anchor=base west,inner sep=0pt, outer sep=0pt, scale=  0.88] at (316.92, 28.49) {$\mathcal{R}^0$};
\end{scope}
\begin{scope}
\path[clip] (  0.00,  0.00) rectangle (469.76,252.94);
\definecolor{drawColor}{RGB}{0,0,0}

\node[text=drawColor,anchor=base west,inner sep=0pt, outer sep=0pt, scale=  0.88] at (316.92, 11.14) {$\mathcal{R}$};
\end{scope}
\end{tikzpicture}

%% file: Benchmark_datasets.tex
\section{Comparisons on benchmark datasets}\label{sec:comparisons on benchmark datasets}

\begin{table} \centering 
\begin{tabular}{@{\extracolsep{4pt}}lccccc} 
\\[-1.8ex]\hline 
\hline \\[-1.8ex] 
Dataset & \multicolumn{1}{c}{$n\times m$} & \multicolumn{1}{c}{Loss function} & \multicolumn{1}{c}{train vs test} & \multicolumn{1}{c}{Source packages} \\ 
\hline \\[-1.8ex] 
Boston & $506 \times 14$  & MSE & $50-50$ & 	MASS	  \\
Ozone &  $111\times 4$  & MSE &  $50-50$   & 	ElemStatLearn \\
Auto & $392\times 311$  & MSE &  $70-30$   & 	ISLR  \\
Carseats & $400\times 12$  & MSE & $70-30$   & 	ISLR	  \\
College & $777\times 18$  & MSE &  $70-30$   & 	ISLR	  \\
Hitters & $263\times 20$ & MSE &  $70-30$    & 	ISLR	  \\
Wage & $3000\times 26$  & MSE & $70-30$   & 		ISLR  \\
\hline
Caravan & $5822\times 86$	& Logloss & $70-30$		 & 	ISLR	  \\
Default & $10000\times 4$	& Logloss & $70-30$			 & 	ISLR  \\
OJ & $1070\times 18$	& Logloss & $70-30$		 & ISLR	  \\
Smarket & $1250\times 7$	& Logloss & $70-30$		 & 	ISLR  \\
Weekly & $1089\times 7$	& Logloss & $70-30$		 & 	ISLR  \\
\hline \\[-1.8ex] 
\end{tabular} 
\caption[]{
All regression and classification datasets from the books \citet{friedman2001elements, james2013introduction}, their dimensions, loss functions (MSE corresponds to regression, Logloss to classification), the percentage split to training and test, and source.
Dimensions are after using the R function \texttt{model.matrix()}, which performs one-hot encoding on the data, and remove NA values.
See Table 1.1 in \citet{james2013introduction} for further descriptions of the datasets.
} 
\label{tab:datasets} 
\end{table}

To further illustrate the validity of the modified boosting algorithm implemented in aGTBoost, we test it on all regression and classification datasets in \citet{friedman2001elements} and \citet{james2013introduction}.
These datasets represent a relatively broad spectrum of model-types (Table \ref{tab:datasets}).

\subsection{Algorithms}\label{subsec:real-data-methods}

Our algorithm is compared against the \texttt{xgboost} implementation.
Our hypothesis is that the two algorithms will give similar predictions, but will differ in 
computation time and ease of use. 
To ensure comparability, we avoid L1 and L2 regularization of the loss and stochastic sampling in \texttt{xgboost}.
In addition, we include random forest and generalized linear models in the comparisons.
Lastly, we include a version of our proposed algorithm restricted to a single ($K=1$) unscaled ($\delta=1$) tree, and a CART tree learned with CV and cost complexity
pruning. This gives additional validation of the root-stump criterion \eqref{eq:stopping criterion splitting}.

\subsection{Computation}\label{subsec:real-data-computation}

Computations are done in \texttt{R} version 3.6.1 on a Dell XPS-15 computer running 64-bit Windows 10, utilizing only a single core for comparability of algorithms.
We run \texttt{xgboost} 0.90.0.2, \texttt{randomForest} 4.6-14 and \texttt{tree} 1.0-40 which contain the CART algorithm.
GLM algorithms are found in the base-\texttt{R} \texttt{stats} library, through the functions \texttt{lm()} for linear regression, and \texttt{glm()} with specified \texttt{family=binomial} for logistic regression.
For \texttt{randomForest} we use the default parameter values. 
The same is the case for \texttt{lm} and \texttt{glm}, while \texttt{tree} is trained using pruning on a potentially deep tree.

For the results in Table \ref{tab: dataset results}, \texttt{xgboost} is trained with a learning rate of $\delta=0.1$, the same as aGTBoost, and importantly, L2 regularization are removed from the boosting objective by setting the (by-default non-zero) \texttt{lambda} parameter to zero. The number of trees, $K$, for \texttt{xgboost} models are found by
10-fold CV, where we check if the 10 consecutive trees improve overall CV-loss, selected by setting
\texttt{early\_stopping\_rounds}=10.
The configuration of \texttt{xgboost} in Table \ref{tab: cpu times} is identical to Table \ref{tab: dataset results}, except for the learning rate set to $\delta = 0.01$ (same as for aGTBoost).
The different variants of \texttt{xgboost} in Table \ref{tab: cpu times} differ in the CV profiling over the hyperparameters \texttt{max\_depth} and \texttt{gamma}. Also, a variant using 30\% of the training data as a validation set for selecting $K$ is included.

Each dataset is split randomly into a training set and a test set (see Table \ref{tab:datasets}). 
All algorithms train on the same training set, and report the loss over the test set.
This is done for 100 different splits, and the mean and standard deviation of relative  test loss (to \texttt{xgboost}) is calculated across these 100 datasets.

\begin{table}
\begin{tabular}{@{\extracolsep{4pt}}lccccccc} 
\\[-1.8ex]\hline 
\hline \\[-1.8ex] 
Dataset & \multicolumn{1}{c}{xgboost} & \multicolumn{1}{c}{aGTBoost} & \multicolumn{1}{c}{random forest} & \multicolumn{1}{c}{glm} & \multicolumn{1}{c}{CART} & \multicolumn{1}{c}{gbtree} \\ 
\hline \\[-1.8ex] 
Boston  & 1 (0.173)  & 1.02 (0.144)  & 0.877 (0.15)  & 1.3 (0.179)  & 1.55 (0.179)  & 1.64 (0.215)  \\  
Ozone  & 1 (0.202)  & 0.816 (0.2)  & 0.675 (0.183)  & 0.672 (0.132)  & 0.945 (0.225)  & 1.13 (0.216)  \\  
Auto  & 1 (0.188)  & 0.99 (0.119)  & 0.895 (0.134)  & 11.1 (14.6)  & 1.45 (0.185)  & 1.45 (0.201)  \\  
Carseats  & 1 (0.112)  & 0.956 (0.126)  & 1.16 (0.141)  & 0.414 (0.0433)  & 1.84 (0.212)  & 1.9 (0.195)  \\  
College  & 1 (0.818)  & 1.27 (0.917)  & 1.07 (0.909)  & 0.552 (0.155)  & 1.46 (0.881)  & 1.71 (1.08)  \\  
Hitters  & 1 (0.323)  & 0.977 (0.366)  & 0.798 (0.311)  & 1.21 (0.348)  & 1.23 (0.338)  & 1.21 (0.408) \\  
Wage  & 1 (1.01)  & 1.39 (1.64)  & 82.5 (21.4)  & 290 (35.5)  & 109 (6.78)  & 2.41 (1.91)  \\  
\hline
Caravan  & 1 (0.052)  & 0.983 (0.0491)  & 1.3 (0.167)  & 1.12 (0.115)   \\  
Default  & 1 (0.0803)  & 0.926 (0.0675)  & 2.82 (0.508)  & 0.898 (0.0696)   \\  
OJ  & 1 (0.0705)  & 0.966 (0.0541)  & 1.17 (0.183)  & 0.949 (0.0719)   \\  
Smarket  & 1 (0.00401)  & 0.997 (0.00311)  & 1.04 (0.0163)  & 1 (0.0065)   \\  
Weekly  & 1 (0.00759)  & 0.992 (0.00829)  & 1.02 (0.0195)  & 0.995 (0.0123)   \\ 
\hline \\[-1.8ex] 
\end{tabular} 
\caption[]{
Average relative test-loss and standard deviations (parentheses) across 100 random splits of the full datasets into training and test, for the 
datasets in \ref{tab:datasets}.
The reported values are relative to the average \texttt{xgboost} test-loss values.
\texttt{aGTBoost} is the modified boosting algorithm \ref{alg:gradient-tree-boosting}, gbtree is a regression tree stopping according to \eqref{eq:stopping criterion splitting}, CART is from the R package "tree", GLM uses a linear regression model for MSE-loss and logistic regression for classification. Random forest uses the default settings in the "randomForest" R package, while \texttt{xgboost} is trained deterministically with CV on the number of trees with maximum depth 6 but no L1 or L2 regularisation.
The learning rate, $\delta$, is set to 0.1 for both \texttt{aGTBoost} and \texttt{xgboost}.
}
\label{tab: dataset results} 
\end{table}
\begin{table}
\begin{tabular}{@{\extracolsep{4pt}}lcccccc} 
\\[-1.8ex]\hline 
\hline \\[-1.8ex] 
& aGTBoost & \multicolumn{5}{c}{xgboost} \\
\cmidrule(lr){3-7} 
Variant & 	& $30\%$ Validation & 
$K$ & $K$, gamma & $K$, max depth & $K$, gamma, max depth\\
\hline \\[-1.8ex] 
Runtime (seconds) &	1.46 & 1.3 & 8.55 & 190 & 90.6 & 2033   \\
Test loss & 0.3792 & 0.4229 & 0.3985 & 0.3839 & 0.3743 & 0.3983 \\
\hline \\[-1.8ex] 
\end{tabular}
\caption{CPU computations time in seconds for the training of \texttt{aGTBoost} versus different variants (Section \ref{subsec:real-data-computation}) of \texttt{xgboost} for the "OJ" dataset. 
\texttt{gamma} takes values on integers from 0-9, and max depth takes values on integers 1-10.	
Also reported is the loss on $30\%$ test data. The naive test loss (constant prediction) is 0.662. 
}
\label{tab: cpu times} 
\end{table}

\subsection{Results}\label{subsec:real-data-results}
Consider first the two rightmost columns in Table \ref{tab: dataset results}, reporting the 
results from the CART and gbtree single-tree models.
These constitute the building blocks of \texttt{xgboost} and \texttt{aGTBoost}, respectively, and might therefore indicate an explanation for potential differences in 
the results of \texttt{xgboost} and \texttt{aGTBoost}.
Overall, the results are fairly similar with a slight advantage for CART, but well within the 
standard deviations of Table \ref{tab: dataset results}, except for the Wage
data. 
The fundamental difference of the CART trees and gbtree lies in the tree-building method of CART which performs consecutive splitting, also after encountering the first split giving a negative reduction in loss, until 
a pre-defined depth is reached and then a pruning process is initiated.
The gbtree method, on the other hand, and by extension aGTBoost, stops splitting immediately when encountering the first split giving a negative loss reduction in approximate generalization loss.
Most of the results favour slightly the cost-complexity pruning done by CART.
However, the wage data strongly favour gbtree, showing that the adaptiveness of gbtree
has other advantages than just speed and ease-of-use. The CART trees are constrained by their 
default setting for tree-depth, which is likely to cause the inferior performance for this dataset. The adaptive gbtrees, on the other hand, are able to 
build rather deep trees. 
Overall, the results are so similar that we would be hard pressed to attribute potential large differences in \texttt{xgboost} and \texttt{aGTBoost} to their individual tree building algorithms.

We then turn to the comparison of \texttt{xgboost} and \texttt{aGTBoost} in Table \ref{tab: dataset results}.
aGTBoost outperforms \texttt{xgboost} on 9 out of 12 datasets, although the average test losses are within the Monte-Carlo (permutation) uncertainty of each other.
The results for the other methods, random forest and GLM, gives an additional perspective on difference between \texttt{xgboost} and \texttt{aGTBoost}.
For some datasets the GLM and random forest have slightly lower test-loss, but for other significantly higher test-loss.

Having demonstrated similar performance as regularized un-penalized \texttt{xgboost}, the vantage point of \texttt{aGTBoost} is its automatic properties and as a consequence, speed.
Table \ref{tab: cpu times} tells a story of computational benefits to this adaptivity: What took 1.46 seconds for \texttt{aGTBoost} took a regularized \texttt{xgboost} ($K$, \texttt{gamma}, \texttt{max\_depth} variant) 2033 seconds.
Furthermore, this adaptivity does not only have computational benefits, but also decreases the threshold for users that are new to tree-boosting: By eliminating the need to set up a search grid for the \texttt{gamma} and the \texttt{max\_depth} hyperparameters in \texttt{xgboost}, \texttt{aGTBoost} lowers the bar to employ gradient tree boosting as an off-the-shelf method for practitioners.
Notice also that of all the different variants of \texttt{xgboost}, only one (tuning $K$ and maximum depth), slightly outperformed \texttt{aGTBoost} in terms of test-loss.
A final observation is that simultaneously tuning $K$, \texttt{gamma} and \texttt{max\_depth}, gives higher test-loss than only tuning $K$ and \texttt{max\_depth} in \texttt{xgboost}.
This is likely due to the high variation inherent in CV.

%% file: Discussion.tex
\section{Discussion}\label{sec:discussion}

This paper proposes an information criterion for the individual node splits in gradient 
boosted trees, which allows for a modified and more automatic gradient tree 
boosting procedure as described in Algorithm \ref{alg:gradient-tree-boosting}.
The proposed method (\texttt{aGTBoost}), and its underlying assumptions, were tested on both 
simulated and real data, and were seen to perform well under all testing regimes.
In particular, the modifications allow for significant improvements in
computational speed for all variants of \texttt{xgboost} involving hyperparameters.
Additionally, \texttt{aGTBoost} lowers the bar for employing GTB as an off-the-shelf algorithm, 
as there is no need to specify a search grid and set up $k$-fold CV for hyperparameters.

One potential problem with \texttt{aGTBoost} is the tendency of early
trees being too deep in complex datasets, as illustrated in Figure \ref{fig:boosting convergence}.
This is because \texttt{aGTBoost} does not have a global hyperparameter for the maximum
complexity of trees (\texttt{max\_depth} as in \texttt{xgboost}, or a maximum number of leaves hyperparameter).
The problem of too deep trees in GTB was first noted in \citet{friedman2000additive}, who suggested
to put a bound on the number of terminal nodes for all trees in the ensemble.

The leading implementations of GTB come with options to modify the algorithm
with stochastic sampling and L1 and L2 regularization of the loss, 
modifications that often improve generalization scores.
This differ from the deterministic un-penalized GTB flavour discussed in this paper,
and which the theory behind the information criterion assumes.
Further work will try to accommodate these features, and allow for 
automatic tuning of sampling-rates and severity of loss-penalization.

%% file: Appendix.tex
\newpage
\appendix
\setcounter{page}{1}
\begin{center}
\begin{huge}
Online Appendix to "An information criterion for automatic gradient tree boosting" by Lunde, Kleppe and Skaug
\end{huge}
\end{center}
\section{Derivation of Equation \ref{eq:varying optimism as cir}}
\label{app:regularity conditions}
This section derives the CIR limit of stump optimism, as function
of split point $s$. All equation references $<$ \ref{eq:gen_optimism_restated} are for equations in the main paper.

The derivation relies on results for M-estimators. These results rely on certain regularity conditions, which may be found in \citet{vanDerVaart} for Theorem 4.21 page 52, but are restated here for convenience. 
The parameter vector $\theta$ is assumed finite-dimensional and to take values
in an open subset of Euclidian space,
$\theta\in\Theta\subset \mathbb{R}^d$,
further, assume $z_1,\dots,z_n$ to be a sample from some 
distribution $P$. The loss function $l(z,\theta)$ needs to be twice
continuously differentiable,
and we denote its first derivative, the score, as $\psi_\theta(z_i)=\nabla_\theta l(z_i,\theta)$.
Parameter estimates, $\hat{\theta}$, are assumed to solve the following
estimating equations
\begin{align*}
\frac{1}{n}\sum_{i=1}^n \psi_{\hat{\theta}}(z_i) = 0
\end{align*}
and further consistency with $\hat{ \theta}_n\underset{p}{\to} \theta_0$, where
$\theta_0$ is the population minimizer, i.e. 
$E\left[\psi_{\theta_0}(Z)\right] = 0.$
Finally we impose conditions on the score. 
First a Lipschitz condition: 
For all $\theta_1$ and $\theta_2$ in a neighbourhood of $\theta_0$ and a 
measurable function $H$ with $E[H^2]\leq\infty$, we assume
\begin{align*}
\norm{ \psi_{\theta_1}(z) - \psi_{\theta_2}(z)} 
\leq H \norm{ \theta_1-\theta_2}.
\end{align*}
Lastly that $ E[\norm{\psi_{\theta_0}}^2] < \infty $,
and that the map $\theta \mapsto E[\psi_\theta]$ is differentiable at 
$\theta_0$ with a nonsingular derivative matrix \citep{vanDerVaart}.

Note that it is possible to loosen these conditions and still obtain
asymptotic normality (needed in Section \ref{app:asymptotic normality of score} and 
\ref{app:asymptotic normality of estimator}), for example with regards
to the differentiability of the score function, the estimating equation need not be
exactly zero, but $ o_p(n^{-\frac{1}{2}})$, the Lipschitz condition is too stringent,
and $\theta$ need not be finite dimensional.

However, the gradient boosting approximate loss function we work with, $\hat l$, is 
appropriately differentiable, and allows solutions $\hat{w}$ that are exact zeroes
of the estimating equations. While the set of score functions 
$\{\psi_{\theta_0}(z),~ -\infty < s < \infty \}$ can be established to be a Donsker
class \citep{vanDerVaart}.

\subsection{Insights behind AIC/TIC/NIC}
\label{app:insights behind aic-tic-nic}

When parameter estimates $\hat{\theta}$ satisfy the regularity conditions
in Section \ref{app:regularity conditions},
importantly, the loss $l$ is differentiable in $\theta$ and estimates are found by minimizing the loss over data
$$\hat{\theta} = \arg\min_\theta \sum_{i=1}^{n}l(y_i,f(x_i;\theta)),$$
then the Akaike Information Criterion (AIC) \citep{akaike1974new}, Takeuchi Information Criterion (TIC) \citep{takeuchi1976distribution} or Network Information Criterion (NIC) \citep{murata1994network} all result in the optimism estimate (\ref{eq:gradient boosting bias approximate}), for convenience given again here: 

\begin{align}
\hat{C} = \texttt{tr}\left(E[\nabla_\theta^2 l(y_1,f(x_1;\theta_0))]Cov(\hat{\theta})\right).\label{eq:gen_optimism_restated}
\end{align}
In the case of TIC and NIC, using the asymptotic normality of $\hat{\theta}$ (see e.g. \citet[Eq. 5.20, p.52]{vanDerVaart}) and the empirical estimator of the Hessian.

AIC follows from assuming that the true data-generating-process is in the family of models being optimized over, and hence asymptotically the $E[\nabla_\theta^2 l(y_1,f(x_1);\theta_0)]Cov(\hat{\theta})=n^{-1}I$. Finally, this result in estimate of the optimism being simply $n^{-1}d$ where $d$ is the number of parameters.

A full derivation of (\ref{eq:gen_optimism_restated}) found in \citet[Chapter 7]{burnham2003model},
and we refer to AIC/TIC/NIC for the original articles and derivations. 
Some insight behind this result is however needed. First, the derivation of (\ref{eq:gen_optimism_restated}) relies on the following approximation which
according to Slutsky's theorem is valid for large $n$:
\begin{align}\label{eq:quadratics independence assumption}
n
\nabla_\theta^2 l(y,f(x;\hat{\theta}))
(\hat{\theta}-\theta_0)(\hat{\theta}-\theta_0)^T
&\approx
n
\left[\nabla_\theta^2 l(y_1,f(x_1;\theta_0))\right]
(\hat{\theta}-\theta_0)(\hat{\theta}-\theta_0)^T,
\end{align}

Further, an approximation expressing the difference in test- and training loss is also derived in  \citep{burnham2003model}:
\begin{align}\label{eq:test_train_loss_approx}
l(y^0, f(x^0;\hat{\theta})) - l(y_1,f(x_1;\hat{\theta})) \approx 
(\hat{\theta}-\theta_0)^T\nabla_\theta^2 l(y^0,f(x^0;\theta_0))(\hat{\theta}-\theta_0).
\end{align}
In the case of a stump CART with fixed split point $s$, (\ref{eq:test_train_loss_approx}) reduces to
\begin{align}\label{eq:test_train_loss_approx_CART}
l(y^0, f(x^0;\hat{\theta})) - 
l(y_1,f(x_1;\hat{\theta})) \approx 
1_{(x^0\leq s)}\frac{\partial^2}{\partial w_{l,0}^2}l(y^0,w_{l,0})(\hat{w}_l-w_{l,0})^2 + 
1_{(x^0> s)}\frac{\partial^2}{\partial w_{r,0}^2}l(y^0,w_{r,0})(\hat{w}_r-w_{r,0})^2,
\end{align}
due to the diagonal Hessian of CART in this case.

In order to characterize the distribution of the right hand side of (\ref{eq:test_train_loss_approx_CART}) also under optimization over $s$, conventional M-estimator asymptotic theory as used in TIC and NIC does not apply directly. 
This is due to the multiple-comparison problem 
for different split-points and subsequent selection of $\hat{w}=(\hat w_l, \hat w_r)$ w.r.t. the 
training loss which effectively changes the distributions of $\hat{w}_l^2$, $\hat{w}_r^2$ relative to those obtained for fixed $s$. The next 
section discuss the distributional change in squares of $\hat{w}$ under 
profiling.

\subsection{A loss function for the deviation from the null-model}
Recall that, conditioned on being in a region with prediction $w$, the relevant 
Taylor expanded loss (\ref{taylor2}), modulus unimportant constant terms, is given
$$\hat{l}(y_1,w) = g(y_1,\hat{y}_1)w + \frac{1}{2}h(y_1,\hat{y}_1)w^2.$$
For simplicity we write $g(y_1,\hat{y}_1)$ and $h(y_1,\hat{y}_1)$, with 
dependence in $y_1$ and $\hat{y}_1$ as $g_1$ and $h_1$ respectively.
Let $w_t$ be the constant prediction in the root-node and $(w_l, w_r)$ be the 
prediction in the left and right descendant nodes. We then write 
$f_{stump}(x_1;\theta)$ for a stump-model, where the parameter $\theta$ holds 
all relevant information of the tree-stump, namely the split-point, and the left 
and right weights $\theta=\{s,w_l,w_r\}$.

We start off with rewriting $\hat{l}(y_1,f_{stump}(x_1,\theta))$, such that
\begin{align}\label{eq:app_omega_def}
\omega_i := \hat{l}(y_i,f_{stump}(x_i,\theta)) - \hat{l}(y_i,w_t),~
i=1,\dots,n,
\end{align}
where $\hat{l}(y_1,w_t)$ is the loss of the root model with constant prediction
$w_t$, and hence $\omega$
is a measure of deviation from the root model. 
Loosely speaking, the idea is to calculate how much deviation from the root model 
we are to expect from pure randomness, and let the split no-split decision 
calculates w.r.t. this threshold.
Further, it is convenient to introduce deviation from root parameters $\tilde w_l=w_l-w_t$ and $\tilde w_r = w_r-w_t$, and modified first order derivatives
$\tilde{g}_i=g_i+h_iw_t$. 
Then $\omega$ might be written 
\begin{align}\label{eq:app:modified loss}
\frac{1}{n}\sum_{i=1}^n \omega_i = 
\frac{1}{n}\sum_{i\in I_l} \left( \tilde{g}_i\tilde{w}_l+\frac{1}{2}h_i\tilde{w}_l^2\right) + 
\frac{1}{n}\sum_{i\in I_r} \left(\tilde{g}_i\tilde{w}_r+\frac{1}{2}h_i\tilde{w}_r^2 \right). 
\end{align}
Notice importantly, that 
$\sum_{i\in I_t}\tilde{g}_i=0$, which for those familiar with Wiener processes 
and the functional convergence of estimators might give immediate associations to 
the Brownian bridge, which indeed follows shortly. Viewing $\omega$ as a loss function,
the estimates of $\tilde{w}_l$ 
and $\tilde{w}_r$ are found directly from the score function / estimating equation
\begin{align}
0 = \nabla_{\tilde{w}} \frac{1}{n}\sum_{i = 1}^n \omega_i
= \frac{1}{n}\sum_{i = 1}^n \tilde{\psi}_i(\tilde{w}),
~
\tilde{\psi}_i(\tilde{w}) := \nabla_{\tilde{w}} \omega_i,
\end{align}
which we for convenience split into the score function for the left and right 
estimators $\tilde{\psi}_{i,l}(w)$ and $\tilde{\psi}_{i,r}(w)$.
Direct calculation gives
\begin{align}
\hat{\tilde{w}}_l = \frac{\sum_{i\in I_l}\tilde{g}_i}{\sum_{i\in I_l}h_i} = \hat{w}_l - \hat{w}_t,~
\hat{\tilde{w}}_r = \frac{\sum_{i\in I_r}\tilde{g}_i}{\sum_{i\in I_r}h_i} = \hat{w}_r - \hat{w}_t,
\end{align}
which verifies that $\tilde w_l = w_l - w_t$, and correspondingly for $\tilde w_r$.

To directly restate the importance of this specification of the loss: The training loss 
reduction, $R$, might now be written only as a function of $\omega_i$'s:
\begin{align}
R = \frac{1}{n} \sum_{i=1}^n \hat{l}(y_i,w_t) - \hat{l}(y_i,f_{stump}(x_i,\theta)) = 
-\frac{1}{n}\sum_{i=1}^n \omega_i,
\end{align}
and therefore, by using the adjustment factor of $R$ given in \eqref{eq:test_train_loss_approx_CART}, to obtain an estimate of $R^0$, gives
\begin{align}\label{eq:diff text-train as deviance quadratics}
\hat{R}^0 = R - 
\hat{\tilde{w}}_l^2
\frac{\partial^2\omega^0}{\partial \tilde{w}_l^2}
-
\hat{\tilde{w}}_r^2 
\frac{\partial^2\omega^0}{\partial \tilde{w}_r^2},
\end{align}
where it is understood that $\omega^0$ 
obtains as (\ref{eq:app_omega_def}) but with $(y^0,x^0)$ in the place
of $(y_i,x_i)$, and with parameters at the population minimizer.
Note that under the true root model, the population minimizers of $\tilde w_l$ and $\tilde w_r$ are zero.

Now, an estimate of reduction in generalization loss may be obtained by 
estimating the expected value. 
To this end, we need to characterize the joint distribution of 
the estimator $\hat{\tilde{w}}$ for any given split-point.
This distribution is obtained in the preceding sections.

\subsection{Asymptotic normality of modified score/estimating equation}
\label{app:asymptotic normality of score}
\begin{figure}
\centering
\includegraphics[width=0.9\textwidth,height=7cm]{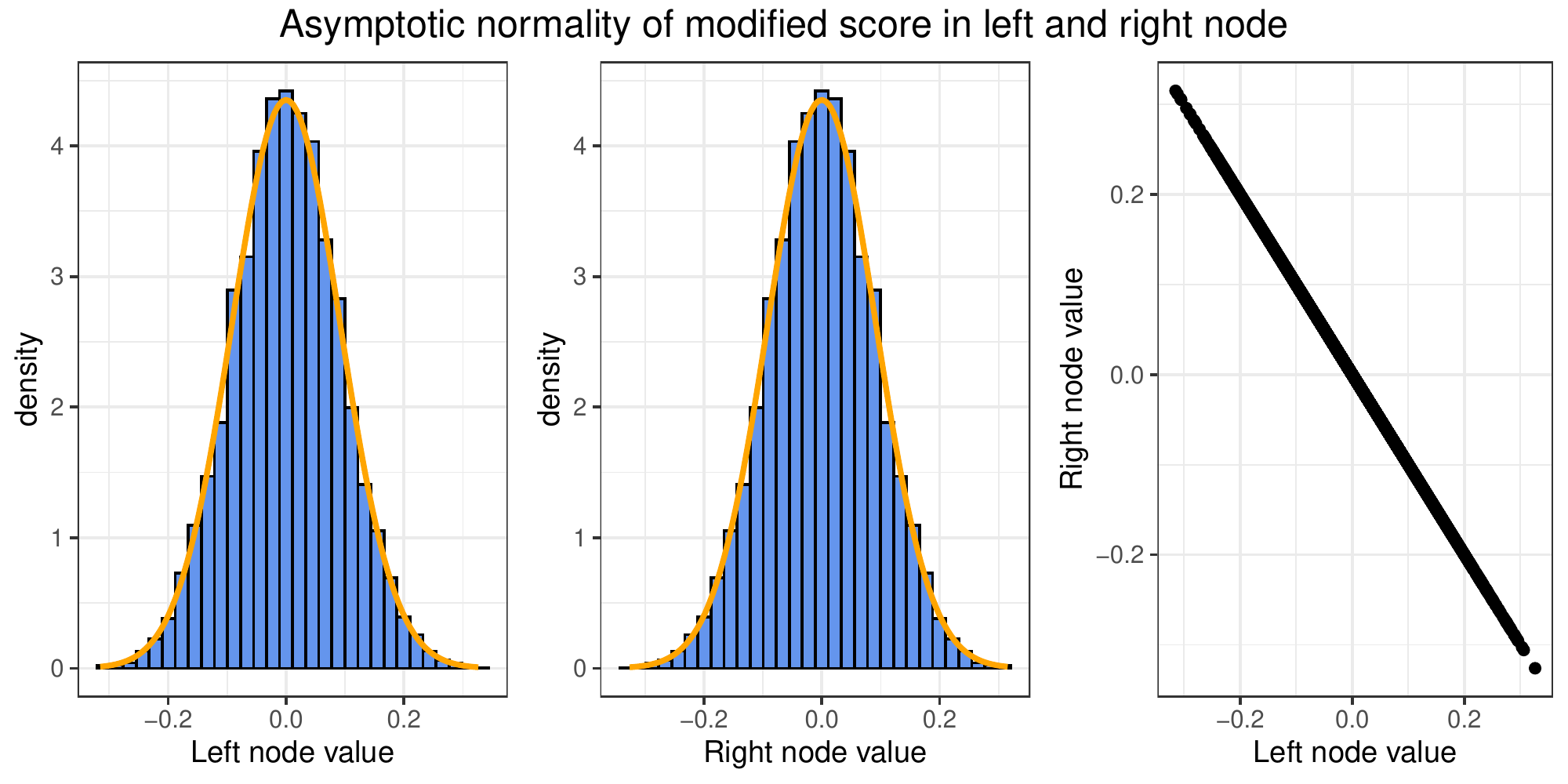} 
\rule{25em}{0.5pt}
\caption[]{
Simulation of the left side of \eqref{eq:app score AN}, and comparison to 
the normal on the right. Simulate $n=100$ observations of $y_i\sim N(3,1)$
and $x_i\sim U(0,1)$, to compute derivatives $g_i$ and $h_i$ using the 
prediction $\hat{y}=2$ with MSE loss. From this, the modified gradients 
$\tilde{g}_i$ are computed, and the two (for finite $n$) quantities on 
the left in \eqref{eq:app score AN} are calculated, using index sets 
$I_l=\{i:x_i\leq u\}$ and $I_r=\{i:x_i>u\}$ for $u=0.3$. This experiment 
is repeated 10000 times to create the observations behind the histogram.
Note that the values are completely dependent, as $\sum_{i\in I_l}\tilde \psi_{i,l}(\tilde w_{l,0}) + \sum_{i\in I_r}\tilde \psi_{i,r}(\tilde w_{r,0}) = 0$.
}
\label{fig: appendix score AN}
\end{figure}

The asymptotic normality of $\hat{\tilde{w}}_l$ and $\hat{\tilde{w}}_r$ 
follows from the convergence of M-estimators to an empirical process. 
We will make use of the following asymptotic result \citep[Theorem 5.21]{vanDerVaart} (Huber, Van der Vaart):
Let $\theta$ be a differentiable parameter satisfying the regularity conditions
in Section \ref{app:regularity conditions}, then
\begin{align}\label{eq:app:parameter asymptotic normality}
\sqrt{n}(\hat{ \theta}-\theta_0) 
\underset{p}{\to}
E[\nabla_\theta^2 l(y,f(x;\theta_0))]^{-1}
E[\psi_i(\tilde{w}_0)\psi_i(\tilde{w}_0)^T].
\end{align}

The remaining part of this subsection finds the (joint) empirical process the 
score converges to.
Specifically, the score of $\hat{\tilde{w}}_l$ can be expanded and written as
\begin{align}\label{eq:app_modified_score}
\tilde{\psi}_{i,l}(\tilde{w}_{l,0}) = \psi_{i,l}(w_{l,0}) - \frac{h_i}{\sum h_i}\psi_{i,t}(w_{t,0}),
\end{align}
where
\begin{align}
\psi_{i,l}(w_l) = (g_i + h_iw_l)1_{(x_i \leq s)}
,~
\psi_{i,t}(w_t) = g_i + h_i w_t,
\end{align}
and completely analogous for $\psi_{i,r}$.
Let $u\in[0,1]$ and define the rescaled partial sum
\begin{align}
S_u = \frac{1}{n}\sum_{i=1}^{\lfloor nu \rfloor} \psi_{i,t}(w_{t,0}).
\end{align}
The CLT gives asymptotic convergence of $\sqrt{n}S_u$ to $N\left(0, uE[\psi_{i,t}(w_{t,0})^2]  \right)$
for any $u\in[0,1]$
However, in our application we need the distribution of $S_u$ 
for an infinite collection of $u$s.
For this purpose, as $\psi_{i,t}(w_{t,0})$s are i.i.d. with finite mean and variance,
we may apply Donsker's 
invariance principle 
that extends the convergence uniformly and simultaneous over all $u\in [0,1]$.
This allow us to write
\begin{align}\label{eq:app_partial_sum_convergence}
\sqrt{n} S_u \to_d \sqrt{E[\psi_{i,t}(w_{t,0})^2]}W(u),
\end{align} 
where $W(u)$ is a standard Brownian motion on $u\in [0,1]$.
Now, for the index $i$ sorted by $x$, and defining $u$ from $u=p(x\leq s)$, then $n^{-1}\sum_{i\in I_l}\psi_{i,l}=S_u$ and $n^{-1}\sum_{i\in I_t}\psi_{i,t} = S_1$.
Furthermore, from the time reversibility property of the Brownian motion, the same result
applies to the right node and $\psi_{i,r}$ but with $(1-u)$ in place of $u$ and perfect
negative dependence with that of the left node.
Lastly, notice that from the law of large numbers, 
$\sum_{i\in I_l}\frac{h_i}{\sum_{i\in I_1} h_i}\to_p u$.
Thus, when inspecting the asymptotic normality of the score of $\omega$, we might use 
\eqref{eq:app_modified_score} together with \eqref{eq:app_partial_sum_convergence} 
to obtain
\begin{align}
\sqrt{n}\sum_{i\in I_l} \tilde{\psi}_{i,l}(\tilde{w}_{l,0})
\to_d \sqrt{E[\psi_{i,t}(w_{t,0})^2]}(W(u)-uW(1)) 
= \sqrt{E[\psi_{i,t}(w_{t,0})^2]}B(u)
\end{align}
where $B(u)$ is a 
standard Brownian bridge on $[0,1]$, i.e. $B(u)\sim N(0,u(1-u))$ and 
$Cov(B(u),B(v))=\min\{u,v\}-uv$.
Necessarily, the standardized sum of scores of $\omega$ in the left and right nodes 
has the same marginal asymptotic distribution
\begin{align}\label{eq:app score AN}
\lim_{n\to\infty}\sqrt{n}\sum_{i\in I_l} \tilde{\psi}_{i,l}(\tilde{w}_{l,0})
\sim \lim_{n\to\infty}\sqrt{n}\sum_{i\in I_{r}} \tilde{\psi}_{i,r}(\tilde{w}_{r,0})
\sim N\left(0,u(1-u)E\left[\psi_{i,t}(w_{t,0})^2\right]\right),
\end{align}
and have perfect negative dependence
\begin{align}
\sqrt{n}
\begin{pmatrix}
\sum_{i\in I_l} \tilde{\psi}_{i,l}(\tilde{w}_{l,0}) \\
\sum_{i\in I_{r}} \tilde{\psi}_{i,r}(\tilde{w}_{r,0})
\end{pmatrix}
\underset{p}{\to}
\sqrt{E[\psi_{i,t}(w_{t,0})^2]}
\begin{pmatrix}
B(t) \\
-B(t)
\end{pmatrix},
\end{align}
as $\sum_{i\in I_l}\tilde \psi_{i,l}(\tilde w_{l,0}) + \sum_{i\in I_r}\tilde \psi_{i,r}(\tilde w_{r,0}) = 0.$

\subsection{Asymptotic normality of modified estimator }
\label{app:asymptotic normality of estimator}
\begin{figure}
\centering
\includegraphics[width=0.9\textwidth,height=7cm]{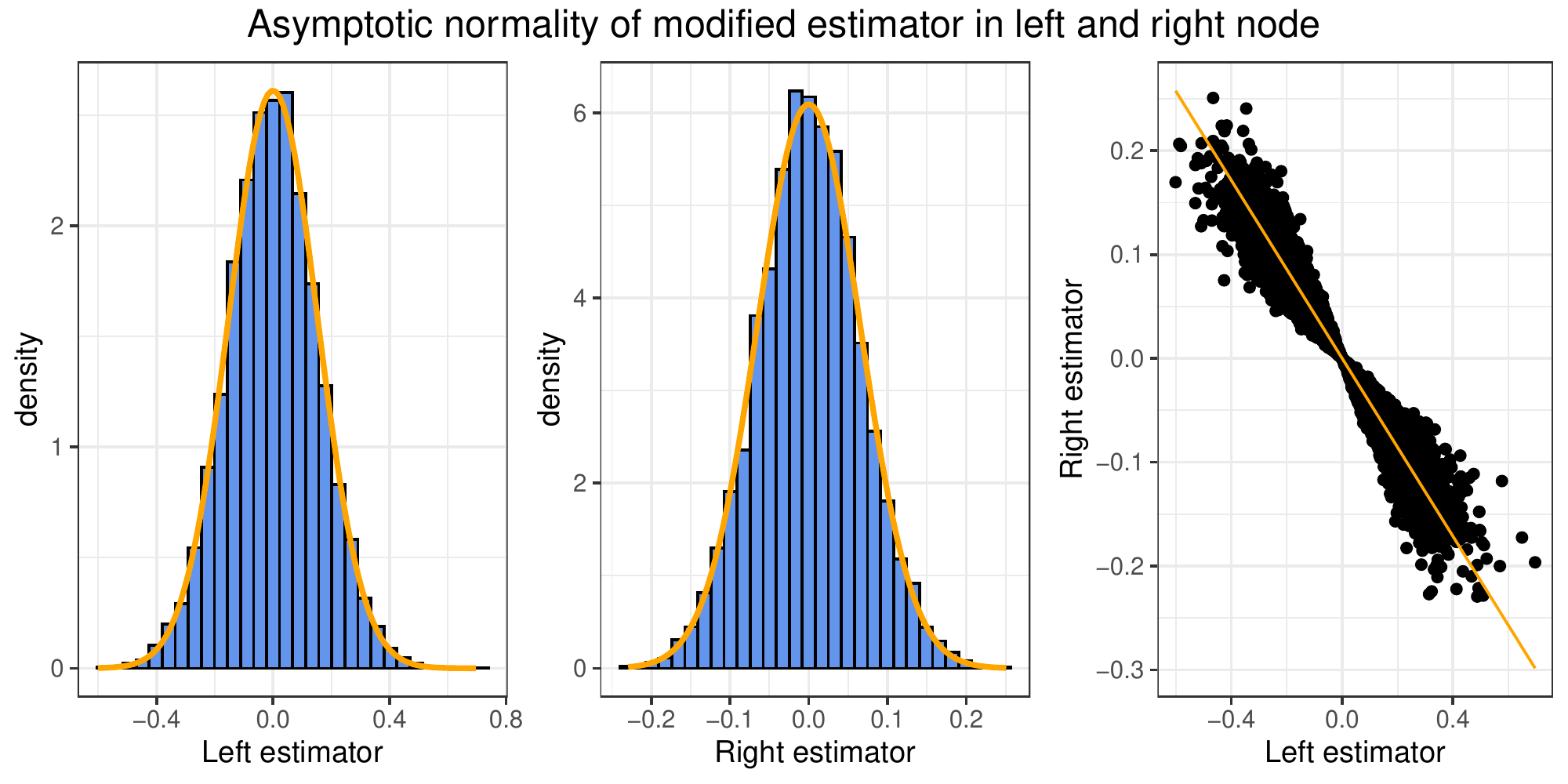} 
\rule{25em}{0.5pt}
\caption[]{
Simulation of \eqref{eq:app:joint_asymptotic_normality_of_modified_w},
where $\hat{\tilde{w}}_l$ and $\hat{\tilde{w}}_r$ are extracted from the simulation experiment explained in the caption of 
Figure \ref{fig: appendix score AN}, and comparisons with the marginal
normal distributions in \eqref{eq:app left estimator AN} and 
\eqref{eq:app right estimator AN}.
Right: Scatter plot of the simulated estimators.
Notice that the exact dependence described in \eqref{eq:app:joint_asymptotic_normality_of_modified_w}
is illustrated with an orange line, and that this is of an asymptotic nature. 
As $n=100<\infty$ for this experiment, the scatter plot emits some randomness
about the dependence line, but for higher $n$ this deviation from the
dependence line tends to zero.
}
\label{fig: appendix estimator AN}
\end{figure}
The remaining part to characterize in \eqref{eq:app:parameter asymptotic normality}
is the expected Hessian. This is rather straightforward, as the population
equivalent of \eqref{eq:app:modified loss}
might be written using indicator functions. Necessarily, the Hessian
is diagonal, expectations over indicator functions are probabilities, 
and its inverse a diagonal matrix with the reciprocal of the diagonal
elements of the expected Hessian.

The expected Hessian of the 
loss in the left node is 
$$E\left[\frac{\partial}{\partial \tilde{w}_l}\tilde{\psi}_{i,l}\right] = uE[h]$$
and the right node
$$E\left[\frac{\partial}{\partial \tilde{w}_r}\tilde{\psi}_{i,r}\right] = (1-u)E[h].$$
Further, the off-diagonal elements of the Hessian are zero.
The asymptotic distribution therefore may be characterized by
\begin{align}\label{eq:app:joint_asymptotic_normality_of_modified_w}
\sqrt{n}
\begin{pmatrix}
\hat{\tilde{w}}_l \\
\hat{\tilde{w}}_r
\end{pmatrix}
\to_p
\begin{pmatrix}
\frac{\sqrt{E\left[\psi_{i,t}(w_{t,0})^2\right]}}{uE[h]} B(u) \\
-\frac{\sqrt{E\left[\psi_{i,t}(w_{t,0})^2\right]}}{(1-u)E[h]} B(u)
\end{pmatrix}
,~
u\in[0,1].
\end{align}
Notice in particular that \eqref{eq:app:joint_asymptotic_normality_of_modified_w} implies 
the marginal limiting distributions
\begin{align}\label{eq:app left estimator AN}
\sqrt{n}\hat{\tilde{w}}_l\to_d N\left(0,\frac{1-u}{uE[h]^2}E\left[\psi_{i,t}(w_{t,0})^2\right]\right)
\end{align}
and
\begin{align}\label{eq:app right estimator AN}
\sqrt{n}\hat{\tilde{w}}_r\to_d N\left(0,\frac{u}{(1-u)E[h]^2}E\left[\psi_{i,t}(w_{t,0})^2\right]\right),
\end{align}
but \eqref{eq:app:joint_asymptotic_normality_of_modified_w} also provides 
the degenerate dependence structure of $(\hat{\tilde{w}}_l,\hat{\tilde{w}}_r)$.

\subsection{Limiting distribution of loss reduction}

\begin{figure}
\centering
\includegraphics[width=0.7\textwidth,height=7cm]{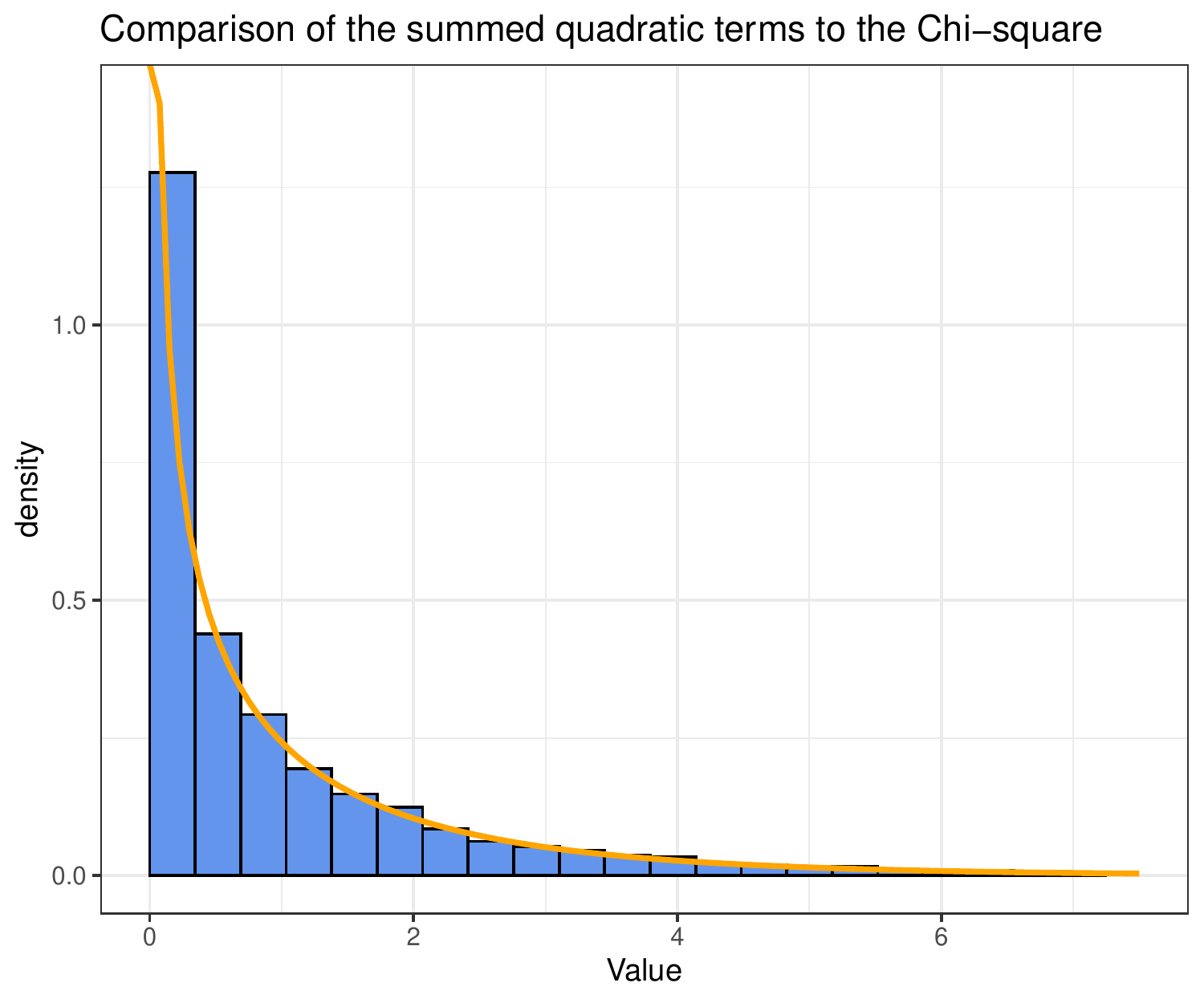} 
\rule{25em}{0.5pt}
\caption[]{
Simulation of the steps in \eqref{eq:app combining squares} using the 
squares of $\hat{\tilde{w}}$ in the simulation experiment explained in 
the caption of Figure \ref{fig: appendix score AN}, and comparison with a 
Chi-square distribution with one degree of freedom.
}
\label{fig: appendix quadratics AsymChi}
\end{figure}

Returning to taking the expectation w.r.t. $(y^0,x^0)$ of Equation (\ref{eq:diff text-train as deviance quadratics}); equipped 
with the joint distribution of $(\hat{\tilde{w}}_l, \hat{\tilde w}_r)$ \eqref{eq:app:joint_asymptotic_normality_of_modified_w}, the two terms of \eqref{eq:diff text-train as deviance quadratics} can be combined and specified in terms of
the single Brownian bridge. To see this, take expectations 
w.r.t. test data $(y^0,x^0)$, and multiply with $\frac{u(1-u)}{u(1-u)}$ to obtain a common 
denominator.
\begin{align}\label{eq:app combining squares}
R - E_{y^0,x^0}[R^0] &\approx 
E_{(y^0,x^0)}
\left[
\frac{\partial^2}{\partial \tilde{w}_l^2}
\omega^0
\hat{\tilde{w}}^2 +
\frac{\partial^2}{\partial \tilde{w}_r^2}
\omega^0
\hat{\tilde{w}}^2
\right]\notag \\
&= 
\frac{E\left[\psi_{i,t}(w_{t,0})^2\right]}{nE[h]}
\frac{\left((1-u)B(u)^2+uB(u)^2\right)}{u(1-u)}\notag \\
&=  
\frac{E\left[\psi_{i,t}(w_{t,0})^2\right]}{nE[h]}
\frac{B(u)^2}{u(1-u)}.
\end{align}
The right hand side of \eqref{eq:app combining squares} gives a convenient asymptotic 
representation of $R-E_{y^0,x^0}[R^0]$.
The subsequent section shows that 
$B(u)^2/(u(1-u))$, subject to a suitable time-transformation $\tau(u)$, $u\in(0, 1)$, is a Cox-Ingersoll-Ross process \citep{cox1985theory} 
which constitutes the right-hand side of \eqref{eq:varying optimism as cir}.
To get from \eqref{eq:app combining squares} to \eqref{eq:varying optimism as cir} 
(modulus the time-transformation) first observe that
\begin{align}
\hat C_{root} 
=E[h]Var(\hat w_t)
= E[h] \left( \frac{E\left[\psi_{i,t}(w_{t,0})^2\right]}{n E[h]^2} \right)
= \frac{E\left[\psi_{i,t}(w_{t,0})^2\right]}{nE[h]},
\end{align}
and thus simply adding the root optimism on both sides of \eqref{eq:app combining squares} gives
\begin{align}
R - E_{y^0,x^0}[R^0] + \hat C_{root} 
&= 
\left[E_{y^0,x^0}\left[\hat{\loss}(y^0, f(\mathbf{x}^0; \hat{w}_l(u),\hat{w}_r(u)))\right]
- \hat{\loss}(y, f(\mathbf{x}; \hat{w}_l(u),\hat{w}_r(u)))\right] \notag \\
&\approx
\hat C_{root} 
\left( 1 + 
\frac{B(u)^2}{u(1-u)}
\right).
\end{align}
The final step of the calculations leading to \eqref{eq:varying optimism as cir} is to show that $B(u)^2/u(1-u)$ is indeed equivalent to the CIR process \eqref{eq: cir process sde, param (2,1,2)}.

\subsection{The process $B(u)^2/(u(1-u))$ is a time-transformed CIR process}
\begin{figure}
\centering
\includegraphics[width=1\textwidth,height=5cm]{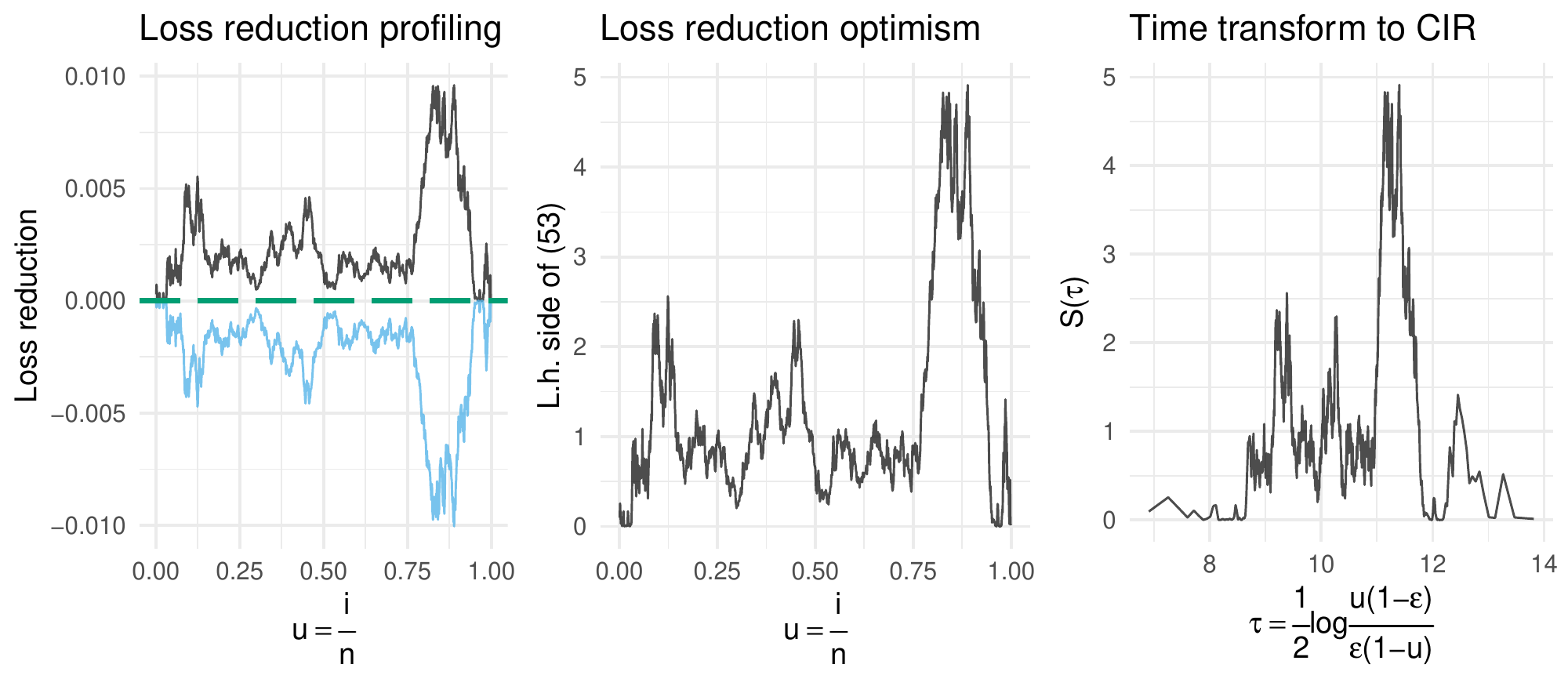} 
\rule{25em}{0.5pt}
\caption[]{The equivalent and same process as in Figure 
\ref{fig: split sequence to cir}, for loss reduction $R$ and $R^0$.
Left: Reduction in training loss $R$ (black) and reduction in 
generalization loss $E_{y^0,x^0}\left[R^0\right]$ (blue) as a function
of $u=\frac{i}{n}$, defined from the sorted order of $x_j$.
Green long-dashed line is the expected loss-value at $\theta_0=\lim_{n\to\infty}\hat{\theta}_n$,
constant and zero as there is no information in $x_j$ for this instance.
Right plot: The transformation of distance between generalization loss
and training loss into a CIR process. In this case with no information
in feature $x_j$, choosing the value of $s$ giving the smallest value
of training loss in the left plot induces an optimism at the value of
the expected maximum of the CIR-process in the right plot. }
\label{fig: split_reduction_sequence to cir}
\end{figure}
It was previously mentioned, and used in notation, that $B(u)^2/(u(1-u))$ is a 
CIR process over time $\tau(u)$, where $u\in(0,1)$.
Note that the interval is $u\in(0, 1)$ and not $[0,1]$ as for the functional 
convergence of $\psi$. This is due to the denominator in $B^2/(u(1-u))$ which almost certainly blows up the value at the endpoints. For 
this reasons, \citet{miller1982maximally} approximates the search over $[0,1]$ 
by $(\epsilon,1-\epsilon)$ for $\epsilon n>1$ and $(1-\epsilon)n>1$, which is of 
little practical importance, as it makes sense to at least have a few 
observations when estimating each leaf-weight. \citet{gombay1990asymptotic} 
relaxes this assumption, and shows that the supremum of $B^2/(u(1-u))$ over $(0,1)$
asymptotically have a Gumbel distribution. This result is in alignment with the 
use the Gumbel distribution in the simulation approach discussed in Section 
\ref{subsec: evaluation and implementation}.

We show that the sum of the scaled-squared Brownian bridge is a 
Cox-Ingersoll-Ross process.
As this paper eventually takes a simulation approach to obtain the distribution 
of $\max Y(\tau(u)),~ u\in\{u_1,\cdots,u_a\}$, the exact same results would be 
obtained by simulating $\max \frac{B^2}{u(1-u)},~u\in\{u_1,\cdots,u_a\}$. 
Here $u\in\{u_1,\cdots,u_a\}$ are the time-points and probabilities on $(0,1)$, $p(x\leq s)$, for which we observe the process.
The specification of the scaled-squared Brownian bridge, through a 
time-transform, as a CIR is therefore not strictly necessary.
However, for completeness, and for the purpose/benefit of working with a 
time-homogenous stationary process that is well known and studied, we show that 
this is indeed the case.
Important is also the CIR's stationary Gamma distribution, which implies that the 
CIR is in the maximum domain of attraction of the Gumbel distribution, and 
warrants its use as an asymptotic approximation to supremums of the CIR.

\citet{anderson1952asymptotic} shows that 
\begin{align}\label{eq:app:bridge to ou}
\frac{|B(u)|}{\sqrt{u(1-u)}} = |U(\tau(u))|,~
\tau(u) = \frac{1}{2}\log \frac{u(1-\epsilon)}{\epsilon(1-u)}.
\end{align}
where $U(\tau)$ is an Ornstein-Uhlenbeck process which solves the 
stochastic differential equation 
\begin{align}
dU(\tau) =  -U(\tau)d\tau + \sqrt{2} dW(\tau).
\end{align}
Notice that \eqref{eq:app:bridge to ou} is the square root of 
$B(u)^2/(u(1-u))$ 
appearing in right-hand side of \eqref{eq:app combining squares}.
Hence, obtaining a stochastic differential equation for 
$B(u)^2/(u(1-u))$ 
simply amounts to applying Ito's Lemma \citep{oksendal2003stochastic}
to obtain the stochastic differential equation for the square of $U(\tau)$.
More precisely, define
$S(\tau) = U(\tau)^2,$ which gives the stochastic differential equation
given in Equation \eqref{eq: cir process sde, param (2,1,2)}, namely
\begin{align}
dS(\tau) = 2\left(1-S(\tau)\right)d\tau+2\sqrt{2S(\tau)}dW(\tau).
\end{align}
This is recognized as a Cox-Ingersoll-Ross (CIR) process \citep{cox1985theory}, 
with speed of adjustment to the mean $a=2$, long-term mean $b=1$, and 
instantaneous rate of volatility $2\sqrt{2}$.

The profiling over loss reduction $R$ for different split points $s$, the optimism, and the time-transform to the CIR process is illustrated in Figure \ref{fig: split_reduction_sequence to cir}.
This is the same experiment as in Figure \ref{fig: split sequence to cir}, but with loss reduction profiling instead of stump-loss profiling.

\section{Maximal CIR as a bound on optimism}
\label{app:max cir as a bound on optimism}
Section \ref{app:regularity conditions} shows that $R-E[R^0]$ behaves
asymptotically as a CIR process, $S(\tau)$, when profiling over a continuous feature. 
It immediately follows that a bound
on this optimism is given as the expected maximal element of the CIR process
\begin{align}\label{eq:app:optimism cir bound}
E[R-R^0] \leq \hat C_{root} E[\max_{u}S(\tau(u))].
\end{align}
If we are comparing maximum reductions of multiple features, then we would
instead be interested in the distribution, $p(\max S(\tau)\leq s)$, for 
its use in Equation \eqref{eq:final approximate stump optimism}, which
reduces to the equation above when $m=1$.

However, more can be said, namely that this bound is tight when the feature being profiled over is independent of the response.
To see this, Taylor expand $\omega_i$ about its estimate $\hat{\tilde{w}}$ and again make use of the approximation in \eqref{eq:quadratics independence assumption}
\begin{align}
0=\frac{1}{n}\sum_{i=1}^{n} \left(\tilde{g}_i+h_i\tilde{w}_0\right)
\approx
\left[\frac{1}{n}\sum_{i=1}^{n}\omega_i\right] + 
\frac{1}{2}E[h]\left(u\hat{\tilde w}_l^2 + (1-u)\hat{\tilde w}_r^2\right) 
\end{align}
since both $\tilde{w}_0$ and $\sum_{i=1}^n \tilde{g}_i$ are zero.
Rearranging, we may re-express the training loss-reduction 
\begin{align}
R = -\frac{1}{n} \sum_{i=1}^n \omega_i 
\approx \frac{1}{2}E[h]\left(u\hat{\tilde w}_l^2 + (1-u)\hat{\tilde w}_r^2\right). 
\end{align}
By recognizing that the term on the right is exactly half the value of \eqref{eq:app combining squares}, it is evident that maximizing $R$  corresponds to selecting split-point and leaf-weights that 
are at the time-point where the CIR process, $S(\tau(u))$, attains its maximum. 
Consequently, we obtain equality in Equation \eqref{eq:app:optimism cir bound}, i.e. 
\begin{align}
E[\hat{R}^0] = E[R]-\frac{E\left[\tilde{\psi}_{i,t}(\tilde{w}_{t,0})^2\right]}{nE[h]}E\left[\max_u S(\tau(u))\right].
\end{align}
Finally, notice that $E[R-R^0]$ might also be expressed in terms of the 
optimism of the root and stumps models, so that $E[R-R^0]=\hat{C}_{stump}-\hat{C}_{root}$.
Thus, rearranging, we immediately obtain the pure stump optimism,
expressed as an adjustment of the root optimism
\begin{align}
\hat{C}_{stump}
= E[R-R^0]+\hat{C}_{root}
=\frac{E\left[\tilde{\psi}_{i,t}(\tilde{w}_{t,0})^2\right]}{nE[h]}\left(1+E\left[\max_u S(\tau(u))\right] \right).
\end{align}

A check: In likelihood theory, we would expect one additional degree of freedom, 
thus $R-R^0=1$. Indeed, if we take the final expectation w.r.t. the training 
data, we have $E[B^2]=u(1-u)$, multiply with $n$ to obtain a log-likelihood, and 
assume the expected Hessian equals the variance of the score, then this indeed 
reduces to exactly 1.